\documentclass[10pt,journal,compsoc]{IEEEtran}
\ifCLASSOPTIONcompsoc
  \usepackage[compress]{cite}
\else
  \usepackage{cite}
\fi
\ifCLASSINFOpdf
\else
\fi
\usepackage{graphicx}
\usepackage{multirow}
\usepackage{framed}
\usepackage[ruled,vlined]{algorithm2e}
\usepackage{color}

\usepackage{amsmath,amssymb}
\usepackage{setspace}
\usepackage{makecell}
\usepackage{amsthm}
\usepackage{color}
\usepackage{booktabs}
\usepackage{balance}
\usepackage{array}
\usepackage{flushend}
\usepackage{balance}
\usepackage{caption}
\usepackage{subcaption}
\usepackage{float} 
\usepackage{url}
\usepackage{soul}
\usepackage{xcolor}
\usepackage{soul}
\usepackage{balance}
\usepackage{setspace}

\usepackage{booktabs}
\usepackage{makecell}

\usepackage{enumitem}%
\usepackage{color}
\usepackage{xspace}

\setlist[itemize]{noitemsep, topsep=0pt}

\newcommand{\CkNN}{C{\em k}NN}
\newtheorem{example}{Example}[section]
\newtheorem{myDef}{Definition}
\newtheorem{theorem}{Theorem}[section]

\newcommand\vldbdoi{XX.XX/XXX.XX}

\definecolor{shadecolor}{rgb}{0.92,0.92,0.92}
\definecolor{Firebrick4}{RGB}{255, 0, 0}

\AtBeginDocument{%
  \providecommand\BibTeX{{%
    \normalfont B\kern-0.5em{\scshape i\kern-0.25em b}\kern-0.8em\TeX}}}

\begin{document}
%
\title{ODIN: Object Density Aware Index for {C$k$NN Queries} over Moving Objects {on Road Networks}}

\IEEEtitleabstractindextext{%

\author{\IEEEauthorblockN{{Ziqiang Yu}$^1$, Xiaohui Yu$^{2*}$\thanks{*Corresponding author}, Tao Zhou$^3$, Yueting Chen$^2$, Yang Liu$^4$, Bohan Li$^5$} \\
\IEEEauthorblockA{\small
$^1$Yantai University 
\hspace{0.2em}$^2$York University
\hspace{0.2em}$^3$University of Science and \\ Technology of China
\hspace{0.2em}$^4$Wilfrid Laurier University
\hspace{0.2em}$^5$Nanjing University of Aeronautics and Astronautics}
}

\begin{abstract}
We study the problem of processing continuous $k$ nearest neighbor ({C$k$NN}) queries over moving objects on road networks, which is an essential operation in a variety of applications. We are particularly concerned with scenarios where the object densities in different parts of the road network evolve over time as the objects move. Existing methods on C$k$NN query processing are ill-suited for such scenarios as they utilize index structures with fixed granularities and are thus unable to keep up with the evolving object densities. In this paper, we directly address this problem and propose {an object density aware index structure called ODIN that is an elastic tree built on a hierarchical partitioning of the road network.} It is equipped with the unique capability of dynamically folding/unfolding its nodes, thereby adapting to varying object densities. We further present the {ODIN-KNN-Init} and {ODIN-KNN-Inc} algorithms for the initial identification of the $k$NNs and the incremental update of query result as objects move. Thorough experiments on both real and synthetic datasets confirm the superiority of our proposal over several baseline methods.
\end{abstract}

}
\maketitle


\vspace{-0.2in}
\section{Introduction}

Given a road network and a set of objects, we study the processing of continuous {\em k} nearest neighbor ({\CkNN}) queries, where a {\CkNN}  query is to find and continuously monitor the $k$ closest objects (as measured by road network distances) to a fixed query point $q$ as the objects move along the road network. Processing {\CkNN} queries on the road network is central to many location-based services~\cite{mouratidis2006continuous, luo2018toain, luo2019mpr, he2019efficient,DBLP:conf/icde/ZhengHJCHLLZ20, DBLP:journals/tkde/ZhengZJHSLZ20,DBLP:journals/pvldb/ZhengBCCF0G0J21,fang2023ghost,DBLP:conf/icde/TongSDWC16} such as ride-hailing, POI discovery, and location-based games (e.g., $Pok\acute{e}mon\;GO$). 

Considering the paramount importance of query processing efficiency in such applications, we focus on main-memory-based solutions in this work. Moreover, we assume that the continuous monitoring of $k$NNs is based on snapshots of object positions taken at a regular interval, {which is} a commonly adopted semantics in the literature~\cite{shen2017v,cao2018sf}. 
 
The problem of $k$NN query processing on road networks has been extensively studied~\cite{ cao2018sf, abeywickrama2016k,demiryurek2009efficient, papadias2003query, mouratidis2006continuous, huang2005islands, huang2007s, sankaranarayanan2005efficient, shen2017v,lee2010road, li2018gpu, he2019efficient, luo2019mpr, luo2018toain, ouyang2020progressive, 10.1155/2021/2050489,li2020efficient}, and a common methodology to find $k$NNs based on a snapshot of object positions is to gradually expand the search scope along the road network until the $k$NNs are found. {When conducting $k$NN search over moving objects, the number of objects in a region (i.e., the {\em object density}{, OD for short}) is a major factor impacting the search efficiency as the search scope is inversely dependent on OD for a given $k$. That is, it is more likely for the search algorithm to be able to find the $k$NNs with a smaller search scope when the query point falls in a denser region than when it belongs to a sparse region.  

Unfortunately, most existing approaches fail to fully consider the impact of {object densities (ODs for short)}, especially when they evolve as the objects move, appear, or disappear over time. In particular, some methods directly explore $k$NNs along the road network with the Incremental Network Expansion (INE for short) strategy \cite{papadias2003query, huang2005islands, mouratidis2006continuous, huang2007s}. They take the query point as the starting point and incrementally expand the search scope using the principle of relaxation similar to the underlying mechanism of Dijkstra's Algorithm. {While such a strategy works well when the objects are more or less evenly distributed over the road network, its performance deteriorates quickly when expansion to sparse regions is needed~\cite{lee2010road, zhong2013g}}.
{ To address this issue, some hierarchical indexes are proposed based on multi-level partitioning of the road network~\cite{lee2010road, zhong2013g, shen2017v}, with the index granularity expected to match OD in each part of the network. Nevertheless, while the OD is constantly evolving as objects move, the layout of the index (including the number of levels and the index granularity at each level) in those hierarchical indexes are fixed. The mismatch between the fixed layout and the evolving OD makes it very difficult, if not impossible, to choose an appropriate index layout with high pruning power across a given network. }

{In this paper, we propose an index called ODIN (for \underline{O}bject \underline{D}ensity Aware \underline{IN}dex) and its associated query processing algorithms ODIN-KNN-Init and ODIN-KNN-Inc to tackle the aforementioned issue}. The main idea is to (1) index the road network and objects in a way that supports highly efficient dynamic update as objects move and {adapts to the evolving ODs with an elastic structure}, and (2) allow the reuse of selected computation results from one snapshot to the next to achieve substantial cost savings over overhaul computation at each snapshot. 

{ODIN} is a tree structure based on hierarchical non-overlapping partitioning of the road network, which can be considered a graph $G$ if we deem each road junction a vertex and a road segment between two junctions an edge. Each node is responsible for indexing a sub-graph of $G$, and the information maintained in a parent node (including selected vertices and their associated objects, etc.) has a coarser granularity than that in its child nodes. 

A salient feature that sets {ODIN} apart from existing hierarchical indexes for {\CkNN} query processing is its elasticity: the tree is unbalanced by design to take into consideration the different ODs in different parts of the road network, and nodes in the tree can be dynamically {\em folded}/{\em unfolded} to adapt to changing OD in part of the network as objects move. When child nodes are folded into their parent caused by decreased OD, subsequent $k$NN query processing can be conducted using the parent node only without touching the child nodes. This reduces the number of nodes accessed and thus improves the query efficiency. Conversely, when a parent node is unfolded due to an increased density, its child nodes {with finer index granularity} will be utilized in subsequent query processing instead, offering higher pruning power than using the parent node.

Based on {ODIN}, we propose the {ODIN-KNN-Init} algorithm for the initial search of $k$NNs for a given query $q$, and the {ODIN-KNN-Inc} algorithm for the incremental update of the $k$NNs from one snapshot to the next. Since critical distance information between selected important vertices is kept in {ODIN}, {ODIN-KNN-Init} is able to find the initial $k$NNs with high efficiency. As for the {ODIN-KNN-Inc} algorithm, a key insight is that since (1) the $k$NNs are still likely to be in the same or similar neighborhood of $q$ as that in the previous snapshot and (2) the shortest distance between vertices does not change over time, we are able to reuse a large portion of the distance information carried over from the previous snapshot. However, the incremental $k$NN search on {ODIN} is more challenging than that on static indexes such as ER-{\em k}NN~\cite{demiryurek2009efficient}, as the elastic structure of {ODIN} makes the explored region prone to change from one snapshot to the next, increasing the difficulty to reuse the computed distance information for incremental evaluation. {We propose methods to tackle this challenge and support the incremental evaluation of $k$NNs with minimal overhead.}

In summary, we make the following contributions.

(1) We propose {ODIN, an OD aware index structure} that dynamically adjusts the indexing granularity based on the varying ODs in different parts of the road network, facilitating the efficient exploration of vertices for $k$NN search. {To the best of our knowledge, this is the first index structure specifically designed to support {\CkNN} query processing over moving objects in a road network with varying ODs.} 

(2) We present the {ODIN-KNN-Init} and the {ODIN-KNN-Inc} algorithms for the initial and incremental search of $k$NNs utilizing {ODIN}. Through the reuse of information between consecutive rounds of query evaluation, the monitoring of $k$NN results can be performed efficiently without recomputing the $k$NNs at each snapshot.

(3) Extensive experiments are conducted on both synthetic and real datasets to fully evaluate the performance of our proposal. Compared with the state-of-the-art method Ten$^*$-Index~\cite{ouyang2020progressive}, {our proposal offers up to 1000$\times$ speedup in query processing. }

The rest of the paper is organized as follows. Section \ref{sec:related-work} discusses the related work. Section \ref{sec:preliminaries} presents some important preliminaries. Section \ref{sec:AMT-Index} describes the {ODIN} index, and Section~\ref{sec:AMT-KNN} discusses the query processing algorithms based on ODIN. Section~\ref{sec:exp} presents the results of our experimental evaluation, and Section~\ref{sec:con} concludes this paper.

\section{Related work}\label{sec:related-work}
 {The search for C$k$NNs over moving objects in Euclidean space has garnered significant interest~\cite{gao2011continuous, 10.1145/342009.335427,yu2005monitoring,6930803,5153337,zhu2020processing,7498339}. Some approaches have leveraged spatial-temporal indexes, such as R-tree-based~\cite{gao2011continuous, 10.1145/342009.335427}, Grid-based~\cite{yu2005monitoring, 6930803}, and Voronoi-based indexes~\cite{5153337, zhu2020processing}, to enhance $k$NN search efficiency by pruning the search space. Extensive research has also gone into incremental updates of $k$NNs using safe region techniques~\cite{7498339}, avoiding recomputation from scratch. However, these methods are not suitable for $k$NN search on road networks due to the difference in distance measures. Thus, numerous works have examined $k$NN search with road network constraints~\cite{cao2018sf, abeywickrama2016k, papadias2003query, huang2005islands, sankaranarayanan2005efficient, huang2007s,demiryurek2009efficient,shen2017v,lee2010road, li2018gpu, he2019efficient, luo2019mpr, luo2018toain, li2020efficient, ouyang2020progressive,10.1155/2021/2050489}. Yet, such existing methods are mostly insensitive to the evolving ODs that significantly impact $k$NN search efficiency.} 


{\bf Non-hierarchical indexes.} Some existing work conducts $k$NN search directly on the original road network or through flat/non-hierarchical indexes. A common problem with this category of approaches is the lack of differentiated expansion granularities during the exploration of $k$NNs to adapt to the distinct ODs in different parts of the network. 

(1) {\em INE-based approaches.} {INE-based approaches such as  INE~\cite{papadias2003query}, Islands \cite{huang2005islands}, IMA~\cite{mouratidis2006continuous}, and S-Grid~\cite{huang2007s}} {perform network expansion with Dijkstra's algorithm starting from the query point and examine vertices in the order they are encountered until the $k$NNs are identified.} While these approaches perform well in parts of the road network densely populated with objects,  
their performance deteriorates significantly in sparse regions with very few objects as they tend to conduct extensive expansions on many vertices irrelevant to finding $k$NNs. 

{(2) {\em Grid-based indexes.} Some methods (e.g., IER~\cite{wang2007location}, ER-$k$NN~\cite{demiryurek2009efficient}, SIM$k$NN~\cite{cao2018sf}, GLAD~\cite{he2019efficient} and G-Grid~\cite{li2018gpu}) 
partition the road network into non-overlapping grid cells according to the latitude and the longitude of vertices~\cite{he2019efficient}, which can help limit the search region by pruning those cells that are guaranteed not to cover $k$NNs. However, after the pruning, such methods still need to compute the network distances from the potential objects to the query point by exploring part of the road network covered by the remaining cells. Similar to INE-based approaches, the exploration suffers from differing ODs across the road network. }

(3) {\em Indexing nearest neighbors.} Some work such as INSQ~\cite{7498339}, TOAIN~\cite{luo2018toain}, and TEN$^*$-Index~\cite{ouyang2020progressive}, indexes the {local} $k$NNs of selected vertices to enhance search efficiency, but their applicability is limited as they are built for a specific $k$ and thus require the value of $k$ a priori. {Moreover,  these work needs to traverse the vertices based on the index to identify the final results among the indexed $k$NNs of the visited vertices. As the index construction only considered the topology of the road network but not the OD,  the traversal based on the index cannot adapt to the various ODs across the road network.}



{\bf Hierarchical indexes.} In order to reduce the impact of {the distinct ODs} on the query efficiency, hierarchical indexes on road networks such as SILC~\cite{sankaranarayanan2005efficient}, ROAD~\cite{lee2010road}, G-tree~\cite{zhong2013g},  V-tree~\cite{shen2017v}, and G$^{*}$-tree~\cite{8731380} are proposed. These methods typically partition the entire road network into hierarchical sub-networks, each serving as an index unit. Different levels of sub-networks offer varying index granularities to accommodate the distinct ODs found across the road network. 

However, such hierarchical indexes are far from ideal. First, the partitioning is done uniformly with the same granularity across the entire road network (for a given level) without consideration to the different ODs in different parts of the network. It is challenging for this ``one-size-fits-all'' approach to provide high pruning efficiency. Second, index granularity is fixed once the index is constructed, but the ODs in parts of road network may evolve as objects move over time. As such, the mismatch between the fixed index granularity and varying ODs is likely to happen, leading to deterioration in the pruning efficiency of the indexes over time. Finally, propagating the object updates through each level of the hierarchy in such indexes may lead to high maintenance cost~\cite{li2018gpu}, again as such update procedures are agnostic to different ODs in different regions.

\vspace{-0.1in}
\section{Preliminaries}\label{sec:preliminaries}
We model the road network as a weighted graph $G$($\mathcal{V}$, $\mathcal{E}$, $\mathcal{W}$), where a vertex $v_i\in \mathcal{V}$ represents a road junction, an edge $e_{i,j}\in\mathcal{E}$ denotes a road segment between junctions $v_i$ and $v_j$, and the weight $w_{i,j}\in\mathcal{W}$ represents the length of road segment (edge) $e_{i,j}$. In what follows, we consider \textit{road network} and \textit{graph} exchangeable unless otherwise specified.
\vspace{-4pt}
\begin{myDef}
[Path] Path $P(s, t)$ from one vertex $v_s$  (the source vertex) to another vertex $v_t$ (the destination vertex) in graph $G$ is a sequence of vertices $\langle \texttt{v}_0$= $v_s$,$\cdots$ $\texttt{v}_l$,$\cdots$,$\texttt{v}_n$ =$v_t\rangle$ such that $\forall l\in [1,n]$, $e_{i,j}\in E$ if $\texttt{v}_{l-1}=v_i$ and $\texttt{v}_l=v_j$. The path distance of $P(s, t)$ is defined as $D(P(s,t))$=$\sum_{l=1}^{l<n}w(v_l,v_{l+1})$.  
\end{myDef}

\begin{myDef}
[The shortest distance] For any two vertices $v_s$ and $v_t$ in a graph $G$, the shortest distance from $v_s$ to $v_t$ is the minimal path distance between $v_s$ and $v_t$, denoted by $SD(v_s, v_t)$.
\end{myDef}

The {ODIN} index is built upon a hierarchical partition of the graph $G$, which is defined as follows.
{
\vspace{-4pt}
\begin{myDef}
[Hierarchical partition] Given a graph $G$, and parameters $m$ and $z$, the {hierarchical} partition of $G$ is obtained using the following procedure:

(1) if graph $G (\mathcal{V}$, $\mathcal{E}$, $\mathcal{W})$ has more than $z$ vertices, it is evenly partitioned into $m$ subgraphs $G_{1} (\mathcal{V}_{1}$, $\mathcal{E}_{1}$, $\mathcal{W}_{1})$, $\cdots$, $G_{m}(\mathcal{V}_{m}$, $\mathcal{E}_{m}$, $\mathcal{W}_{m})$ in the sense that each subgraph {has approximately the same number} of vertices, where $\mathcal{V}$=$\bigcup_{i=1}^m \mathcal{V}_{i}$, $\mathcal{V}_{i}\cap \mathcal{V}_{j}=\phi$ ($i\neq j, j\in[1,m]$), and $\forall e_{x,y}\in \mathcal{E}_{i}$, $e_{x,y}\in \mathcal{E}$. $G$ is called a supergraph of $G_{i}$ ($i\in[1,m]$).

(2) for any subgraph $G_{i}$ with more than $z$ vertices, it is further partitioned into $m$ {subgraphs in the same way that $G$ is partitioned in (1)}. Th partition is recursively applied until no subgraph has more than $z$ vertices.
\end{myDef}\label{def:par}
}

We use the METIS algorithm~\cite{METIS} to conduct graph partition due to its good performance, but any applicable algorithm that can partition a graph evenly may be employed.

The partitioned subgraphs are connected by {\em external edges} between {\em border vertices}, which are described as follows. 

{\bf Border vertex.} A vertex in a subgraph $G_i$ of $G$ is a border vertex if it has at least one adjacent vertex in $G$ that belongs to at least one other subgraph $G_j$ $(j\neq i)$ of $G$.

{\bf External edge.} An edge $e$ ($e\in\mathcal{E}$) is an external edge if it connects two border vertices in  two different subgraphs. \looseness=-1
 

Evidently, border vertices are the entrances of a subgraph in the sense that the shortest path from a vertex in a subgraph $G_i$ to a non-border vertex in a subgraph $G_j$ ($i\neq j$) must pass through one of the border vertices in $G_j$.

{
Moreover, we introduce the notion of {\em live vertex} to ease the computation of the shortest distances between objects and queries.}

{{\bf Live vertex.} When an object $o$ is moving to the vertex $v_i$ along an edge, $v_i$ is a live vertex and $o$ is an associated object of $v_i$. Each live vertex records all of its associated objects.}

{Similar to existing work~\cite{shen2017v}, we also assume that an object has to reach its corresponding live vertex before moving to other vertices. As such, the shortest distance from a vertex $v_d$ to a moving object $o$ associated to $v_i$ ($v_i\neq v_d$) can be represented as $SD(v_d, o)$=$\delta(v_d, v_i)+SD(v_i, o)$, where $\delta(o,v_i)$ is the distance from $o$ to $v_i$.
}




\vspace{-5pt}
\begin{myDef}
[$k$NN query] Given a graph $G$($\mathcal{V}$, $\mathcal{E}$, $\mathcal{W}$), a set of objects $\mathcal{O}$ ($|\mathcal{O}|>k$), and a $k$NN query $q$ ($v_q$, $k$), $k$NNs of $q$ refer to a set of objects $\mathcal{R}(v_q)$ such that 1) $|\mathcal{R}(v_q)|=k$; 2) $\mathcal{R}(v_q)\subseteq \mathcal{O}$; and 3) $\forall o\in \mathcal{R}(v_q)$, ${o'}\in \mathcal{O}/\mathcal{R}(v_q)$, $SD(v_q, o)\leq SD(v_q, o')$.
\label{def:knn-query}
\vspace{-0.1in}
\end{myDef}

In Definition~\ref{def:knn-query}, $v_q$ denotes the query point of $q$. Without loss of generality and in line with existing work such as V-tree \cite{shen2017v} and Ten$^*$-Index \cite{ouyang2020progressive}, we assume that each query point is located at a vertex of the graph representing the road network, which is called the {\em query vertex}. {Cases where the query point $q$ does not fall on a vertex can be easily converted by adding to all shortest distance calculations the distance from $q$ to the immediately adjacent vertex $q$ is currently moving towards.}  

{\bf Query semantics.} We suppose the objects move along graph edges and periodically report their locations with a time interval of $\Delta t$. Correspondingly, we update the snapshot of all objects in the graph per $\Delta t$ {and the live vertices associated with these objects based on their moving direction, determined from their locations at the last and current snapshots.} If the latest snapshot of objects is generated at timestamp $t_s$, all queries received within the time interval ($t_s$, $t_s+\Delta t$] are initially evaluated using this snapshot. Subsequently, the $k$NNs for each query are periodically re-evaluated with each new snapshot created every $\Delta t$, and each evaluation constitutes a search round. Queries following this protocol are termed Continuous $k$ Nearest Neighbor (\CkNN) queries. Whenever a \CkNN query $q$ changes its location, it is treated as a new query.

\section{The ODIN index}\label{sec:AMT-Index}




\subsection{Basic idea}\label{subsec:baic idea}
{ODIN is a OD aware elastic tree structure based on the hierarchical partition of the road network as per Definition~\ref{def:par}}. 
Its following desirable properties enable it to provide strong pruning capability in the presence of evolving ODs.

{\bf (1) Adaptive index granularities.} In {ODIN}, the number of index levels as well as the index units at each level are continuously adjusted to better suit the evolving OD in each part of the road network to enable high pruning efficiency.

{\bf (2) Efficient access to objects.} Objects are associated to live vertices that constantly appear and disappear as objects evolve. To avoid visiting many irrelevant vertices, {ODIN} materializes in real-time the ``shortcuts'' between the live and the {border vertices} within each index unit, facilitating the efficient access to live vertices and their associated objects when the
index units are explored. 

{\bf (3) Minimizing the maintenance cost.} Maintaining {ODIN} involves propagating the index information update through levels of the hierarchy as objects evolve, which depends heavily on the number of index levels. {ODIN} maintains only the essential index levels adaptive to the evolving ODs in different parts of the road network and deactivates the redundant levels to minimize the maintenance cost.

\vspace{-0.3cm}
\subsection{Preprocessing}\label{sec:preprocessing}
Prior to building {ODIN}, we compute a hierarchical partition of the road network graph $G$ as described in Definition~\ref{def:par}, which forms the basis for subsequent index construction. 
{While performing the hierarchical partitioning, we keep track of the parent-child relationships between partitioned subgraphs on different levels utilizing an M-ary tree. More specifically, the original graph corresponds to the root of the M-ary tree and the partitioned subgraphs correspond to nodes of the tree at different levels. {A node $n_c$ is a child of another node $n_p$ if the subgraph corresponding to $n_c$ is obtained from partitioning the (sub)graph corresponding to $n_p$}. Each node (except for the root) in the M-ary tree has a reference to its parent node. Note that the construction of {ODIN} utilizes only the parent-child relationships of the partitioned subgraphs as well as the the subgraphs corresponding to the leaves. {Thus, there is no information materialized at each non-leaf node other than the parent-child relationships.} For the leaf nodes, however, we also store the partitioned subgraphs as well as the external edges connecting those subgraphs.} In the subgraph associated with each leaf node, the shortest distances between any pair of vertices are precomputed {and stored}. Moreover, each vertex keeps the identifier of the leaf node it belongs to, which facilitates the identification of the affiliated leaf node for any vertex. The M-ary tree thus has obtained all the information we need to construct {ODIN}.%

\begin{example}
\vspace{-0.1in}
Fig.~\ref{fig:graph-partition} shows a graph that is recursively partitioned into subgraphs with $m$ and $z$ being 4 and 15 respectively. The M-ary tree corresponding to the graph partitioning is shown in Fig.~\ref{fig:m-ary-tree}, where {each dotted arrow denotes the reference from the corresponding node to its parent node.}
\vspace{-0.05in}
\end{example}
{

}

\subsection{Construction of ODIN}\label{sec:AHP-Index-construction}


{

    


\begin{figure}[ht!]
  \centering
    \begin{subfigure}{0.35\textwidth}
      \centering         \includegraphics[width=\linewidth]{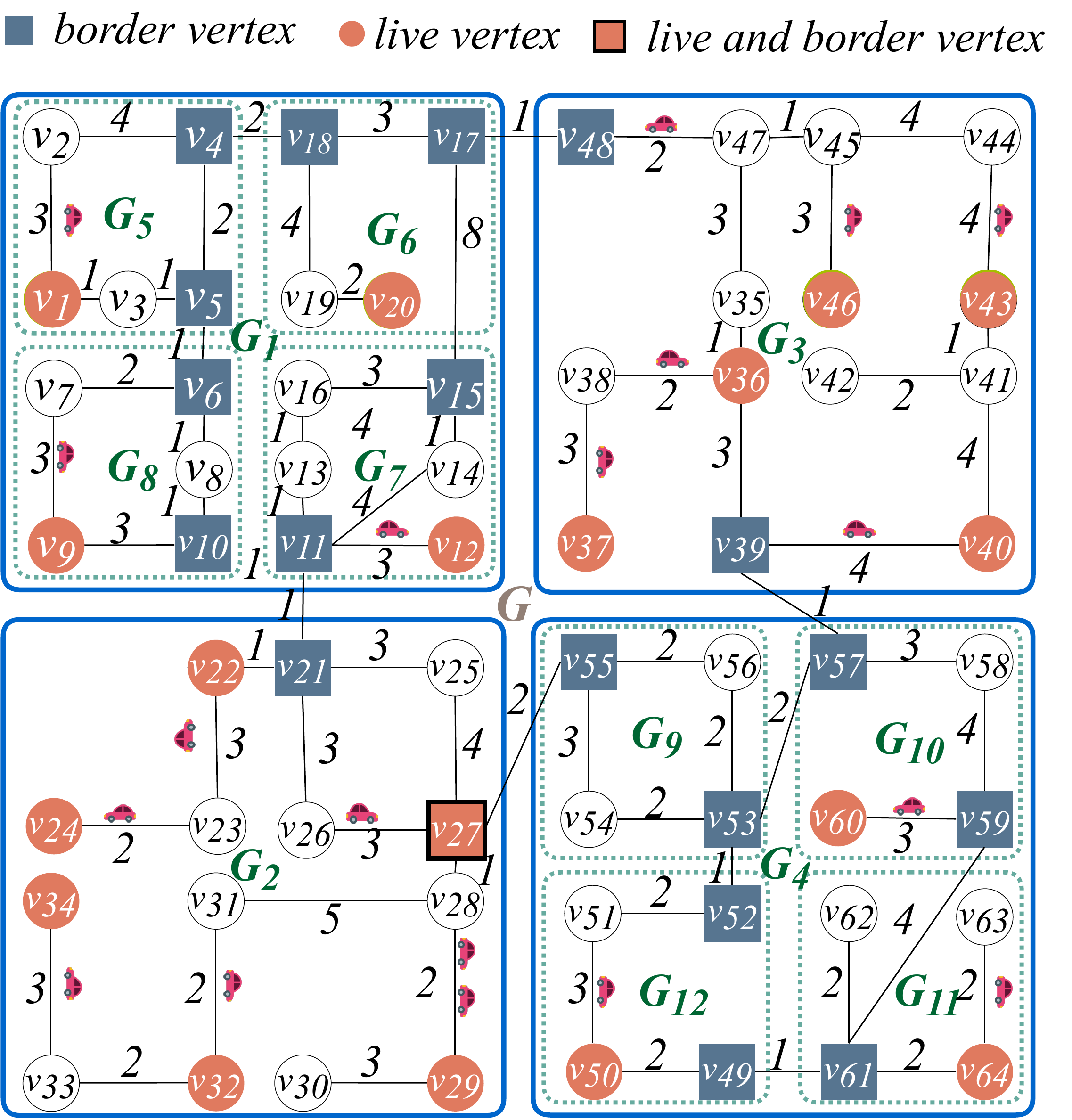}
      \captionsetup{width=3in}
\caption{\scriptsize{{Hierarchical Graph Partitioning} ($z=15$). Graph $G$ is partitioned into $G_1$, $G_2$, $G_3$, and $G_4$ subgraphs. $G_1$ and $G_4$ with more than 15 vertices are further divided into four smaller subgraphs respectively.}}
        \label{fig:graph-partition}
    \end{subfigure}       
    \begin{subfigure}{0.44\textwidth}
      \centering   
\includegraphics[width=\linewidth]{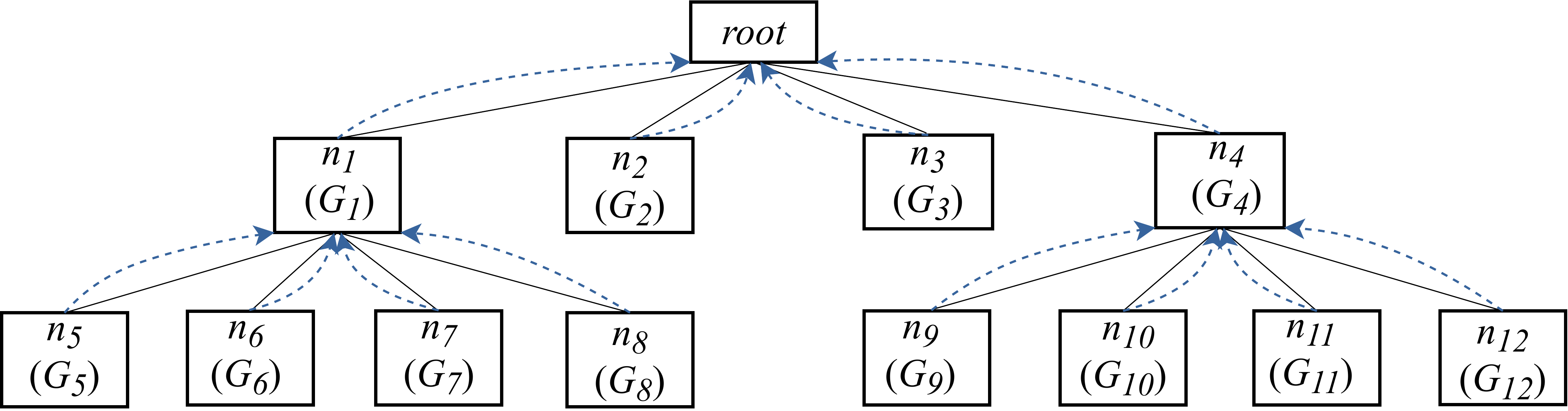}
    \caption{\scriptsize{{Storing Parent-Child Relationships in an M-ary Tree}}}
        \label{fig:m-ary-tree}
    \end{subfigure}
\vspace{-0.05cm}
\caption{Preprocessing}
\end{figure}

{We now describe how to construct {ODIN} using the M-ary tree as the starting point. {ODIN} inherits the same tree structure as the M-ary tree, but the information stored in each node is changed to support efficient query processing. Next, we first discuss what to keep in each node of the tree, and then present the procedure to iteratively construct the nodes at different levels, taking into consideration the varying density of objects in different parts of the network.}

\vspace{-6pt}
 \subsubsection{Node Structure of {ODIN}}

Each node in {ODIN} stores the following attributes: 1) a Boolean state that can be set to \textit{active} or \textit{inactive} to indicate if the node is to be used by the search algorithm; {2) the set of live vertices {belonging to} the node; 3) {the {\em border vertices} in the node (Note that a vertex can be a live vertex and a border vertex at the same time);} 4) the {\em external edges} associated to each border vertex in this node; 
5) an auxiliary data structure named {\em skeleton graph} that is built based on the subgraph corresponding to the node but contains those live vertices and {border vertices} only, which is defined as follows.}\looseness=-1

\vspace{-4pt}
\begin{myDef}
[Skeleton graph]Let $\mathcal{V}^{x}_L$ and $\mathcal{V}^{x}_B$ be the sets of live vertices and border vertices in a node $n_x$ respectively. The skeleton graph of $n_x$ is represented as ${SG}_x$ ($\mathcal{V}_x$, $\mathcal{E}_x$, $\mathcal{W}_x$), 
{where 1) $\mathcal{V}_x=\mathcal{V}^x_B \cup \mathcal{V}^x_L$, 2) $\mathcal{E}_x=\lbrace e_{i,j} \vert v_i\in \mathcal{V}^x_B \wedge v_j\in (\mathcal{V}^x_B\cup \mathcal{V}^x_L) \wedge v_i\neq v_j\rbrace$, and $\mathcal{W}_x=\lbrace w_{i,j} \vert e_{i,j}\in\mathcal{E}_x\rbrace$, where $w_{i,j}$ is the shortest distance between $v_i$ and $v_j$ in the partitioned subgraph corresponding to the node ${n}_x$. 
\label{def:skeleton-subgraph}
}
\vspace{-5pt}
\end{myDef}

{Assisted by the shortcuts between live and border vertices ($\mathcal{E}_x$ and $\mathcal{W}_x$) in the skeleton graph, the search algorithm can directly visit the live vertices from any border vertex when exploring inside each node.}

\subsubsection{Constructing ODIN}\label{sec:build-amt-index} 
{Since {ODIN} inherits the structure from M-ary tree, the primary task in constructing {ODIN} is to {activate the part of nodes that is essential to the search algorithm and} compute what to store in each activated node. {Intuitively, if a sufficient number of objects are covered by a subset of the child nodes of a parent node for a given $k$, these child nodes offer a suitable index granularity already and the search algorithm can just focus on them without considering their parent node. However, if this subset of nodes contain very few objects, their parent node covering more objects has to be examined and thus has to be activated for the search algorithm. 
This forms the basic idea of the construction.}  

Initially, all nodes are set to inactive. The construction process proceeds in a bottom-up fashion, starting from the leaf nodes and iteratively progressing onto upper non-leaf nodes, {and the procedure is shown in Algorithm~\ref{al:construct-amt-index}.} 
More specifically, we start with activating the leaf nodes and materialize their skeleton graphs {(Lines 1-3)}. This can be done easily as the shortest distances between any two vertices in any given partitioned subgraph have been precomputed for each leaf node.} 
{As we move up to the non-leaf nodes, a set of nodes $C(n_p)$ with the same parent node $n_p$ may will be {\em folded} when the {\em Folding Criteria} are met: (1) all nodes in $C(n_p)$ are presently active; and (2) they are {\em underfilled}, i.e., in total they have less than $\mu$ live vertices. During folding, all nodes in $C(n_p)$ are set to {\em inactive} and their parent node is activated with its skeleton graph materialized as needed, according to the activation procedure to be presented in Section~\ref{sec:folding-operation}. }
{Such a combination of activation and deactivation of nodes is to ensure that at any time, {\em any live vertex is present in only one active node}.} 
{ We examine the non-leaf nodes level by level in a bottom-up fashion, enacting the folding operation whenever the Folding Criteria are met, until the root has been processed {(lines 6-13)}. As a special case, the folding operation does not apply to the root and its child nodes even if the folding criteria is met. }

}


\begin{algorithm}[htbp]
\caption{Construction of ODIN}\label{al:construct-amt-index}
\LinesNumbered
\begin{footnotesize}
\KwIn{M-ary Tree}
\KwOut{the ODIN index}
\For{$n_l$: all leaf nodes} 
{Activated $n_l$ and materialize $SG_l$\;
} 
Suppose the leaf level is the {level 0}\;
 $r$=the height of M-ary tree\;
\For{$i=1$; $i<r$; $i++$}
{
\For{$n_p$: the nodes on the $i^{th}$ level}
{
\If{$C(n_p)$ meets the Folding Criteria}
{$n_p$ is activated and $C(n_p)$ are deactivated\;
$SG_p$ is materialized with the activation procedure\;
}
}
\If{no node on the $i^{th}$ level is activated}{break\;}
}
Return ODIN;

\end{footnotesize}
\end{algorithm}

\vspace{-0.1in}
\subsubsection{The Activation Procedure\label{sec:folding-operation}}
{Activating a non-leaf node $n_p$ for the first time involves materializing the skeleton graph $SG_p$ of $n_p$, including its sets of vertices, edges, and weights of edges.
Recall that there are two types of vertices in $SG_p$: live vertices and border vertices. The set of live vertices for $n_p$, $\mathcal{V}^{p}_{L}$, can be constructed by taking the union of all live vertices in the child nodes $C(n_p)$. However, the set of border vertices for $n_p$, $\mathcal{V}^{p}_{B}$, is constituted by only a subset of the union of border vertices of nodes in $C(n_p)$, because some of those border vertices in the child nodes are only connected to internal vertices in $n_p$ now. 
The edge set of $SG_p$ is composed of the edges that connect live and border vertices in $SG_p$ as per Definiton~\ref{def:skeleton-subgraph}. That is, every live vertex is connected by an edge to every border vertex, and there exists an edge connecting any pair of border vertices as well. For the edge weights, recall that the weight of an edge is the shortest distance between the two vertices involved {in the partitioned subgraph of $n_p$. Such shortest distance is not readily available, {but can be computed efficiently based on the combination of skeleton graphs of the child nodes $C(n_p)$ as described by Definition~\ref{def:combine-skeleton-graph}, as such shortest distances are identical in the partitioned subgraph of $n_p$ and $\mathcal{U}(C(n_p))$ as per  Theorem~\ref{theo:edge-wweight-computation}.
The detailed computation procedure will be discussed in Section~\ref{sec:AMT-Index-mpbs}.}}}

{
\vspace{-4pt}
\begin{myDef}
[Combination of skeleton graphs] Given a non-leaf node $n_p$ and the set of skeleton graphs $\mathcal{S}=\lbrace SG_1,\cdots,SG_m\rbrace$ of its child nodes $C(n_p)$, the combination of $\mathcal{S}$ is also a graph, denoted by ${\mathcal{U}_g}(C(n_p))$ ($\mathcal{V}^u$, $\mathcal{E}^u$, $\mathcal{W}^u$), that satisfies {\bf 1)} $\mathcal{V}^u$=$\bigcup_{i=1}^{m}\mathcal{V}_i$; {\bf 2)} $\mathcal{E}^u$=$\bigcup_{i=1}^{m}\mathcal{E}_i\cup \mathcal{E}^{c}$; {\bf 3)} {$\mathcal{E}^{c}$ is the set of external edges connecting different skeleton graphs in $\mathcal{S}$; and {\bf 4)} $\mathcal{W}^u=\lbrace w_{a,b} \vert e_{a,b}\in\mathcal{E}^u\rbrace$.}  
\label{def:combine-skeleton-graph}
\end{myDef}}


\begin{theorem}\label{theo:edge-wweight-computation}
{
For any two vertices $v_x$, $v_y\in$  ${\mathcal{U}_g}(C(n_p))$, if at least one of them is a border vertex, the shortest distance between $v_x$ and $v_y$ in $\mathcal{U}_g(C(n_p))$ is identical to that between $v_x$ and $v_y$ in the partitioned subgraph corresponding to $n_p$.}
\end{theorem}


\vspace{-6pt}
\vspace{4pt}
\begin{figure}[ht!]
  \centering
     \begin{subfigure}{0.33\textwidth}
      \centering   
\includegraphics[width=\linewidth]{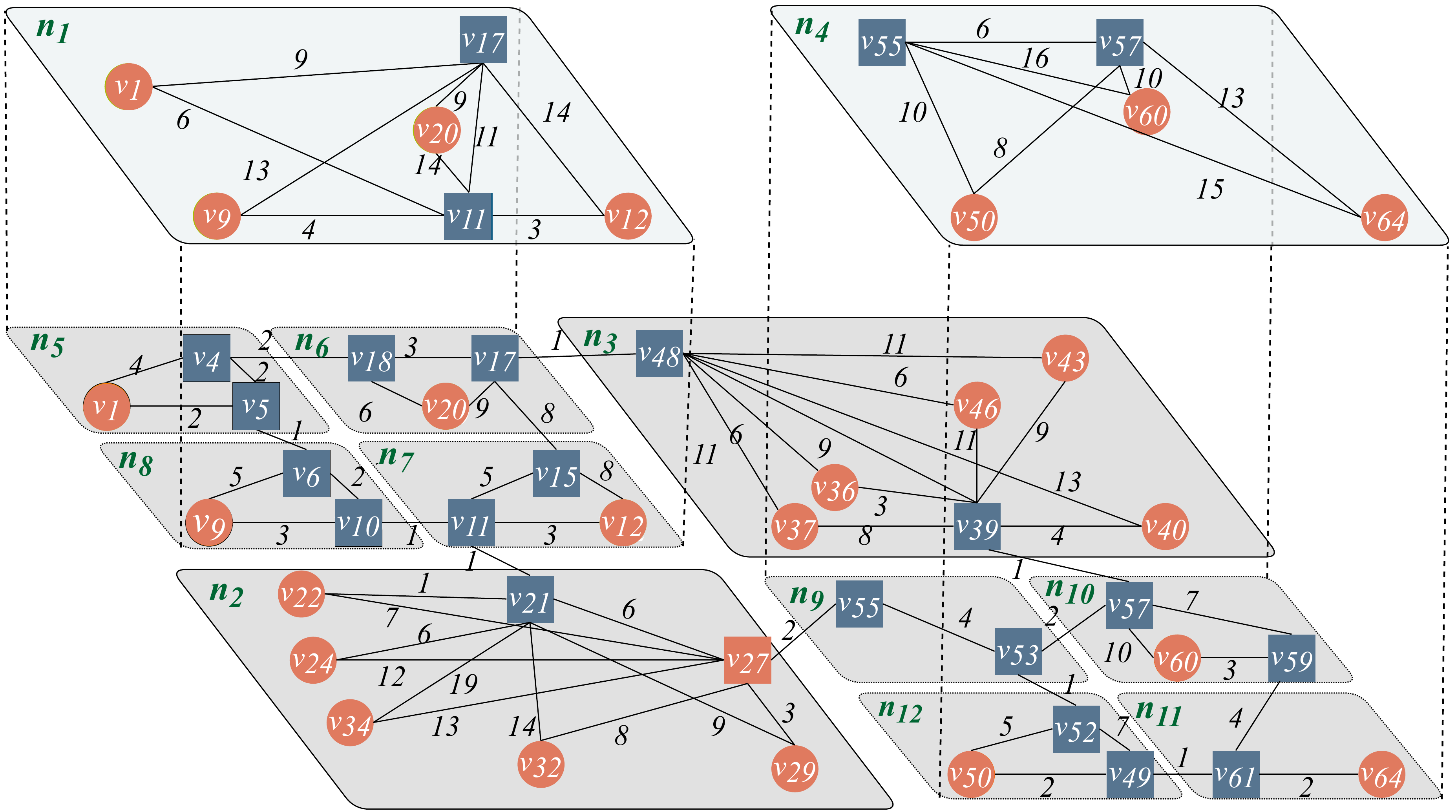}
      \caption{\scriptsize{A Sample {ODIN}}}
      \label{fig.amt-index}
    \end{subfigure}
     \begin{subfigure}{0.15\textwidth}
      \centering   
\includegraphics[width=\linewidth]{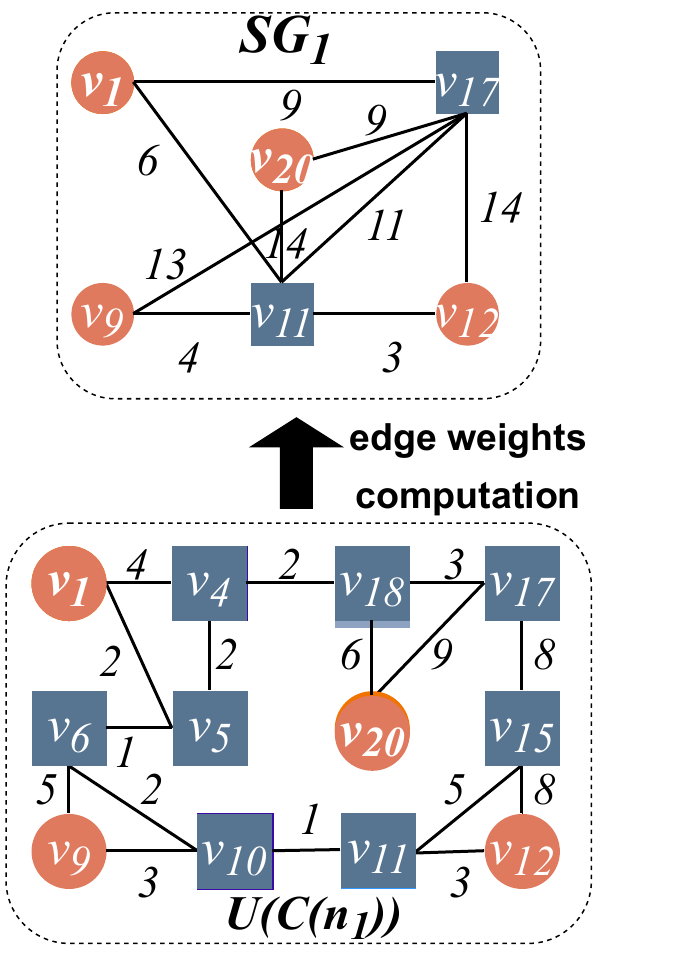}
        \caption{\scriptsize{$SG_1$ generation}}\label{fig:combination-sg}
    \end{subfigure}
    \vspace{-0.05cm}
\caption{An instance of {ODIN}} \label{fig:skeleton-graphs}
\end{figure}
\begin{example}
{Fig.~\ref{fig.amt-index} shows {ODIN} built based on the graph partitioning shown in Figure~\ref{fig:graph-partition}. Initially all nodes are inactive. We start by activating the leaf nodes ($n_2$, $n_3$, $n_5$, $n_6$, $n_7$, $n_8$, $n_9$, $n_{10}$, $n_{11}$, $n_{12}$) and materializing their skeleton graphs based on the partitioned subgraphs. We then proceed upwards in the tree and examine the parent nodes of the leaf nodes. Suppose that $\mu$ is set at 5. {The child nodes of $n_1$ (i.e., $n_5, n_6, n_7,$ and $n_8$) should be folded since they are all active and underfilled. So $n_1$ is activated while its child nodes set to inactive.}  The same goes for $n_4$ and its child nodes. } We take $n_1$ as an example to explain the activation procedure {as shown in Fig.~\ref{fig:combination-sg}, where the primary task is materializing the skeleton graph $SG_1$ for $n_1$ }. 
In particular,  we extract all live vertices ($v_1$, $v_9$, $v_{12}$, $v_{20}$) as well as two border vertices $v_{11}$ and $v_{17}$ from the child nodes of $n_1$ to form the vertex set of $SG_1$. {The other border vertices in the child nodes, such as $v_5$ and $v_{6}$, are not included, 
since they do not connect to any external vertices covered by nodes other than the child nodes of $n_1$.
We form the edge set of $SG_1$ by creating an edge between the border vertices ($v_{11}$ and $v_{17}$) and an edge from each border vertex to every live vertex in $n_1$. {The edge weights in $SG_1$ can be identified based on the combination of skeleton graphs of child nodes $C(n_1)$, $\mathcal{U}(C(n_1))$, and the details will be presented in Section~\ref{sec:AMT-Index-mpbs}.}} 
\end{example}


{{\bf Space cost.} 
{The storage overhead of ODIN 
is proportional to the number} of all materialized nodes. Suppose ODIN has $l$ levels and 
{a fraction} $\rho$ of nodes at each level $i$ ($i\in\lbrace 0, \cdots, l-1\rbrace$) are materialized. The number of all materialized nodes, i.e., the space cost of ODIN is $\sum_{i=0}^{l-1}{\rho\cdot m^i}$, which simplifies to {$O(\frac{\rho(m^l-1)}{m-1})$}. }

\vspace{-6pt}
\subsection{MPBS algorithm}\label{sec:AMT-Index-mpbs}
For a node $n_p$ being activated for the first time, {the edge weights in the skeleton graph $SG_p$ of $n_p$ should be the shortest distances from each border vertex to all live vertices and other border vertices  in the graph $\mathcal{U}_g(C(n_p))$ (abbr. $\mathcal{U}_g$) as per Definition~\ref{def:combine-skeleton-graph}.}  It is however non-trivial to compute these shortest distances efficiently. 

We propose a Multi-Source Parallel Bidirectional Search ({\em MPBS}) algorithm to efficiently compute the edge weights.
MPBS considers each border or live vertex as a source and creates a search instance to compute its shortest distance to other border and/or live vertices utilizing Dijkstra's Algorithm (if the source is a live vertex, we only need to compute its distance to other border vertices). Once a vertex is encountered by two search instances, a path between their respective sources is identified, which could be the shortest between them. The key is to quickly tell when the shortest distance has already been found so that the search can be terminated early. 



To achieve this goal, during the traversal for each source vertex $v_o$, a vertex is labeled as {\em processed} w.r.t. $v_o$ if its shortest distance to $v_o$ has been identified, or else it is considered {\em unprocessed}. Moreover, we use {\em bound vertex} for $v_o$ to denote a processed vertex that has at least one adjacent vertex that is unprocessed for $v_o$. Let $\mathcal{V}_r$ be the set of bound vertices for $v_o$ and the {\em bound distance} be the minimum shortest distance from any $v' \in \mathcal{V}_r$ to $v_o$, i.e., $\min_{v' \in \mathcal{V}_r} SD(v_o, v')$. For any two search instances $r_o$ and $r_d$ that simultaneously traverse the graph $\mathcal{U}_g$ from two corresponding sources $v_o$ and $v_d$, Theorem~\ref{the-correct-mpbs} 
 presents the condition for detecting the shortest distance between $v_o$ and $v_d$.

\vspace{-4pt}
\begin{theorem}
For any two sources $v_o$ and $v_d$ in the graph $\mathcal{U}_g$, let ${D_m}(v_o,v_d)$ denote the minimum distance between $v_o$ and $v_d$ identified so far. If ${D_m}(v_o,v_d)\leq (BD(v_o)+BD(v_d))$, then ${D_m}(v_o,v_d)$ is the shortest distance between $v_o$ and $v_d$ based on $\mathcal{S}(C(n_p))$, where $BD(v_o)$ and $BD(v_d)$ are the bound distances of $v_o$ and $v_d$ respectively. \label{the-correct-mpbs}
\end{theorem}

\begin{proof}
 We prove this conclusion by contradiction that ${D_m}(v_s,v_t)$ is not the shortest distance between $v_s$ and $v_t$, that is, the actual shortest distance $SD(v_s,v_t)$ and the corresponding shortest path $P'(s,t)$ have not been identified so far. The path $P'(s,t)$ can be viewed as the joint of three path segments $P(s,x)$+$P(x,y)$+$P(y,t)$ and $SD(v_s,v_t)$=$D(P(s,x))$+$D(P(x,y))$+$D(P(y,t))$, where $P(s,x)$ and $P(y,t)$ are the shortest path from $v_s$ and $v_t$ to their bound vertices $v_x$ and $v_y$ along the path $P'(s,t)$ respectively, and $P(x, y)$ is the shortest path between $v_x$ and $v_y$. Since $P'(s,t)$ has not been identified, then $v_x \neq v_y$. According to the definition of bound distances, $D(P(s,x))$ and $D(P(y,t))$ cannot be smaller than $BD(v_s)$ and $BD(v_t)$ respectively. As $D(P(x, y))$ is not zero, we infer that $SD(v_s, v_t)>BD(v_s)+BD(v_t)$. Also, because ${D_m}({v_s, v_t})\leq BD(v_s)+BD(v_t)$, we then have $SD(v_s, v_t)>{D_m}(v_s, v_t)$ which is contradiction.   
\end{proof}


\vspace{-4pt}
As per Theorem \ref{the-correct-mpbs}, the search instances from $v_o$ and $v_d$ can terminate once the condition is met. For each source, its search instance will terminate if its shortest distances to specified vertices are all identified. When all search instances stop, all required shortest distances between live and border vertices have been obtained. The pseudocode of the MPBS algorithm is illustrated in Algorithm~\ref{al:MPBS} 

\begin{algorithm}[h]
\caption{{MPBS}}\label{al:MPBS}
\LinesNumbered
\begin{footnotesize}
\KwIn{graph $\mathcal{U}_g$, set of border vertices $\mathcal{V}_b$, and set of live vertices $\mathcal{V}_l$;}
\KwOut{the shortest distance from $v_x\in \mathcal{V}_b$ to $v_y\in\mathcal{V}_b\cup\mathcal{V}_l$\; }
Initialize a distance matrix $D$ with $|\mathcal{V}_b|$ rows and $|\mathcal{V}_b|+|\mathcal{V}_l|$ columns to keep the shortest distances between border and live vertices\;
\For{i=0; $i<|\mathcal{V}_b|$; i++}
{
\For{j=0; $j<(|\mathcal{V}_b|+|\mathcal{V}_l|)$; j++}
{
$D_{i,j}=\infty$\;
}
}
\For{$v$ : $\mathcal{V}_g$}
{
Let $v.\mathcal{R}$ keep identifiers of the search instances that visit the vertex $v$\;
}

\For{$v_i$: $\mathcal{V}_b$}
  {
    Create a search instance $r_i$ ($v_i$, $\mathcal{U}_g$, $\mathcal{V}_b\cup \mathcal{V}_l$)\;
  }
\For{$v_j$: $\mathcal{V}_l$}
  {
    Create a search instance $r_j$ ($v_j$, $\mathcal{U}_g$, $\mathcal{V}_b$)\;
  }
Return $D$ if all search instances terminate\;
\end{footnotesize}
\end{algorithm}

\begin{algorithm}[h!]
\caption{{Search instance $r_i$}}\label{al:MPBS-Dis}
\LinesNumbered
\begin{footnotesize}
\KwIn{source vertex $v_i$, graph $\mathcal{U}_g$, set of destination vertices $\mathcal{V}_d$; }
A set $\mathcal{V}_p^i$ is to keep the processed vertices w.r.t. $v_i$\;
A priority queue $\mathcal{Q}_u^i$ is to keep the unprocessed vertices w.r.t. $v_i$\;
A set $\mathcal{V}_b^i$ is to keep the bound vertices w.r.t. $v_i$\;
A set $\mathcal{V}_d^i$ is to keep IDs of the destination vertices whose shortest distances to $v_i$ have been determined\; 
Initialize $\mathcal{V}_p^i$, $\mathcal{Q}_u^i$, $\mathcal{V}_b^i$, and $\mathcal{R}_t^i$ as empty sets\;
Put $v_i$ into $\mathcal{V}_p^i$, and add the adjacent vertices of $v_i$ into $\mathcal{V}_u^i$\;
\While{$\mathcal{V}_d^i\neq \mathcal{V}_d$}
{
  \While{$\mathcal{Q}_u^i\neq\phi$}
  {
  $v_f$=$\mathcal{Q}_u^i$.pop()\;
  Add $v_f$ into $\mathcal{V}_p^i$, and update $\mathcal{V}_b^i$ based on $v_f$\;
  $BD(v_i)$=$min_{v'\in \mathcal{V}_b^i} SD(v', v_i)$\;
  \For{$r_j$ : ($v_f.\mathcal{R}$-$\mathcal{V}_d^i$)}
  {
  Generate a path $p_{i,j}$ by combining $p_{i,f}$ and $p_{f,j}$\;  
  \If{$D_{i,j}>p_{i,j}.length()$}
  {
  $D_{i,j}=p_{i,j}.length()$\;
  $D_{j,i}=p_{i,j}.length()$\;
  }
  \If{$BD(v_i)$+$BD(v_j)\geq D_{i,j}$}
  {
  $\mathcal{V}_d^i.add(v_j)$\;
  }
  }
  \For{$v_x$ : unprocessed adjacent vertices of $v_f$}
  {
   \If{$SD(v_i,v_x)>SD(v_i, v_f)+e_{f,x}$}
   {
   $SD(v_i,v_x)=SD(v_i, v_f)+e_{f,x}$\;
   Add $v_x$ into $\mathcal{Q}_u^i$ if $v_x$ does not exist in $\mathcal{Q}_u^i$ previously\;
   }
  }
  }
}
$r_i$ terminates\;
\end{footnotesize}
\end{algorithm}

\vspace{-0.05in}
\subsection{Maintenance of {ODIN}}\label{sec:AHP-Index-maintenance}
{After the construction of {ODIN}, we keep the latest reported positions of moving objects in a buffer}, which are utilized to maintain {ODIN} and support query processing at the next snapshot. {The maintenance for {ODIN} is conducted for each snapshot and includes two major components.} {Firstly}, as live vertices appear and disappear, we need to update the skeleton graphs of the nodes involved, which will be discussed in Section~\ref{sec:maintenance-skeleton-subgraphs}. {Secondly}, due to the variation in live vertices, it is possible for nodes to become underfilled or overfilled. This can be handled by the folding and unfolding operations that will be presented in Section~\ref{subsec:folding-unfolding}. 


\subsubsection{Maintenance of skeleton graphs}\label{sec:maintenance-skeleton-subgraphs} 

When a live vertex $v_l$ appears in or disappears from a leaf node $n_f$, we update the skeleton graphs involving $v_l$ in a bottom-up fashion. Since any live vertex is covered by one and only one active node, 
only the skeleton graph of $n_f$ needs to be updated if $n_f$ is active. Otherwise,
there must exist one and only one active ancestor node of $n_f$, denoted by $\mathcal{A}(n_f)$, that covers all live vertices of $n_f$. 
In this case, we {focus on the sub-tree of {ODIN} rooted at $\mathcal{A}(n_f)$ with $n_f$ being a leaf node, and iteratively update the  skeleton graphs of all the ancestor nodes of $n_f$ in this sub-tree. }

 {\bf Insertion of a new live vertex.} If a normal vertex $v_l$ in $n_f$ turns into a live vertex, it can be easily added into the skeleton graph of $n_f$ as the shortest distance between any two vertices in the partitioned subgraph in $n_f$ has been computed. Next, $v_l$ will be further added into the skeleton graph of $n_p$, the parent node of $n_f$. In the skeleton graph, new edges are added to connect $v_l$ and each border vertex of $n_p$, and the edge weights are the shortest distances from $v_l$ to the corresponding border vertices in $\mathcal{U}_g(C(n_p))$. We repeat this step until $v_l$ is recursively added into the skeleton graph of  $\mathcal{A}(n_f)$. 

{\bf Deletion of an obsolete live vertex.} When a live vertex $v_l$ in the leaf node $n_f$ has no associated objects, it could be either a border vertex or a normal vertex. If $v_l$ is a border vertex, we do not need to update the skeleton graphs of the involved nodes; otherwise, $v_l$ is recursively removed from the skeleton graphs of the nodes from $n_f$ to $\mathcal{A}(n_f)$ along the branch of the sub-tree. The deletion of $v_l$ from a skeleton graph means that $v_l$ and its adjacent edges in the skeleton graph are all removed. 

Suppose there exist $\Gamma_s$ nodes on the branch from $n_f$ to $\mathcal{A}(n_f)$. A newly emerged or disappeared live vertex $v_l$ in $n_f$ will be inserted to or deleted from $\Gamma_s$ skeleton graphs in the above process. Thus, the time complexity of both insertion and deletion of $v_l$ is $O(\Gamma_s)$.

\begin{figure}[h!]
  \centering
     \begin{subfigure}{0.16\textwidth}
      \centering         \includegraphics[width=\linewidth]{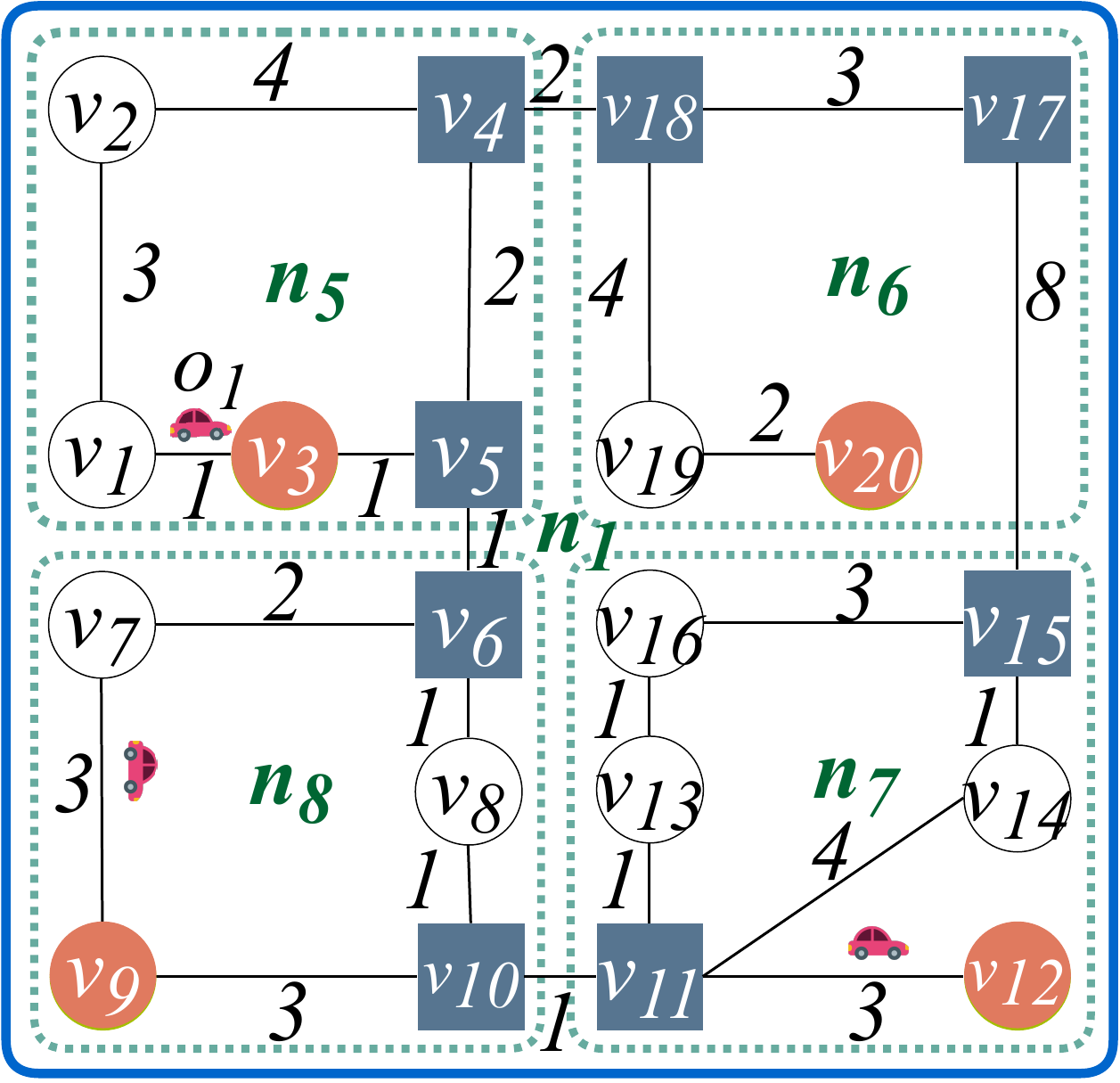}
        \caption{\small{Part of ODIN}}
        \label{fig:sg-live-vertex-change}
    \end{subfigure}
    \hspace{0.1cm}
    \begin{subfigure}{0.29\textwidth}
      \centering   
\includegraphics[width=\linewidth]{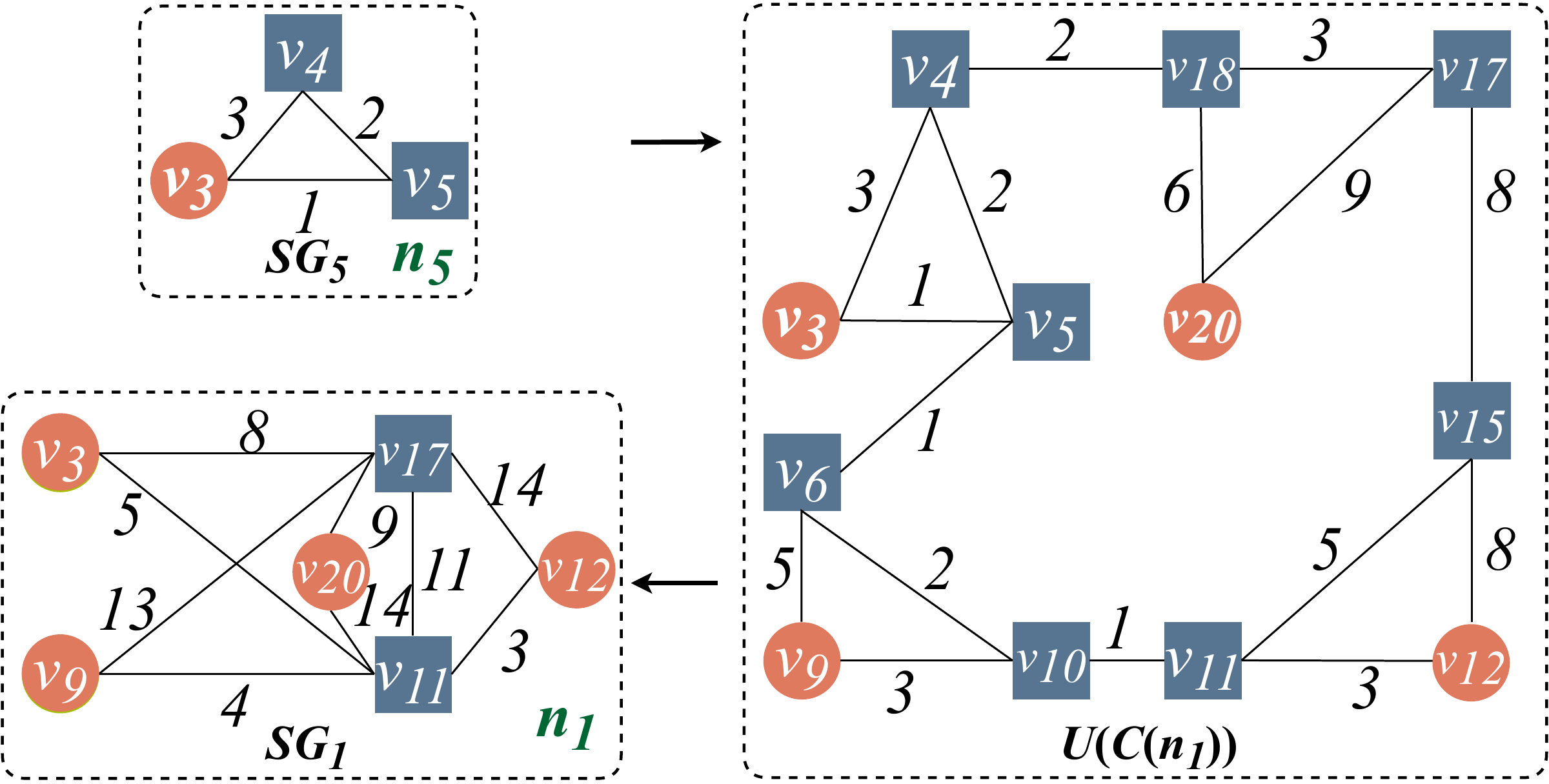}
        \caption{\small{Live vertex insertion process}}
        \label{fig:recursive-add-live-vertex}
    \end {subfigure}
    \vspace{-0.2cm}
\caption{Processing New and Obsolete Live Vertices} \label{exp:update-skeleton-subgraph}
\end{figure}

\begin{example}
{Fig.~\ref{exp:update-skeleton-subgraph} shows an instance of processing varied live vertices to maintain ODIN. 
Suppose object $o_1$ moves to $v_3$ after reaching $v_1$ in the partitioned subgraph of $n_5$ as shown in Fig.~\ref{fig:sg-live-vertex-change}, then $v_1$ is no longer a live vertex while $v_3$ becomes a live vertex. {At this point, suppose $n_1$ is the active ancestor node of $n_5$, then the skeleton graphs of $n_5$ and $n_1$ (i.e., $SG_5$ and $SG_1$) need to be updated.} In particular, $v_1$ is first removed from $SG_5$ and $SG_1$, and then $v_3$ is sequentially inserted into $SG_5$ and $SG_1$ as shown in Fig.~\ref{fig:recursive-add-live-vertex}. During the insertion, $v_3$ is first added to $SG_5$, a straightforward process due to precomputed shortest distances from $v_3$ to $v_4$ and $v_5$. Subsequently, $v_3$ is inserted into $SG_1$ and connected directly to border vertices $v_{11}$ and $v_{17}$ with new edges. The edge weights are the shortest distances from $v_3$ to $v_{11}$ and $v_{17}$ based on $\mathcal{U}_g(C(n_1))$.}
\end{example}

\vspace{-0.1in}
\subsubsection{Folding and unfolding nodes}\label{subsec:folding-unfolding}
After updating the skeleton graphs of the involved nodes, we go through each active node whose skeleton graph has been updated to identify the underfilled and overfilled active nodes, and then process them accordingly using folding and unfolding operations. {The objective is to let the active nodes always cover a moderate number (with a user-configurable range) of live vertices {as the ODs evolve}, which is vital to providing stable pruning power.}





{\bf Folding operation.} When a set of active nodes meet the Folding Criteria discussed in Section~\ref{sec:build-amt-index}, they are folded into their parent node $n_p$. During this process, if any node $n_p$ is activated for the first time, its skeleton graph $SG_p$ will be materialized following the activation procedure presented in Section~\ref{sec:folding-operation}; otherwise, $SG_p$ must have been materialized already, and we just need to incrementally update $SG_p$. As the border vertices of $SG_p$ do not change as objects move, we only need to remove the obsolete live vertices without any associated objects from $SG_p$ and add newly emerged live vertices to $SG_p$ with the insertion procedure discussed in Section~\ref{sec:maintenance-skeleton-subgraphs}. Just like how ODIN is initially constructed, the activation of $n_p$ may trigger further folding operations in its ancestor nodes. We process such changes in a bottom-up fashion until no further changes are necessary. \looseness=-1 

{\bf Unfolding operation.} 
When there are too many objects in a node, its pruning power in the corresponding region for $k$NN query processing is weakened. To address this issue, we introduce the {\em unfolding operation} which is carried out on any non-leaf node that meets the following {\em Unfolding Criteria}: (1) the node is active; and (2) it is overfilled, i.e., {it covers more than $m\times\mu$ live vertices}.


{When an active node $n_d$ is unfolded, $n_d$ is deactivated and its child nodes are activated at the same time. The activated child nodes offer indexing (skeleton graphs) at a finer granularity than $n_d$ does, which is more suitable for densely populated regions. Since the skeleton graph $SG_d$ of $n_d$ will no longer be used in query processing due to its deactivation, we no longer maintain $SG_d$ as objects evolve until $n_d$ is activated again. {As a child node of $n_d$ thus activated may itself be overfilled}, the unfolding operation is applied recursively until there are no more descendants of $n_d$ meeting the Unfolding Criteria or when the leaf nodes are reached. Note that the unfolding operation does not apply to leaf nodes. }



\vspace{-8pt}
 \subsection{Global skeleton graph}\label{subsec:global-skeleton} 
 


{In {ODIN}, since all objects are associated to live vertices which are all kept in the skeleton graphs of the active nodes, theoretically the search algorithm  can identify $k$NNs by only exploring the skeleton graphs of active nodes  instead of all nodes. Hence,} we introduce the notion of {\em global skeleton graph}, denoted by $G^s$, that refers to the combination of the skeleton graphs of all active nodes in {ODIN}. Note that the global skeleton graph is not materialized; rather, it is a conceptual graph that specifies the entire search scope for the query processing algorithms. It covers all live vertices, whose associated objects together form the complete object set. 
It provides sufficient information for the search algorithms to identify the $k$NNs for any given query. 

\vspace{-0.3cm}
 {\subsection{Limitations}
 The elasticity of ODIN equips it with appropriate index granularity in different parts of the road network with varying ODs. {However, }this advantage diminishes if 
 {the entire graph is densely populated by objects such that the index becomes full-fledged, i.e., none of the leaf nodes is underfilled.} In this case, there is no need to aggregate the index information from the leaf level to upper levels with folding operations, {and ODIN is equivalent in effect to a single-level index consisting of all the leaf nodes}.
 }
\section{Processing {C$k$NN} Queries} \label{sec:AMT-KNN}


\subsection{Initial $k$NNs computation}\label{sec:initial-knn} 
\subsubsection{Query vertex preprocessing}
For a given query vertex $v_q$, we can easily locate the leaf node $n_f$ that covers $v_q$ as we maintain the association between any vertex and the leaf node it resides in as discussed in Section~\ref{sec:preprocessing}. In order to identify the $k$NNs of $v_q$ utilizing the global skeleton graph $G^s$, we need to preprocess $v_q$ such that at the end of this processing (1) $v_q$ is in the global skeleton graph, and specifically, the skeleton graph of the active ancestor node $\mathcal{A}(n_f)$ of $n_f$, and (2) there is an edge connecting $v_q$ to every other vertex in the skeleton graph of $\mathcal{A}(n_f)$. {After the preprocessing, the query vertex is effectively a border vertex in $G^s$}. The way the preprocessing is done depends on the type of vertex $v_q$. 

{\bf If $v_q$ is a live vertex in $n_f$}, it must have been in the skeleton graph of $\mathcal{A}(f)$. Further, if $v_q$ is also a border vertex in the skeleton graph of $\mathcal{A}(f)$ at the same time, no preprocessing is needed; otherwise, there would not be edges connecting it to other live vertices in $\mathcal{A}(n_f)$ as per Definition~\ref{def:skeleton-subgraph}. Thus, we add such edges between $v_q$ and other live vertices, and the edge weights can be computed in the same way as when $v_q$ were to be inserted as a new live vertex to $\mathcal{A}(f)$, as described in Section~\ref{sec:maintenance-skeleton-subgraphs}.

{\bf If $v_q$ is a border vertex in $n_f$ but not a border vertex in $\mathcal{A}_f$, or if $v_q$ is neither a live nor a border vertex in $n_f$,} it needs to be inserted into $\mathcal{A}_f$ as well following the same procedure as inserting a new live vertex into $\mathcal{A}_f$. During the insertion, we create edges connecting $v_q$ to all live and border vertices in the skeleton graph of $\mathcal{A}(n_f)$.


After preprocessing, $v_q$ is directly connected to each border and live vertex in the active node it belongs to. We {then} have the following theorem that forms the basis for the search algorithms.

\vspace{-3pt}
\begin{theorem}\label{theo:sd-border-live-vertices}
{For any given query vertex $v_q$ that has been preprocessed, the following holds: (1) the shortest path in $G^s$ from each border vertex to $v_q$ does not contain any live vertex, and (2) the shortest distance from any vertex $v_x$ to $v_q$ in $G^s$ equals to their shortest distance $SD(v_x, v_q)$ in the original graph $G$.}
\end{theorem}

\newcommand{\LS}{{{\mathbf v}_1}\rightsquigarrow {{\mathbf v}_l}}

{\begin{proof}
We firstly prove (1). 
For any border vertex $v_b$ in $G^s$, let $P(b,q)$ denote the shortest path from $v_b$ to $v_q$, and suppose, for the sake of contradiction, that $P(b,q)$ contains at least one live vertex. Thus, there must be a partial path $\langle v_x, {{\mathbf v}_1}\rightsquigarrow {{\mathbf v}_l}, v_y\rangle$ in  $P(b,q)$, such that $\LS$ is a sequence consisting of one or more live vertices only, while $v_x$ and $v_y$ are both border vertices, or one of them is a border vertex and the other is $v_q$. As border vertices are the entrances of each node, we can infer that $v_x$, $v_y$ and $\LS$ must be located in the same node. Therefore, there must exist an edge between $v_x$ and $v_y$ in the skeleton graph, and the edge weight is their shortest distance within the corresponding node. This implies that there does not exist any vertices between $v_x$ and $v_y$ on $P(b, q)$, which contradicts the assumption. 

Next we prove (2). For any given vertex $v_x$, let $P'(x,q)$ and $P(x, q)$ be the shortest paths from $v_x$ to $v_q$ in $G^s$ and $G$ respectively. We consider the following two cases. 

(i) Suppose $v_x$ is a border vertex ($v_x\neq v_q$). Then $P'(x,q)$ can be represented as $\langle v_x, \mathbf{S'}, v_q \rangle$, where $\mathbf{S'}$ is a sequence of border vertices in $G^s$ as proved in (1). Meanwhile, let $\mathbf{S}$  denote the sequence of border vertices in the active nodes between $v_
x$ and $v_q$ in $P(x,q)$. We can conclude that $\mathbf{S}$ is identical to $\mathbf{S'}$, because if it were not the case, $\langle v_x, \mathbf{S}, v_q \rangle$ rather than $\langle v_x, \mathbf{S'}, v_q \rangle$ (i.e., $P'(x,q)$) would be the shortest path from $v_x$ to $v_q$ in $G^s$, which contradicts the assumption. Further, for any two adjacent vertices $\langle v_i, v_j\rangle$ in $\mathbf{S}$ located in the same active node, we can infer that their shortest distance $SD(v_i, v_j)$ in $G$ equals to ${SD^{''}(v_i, v_j)}$, their shortest distance in the original partitioned subgraph of the active node $n_a$ they reside in; otherwise, $v_i$ and $v_j$ could not be adjacent in $\mathbf{S}$. Recall that $SD{''}(v_i, v_j)$ equals to the edge weight between $v_i$ and $v_j$ in the skeleton graph of $n_a$ (a part of $G^s$) as per Definition~\ref{def:skeleton-subgraph}. We can thus conclude that $D(P'(x,q))=D(P(x, q))=SD(v_x, v_q)$, where $D(P'(x,q))$ and $D(P(x, q))$ represent the {lengths} of $P'(x,q)$ and $P(x,q)$ respectively. 

(ii) If $v_x$ is a live vertex but not a border vertex, $P'(v_x, v_q)$ must first pass through {a border vertex $v_b$ adjacent to }$v_x$ in the same node and then reaches $v_q$ from $v_b$. 
If $v_b=v_q$, $D(P'(x,q))$ is the edge weight between $v_x$ and $v_b$ in $G^s$, which equals to $SD(v_x, v_b)$ as proved in case (i); otherwise, $D(P'(x,q))$ can be viewed as $D(P'(x,b))+D(P'(b,q))$. Because $D(P'(x,b))=SD(v_x, v_b)$ and $D(P'(b,q))=SD(v_b, v_q)$ have been {shown in case (i)}, we conclude that $D(P'(x,q))$, the shortest distance from $v_x$ to $v_q$ in $G_s$, equals to $SD(v_x, v_q)=SD(v_x, v_b)+SD(v_b, v_q)$ in the original graph $G$.
\end{proof}}

\subsubsection{The ODIN-KNN-Init Algorithm}\label{subsec:odin-knn-init}
Based on Theorem~\ref{theo:sd-border-live-vertices}, we propose the {ODIN-KNN-Init} algorithm for identifying the initial $k$NNs, as shown in  Algorithm~\ref{al:e-dijkstra}. Here, we employ the same notions of {\em processed} and {\em unprocessed} as used in the MRPS algorithm to label the vertices whose shortest distances to $v_q$ are known or unknown respectively. For a given $v_q$, initially all vertices are labelled unprocessed. The general idea of ODIN-KNN-Init is to iteratively expand the search scope in the global skeleton graph, starting from vertices immediately adjacent to $v_q$, until the $k$NNs are found. For each live vertex visited, the algorithm computes the distances between its associated objects and $v_q$ and updates the $k$NNs identified so far. When the shortest distance from the current $k^{th}$ nearest object to $v_q$ is no greater than that from the nearest unprocessed vertex to $v_q$, the algorithm terminates as the $k$NNs are guaranteed to have been identified.  

In {ODIN-KNN-Init}, we first introduce a max heap $HP_k$ with capacity $k$, where each element in the heap consists of the ID of an object and its distance to $v_q$,  to maintain the $k$NNs found so far (Line 1). For each new object with its distance to $v_q$ computed, we compare it with the element at the top of $HP_k$; if its distance is less than that at the top of $HP_k$, it is inserted into the heap. 
Next, we take $v_q$ as the starting vertex to traverse $G^s$ based on the principal of relaxation similar in spirit to Dijkstra's Algorithm (Lines 2-15). We maintain a priority queue $Q_d$ of vertices to prioritize vertex visits in an ascending order of distance to $v_q$. Initially, {if $v_q$ is live vertex, it is inserted into $Q_d$; otherwise, we label $v_q$ as processed and add its adjacent vertices into $Q_d$.} We then iteratively dequeue the head of $Q_d$ and process this vertex $v_f$ differently depending on its type as follows.  


{\bf If vertex $v_f$ is a border vertex (Lines 4-7)}, its adjacent vertices will be traversed. For each adjacent vertex $v'_f$ of $v_f$, 
if $D(v'_f,v_q)>D(v'_f,v_f)+D(v_f,v_q)$, we update $D(v'_f,v_q)$ to $D(v'_f,v_f)+D(v_f,v_q)$. Next, we insert $v'_f$ into $Q_d$ if $v'_f$ does not exist in $Q_d$ previously. 

{\bf If vertex $v_f$ is a live vertex  (Lines 9-15)}: we just need to examine its associated objects to update the current $k$NNs without visiting its adjacent vertices, as those vertices are all border vertices and their shortest paths to $v_q$ do not pass through $v_f$, as per Theorem~\ref{theo:sd-border-live-vertices}. 
{Once the update on the current $k$NNs kept in $HP_k$ is done}, if the object at the top of $HP_k$ has a smaller shortest distance to $v_q$ than the current head vertex in $Q_d$ to be processed in the next iteration (Line 14), the search procedure can safely terminate as the objects in $HP_k$ are guaranteed to be the $k$NNs of $v_q$ (Line 15). Otherwise, we deal with the next vertex to be processed until the terminal condition is met or $Q_d$ becomes empty.

{\bf If vertex $v_f$ is both a live and a border vertex}, it is first handled as a border vertex and then processed as a live vertex with the above procedures. 

In {ODIN-KNN-Init}, we use a {set $S_v$} to keep the processed vertices as well as their shortest distances to $v_q$ (Line 5). 
For each processed live vertex, its associated objects inserted into $HP_k$ are also {cached} to be reused for the incremental computation of $k$NNs, as stated in Section~\ref{sec:incremental-computation}. 

\begin{example}\label{exam:initial}
In an instance of the global skeleton graph shown in Fig.~\ref{fig:skeleton-graph-previous}, $v_q$ is the query vertex with $k=4$. Algorithm~\ref{al:e-dijkstra} first adds $v_{12}$, $v_{19}$, $v_1$ into $Q_d$, and then iteratively pops and processes the vertex from $Q_d$. At first, the border vertex $v_{12}$ is processed. The algorithm updates the distances from adjacent vertices of $v_{12}$ (i.e., $v_{13}$ and $v_5$) to $v_q$ and insert them into $Q_d$. Next, the live vertex $v_{19}$ is popped and handled, then its associated object $o_2$ is directly put into $HP_k$ as $HP_k$ is not full. In this way, after $v_7$ being processed, the current 4NNs in $HP_k$ are $\lbrace o_2$, $o_1$, $o_4$, $o_3\rbrace$ and the vertices in $Q_d$ are $\lbrace v_9$, $v_{14}$, $v_{17}\rbrace$. 
Since $SD(o_3,v_q)$ and $SD(v_9,v_q)$ both equal to 13, $\lbrace o_2$, $o_1$, $o_4$, $o_3\rbrace$ are the final 4NNs.
\end{example}

\vspace{-0.1cm}
{\bf Correctness of {ODIN-KNN-Init}.} 
While traversing the global skeleton graph $G^s$, {ODIN-KNN-Init} follows Dijkstra's relaxation principle, ensuring it always explores the closest unprocessed vertex to $v_q$ in each expansion. Since the shortest distance from any vertex to $v_q$ in $G^s$ is the same as in $G$ (as per Theorem~\ref{theo:sd-border-live-vertices}), live and border vertices in $G^s$ are visited in ascending order of their shortest distances to $v_q$ in $G$. When {ODIN-KNN-Init} reaches its termination condition, we can conclude that the $k^{th}$ nearest object at the top of $HP_k$ identified so far cannot be farther away than any unprocessed live vertex, including its associated objects. Therefore, the objects in $HP$ represent the final $k$NNs.

\begin{spacing}{0.5}
\begin{algorithm}[htbp]
\caption{{ODIN-KNN-Init}}\label{al:e-dijkstra}
\LinesNumbered
\begin{footnotesize}
\KwIn{Global skeleton graph $G^s$, MaxHeap $HP_k$, Queue $Q_d$, Set $S_v$, query vertex $v_q$}
\KwOut{$k$NNs of $v_q$}
\If{{number of search round==1}}
{
  $HP_k=\phi$;
  $Q_d=\phi$;
  $S_v=\phi$\;
  Put {\em adjacent vertices of} $v_q$ into $Q_d$\;
}
\While{$Q_d\neq \phi$}
{
$v_f=Q_d.dequeue()$, and $S_v.add(v_f)$\;
\eIf{$v_f$ is a border vertex}
{\For{$v'_f$: adjacent vertices of $v_f$}
{
\If{$D(v'_f,v_q)>D(v'_f,v_f)+D(v_f,v_q)$}
{$D(v'_f,v_q)=D(v'_f,v_f)+D(v_f,v_q)$\;}
{Add $v'_f$ into $Q_d$ if $v'_f$ is not in $Q_d$ previously;}}
}
{
$v_p=Q_d.headVertex()$\;
\For{$o_i$ : associated objects of $v_f$} 
{\If{$HP_k$ is not full}
{Insert $o_i$ into $HP_k$\;}
\ElseIf{$SD(o_i,v_q)<SD(o^*,v_q)$}
{remove $o^*$ from $HP_k$ and insert $o_i$ into $HP_k$; {\em//$o^*$ is the top object in $HP_k$}}%
\If{$|HP_k|=k \&\& SD(o^*, v_q)\leq SD(v_p,v_q)$}
{return objects in $HP_k$\;}
}
}
}
\end{footnotesize}
\end{algorithm}
\end{spacing}

\vspace{0.1in}

{{\bf Time complexity.} {Let} $b$ and $l$ denote the number of processed border and live vertices respectively and $d_g$ be the average degree of each border vertex. For each border vertex, {it takes }$O(d_g)$ to update the shortest distance from its each adjacent vertex to $v_q$ and takes $O(\log{(b+l)})$ to select the vertex closest to $v_q$ from the unprocessed vertices. Hence, all processed border vertices are traversed in $O(b\cdot d_g\cdot\log{(b+l)})$. For each live vertex, we view each insertion of the associated objects into $HP_k$ as an unit of operation, leading to a total time complexity of $O(b\cdot d_g\cdot\log{(b+l)}+l)$.}




\vspace{-0.1in}
\subsection{Incremental computation of $k$NNs}\label{sec:incremental-computation} 
As discussed in Section~\ref{sec:preliminaries}, the $k$NNs of each query should be updated at each snapshot as objects move, a task that we call a {\em search round}. 
Instead of recomputing the $k$NNs from scratch in each search round using {ODIN-KNN-Init}, we propose the {ODIN-KNN-Inc} algorithm for incremental update of the search result. It capitalizes on vertices that have already been processed (i.e., the vertices whose shortest distance to $v_q$ are already known) from the previous search round, and performs an incremental computation in the global skeleton graph in the current round to update $k$NNs of $v_q$. {Note that ODIN-KNN-Inc only addresses $k$NNs incremental update {in scenarios} with moving objects but fixed query points. A query will be treated as new one and processed from scratch if it moves.}


The algorithm {ODIN-KNN-Inc}, shown in Algorithm \ref{al:is-algorithm}, accepts four inputs: the global skeleton graph $G^s$; the shortest distances from $v_q$ to the processed vertices in the previous round, kept in $S_v$; the $k$NNs from the previous round $HP_k$; and the query vertex $v_q$. The algorithm consists of three main steps: 
(1) identify those active nodes with processed vertices (from the previous search round) for the current round (Line 2);
(2) explore each identified active node to produce $k$NNs with an incremental strategy that utilizes the processed vertices to reduce the cost (Lines 3-4);
(3) if the $k$NN result cannot be finalized in the above steps (i.e., if there exist unprocessed vertices containing objects that can  potentially affect the $k$NN result), we expand the search scope by traversing $G^s$ with {ODIN-KNN-Init} to produce the final result (Lines 5-8). We will discuss the above three steps in detail in Sections \ref{sec:incremental-computation:step1} to \ref{sec:incremental-computation:step3} respectively.



\begin{algorithm}[htbp]
\LinesNumbered
\caption{{ODIN-KNN-Inc}}\label{al:is-algorithm}
\begin{footnotesize}
\KwIn{$G^s$, $S'_v$, $HP_k$, $v_q$}
\KwOut{$k$NNs of $v_q$}
$HP_k=\phi$\;
Identify the set of candidate nodes ${\mathcal{P}}^{'}$ based on $\mathcal{P}$\;
\While{$n_c:\mathcal{P}_{i+1}$}
{
     Incrementally explore the candidate node $n_c$\;
}
{$G^{s'}$=$G^s$ excluding the accomplished vertices}\;
{Identify the nearest unaccomplished vertex $v_f$}\;
\If{$HP_k$ is full \&\& $SD(o^*, v_q)\leq SD(v_f, v_q)$}{\Return $HP_k$\;}
{
Enqueque all unaccomplished vertices into $Q_d$\;
\Return ODIN-KNN-Init ($G^{s'}$, $HP_k$, $Q_d$, ${S'_v}$, $v_q$)\;

}
\end{footnotesize}
\end{algorithm}


\vspace{-0.1in}
\subsubsection{Identifying candidate nodes.}
\label{sec:incremental-computation:step1}




This step is to identify the active nodes containing processed vertices from the previous round (Line 2). To ease the presentation, we use {\em candidate nodes} to refer to such nodes, and let $\mathcal{P}$ and ${\mathcal{P}}^{'}$ denote the sets of candidate nodes in the previous and current rounds respectively. Considering that the candidate nodes in $\mathcal{P}$ may be folded and unfolded as objects evolve, ${\mathcal{P}}^{'}$ is likely to be different from $\mathcal{P}$. Here, we identify ${\mathcal{P}}^{'}$ by scanning the candidate nodes in $\mathcal{P}$, which handles each candidate node $n_c\in \mathcal{P}$ as follows: (1) If $n_c$ is still an active node, it is included as a candidate node in ${\mathcal{P}}^{'}$; (2) If a folding operation involves $n_c$ in the current round, $n_c$ must have been deactivated and is thus not a candidate node in $\mathcal{P}^{'}$. Meanwhile, its parent node $n_p$ must have been activated and would be a candidate node in ${\mathcal{P}}^{'}$ if $n_p$ contains processed vertices {from the previous round}; (3) If $n_c$ has been unfolded in the current round, it must have been deactivated and its child nodes activated. As such, there must exist some child node(s) that contain the processed vertices in $n_c$ from the previous round and thus become candidate nodes to be included in ${\mathcal{P}}^{'}$. 

\vspace{-5pt}
\subsubsection{Incrementally exploring candidate nodes.} 
\label{sec:incremental-computation:step2}

In this step, we explore each candidate node $n_c \in \mathcal{P}'$ {and update} $HP_k$ based on $S_v$ that contains cached shortest distances from processed vertices to $v_q$ obtained in the previous round {(Lines 3-4)}. 
In each candidate node $n_c$, there are three different types of vertices to be processed in this step, namely 1)  unprocessed live vertices (denoted by $\mathcal{V}^u$) that are either newly emerged in the current round or already existing in the previous round,
2)  processed live vertices from the previous round ($\mathcal{V}^p$) that may either still exist or have disappeared in the current round, and 3) {processed} border vertices ($\mathcal{V}^b$). 

We next discuss how to process each type of vertices. In the following procedure, we will distinguish between two types of processed vertices which we call {\em accomplished vertices} or {\em unaccomplished vertices}, with the implication that the accomplished vertices will no longer need to be visited when traversing the global skeleton graph in the next step. {Note that the processed and unprocessed vertices are not only based on the previous round, but also include those produced in the current round.} More specifically: 1) a processed border vertex or the query vertex will be considered accomplished if it has no unprocessed adjacent vertex; otherwise, such a vertex is considered unaccomplished; 2) a processed live vertex is considered accomplished even if it has unprocessed adjacent vertex considering it is irrelevant in identifying the shortest distances from its adjacent vertices to $v_q$ as per  Theorem~\ref{theo:sd-border-live-vertices}.

{\bf For any unprocessed live vertex $v_l \in\mathcal{V}^u$}, its shortest distance to $v_q$ has to be computed. Depending on the current candidate node $n_c$, two cases arise:

{\bf(1)} If all border vertices in $n_c$ have been processed in the previous round, the shortest distance from $v_l$ to $v_q$ can be easily computed as $SD(v_l, v_q) = \min_{{v_b}\in\mathcal{V}^{b}}\lbrace SD(v_l, v_b)+SD(v_b, v_q)\rbrace$. Note that $SD(v_l, v_b)$, i.e., the shortest distance from $v_l$ to any border vertex $v_b$ in $n_c$, must have been identified during the maintenance of ODIN that precedes the execution of {ODIN-KNN-Inc}, while $SD(v_b, v_q)$, the shortest distance from any border vertex $v_b$ to $v_q$, must have also been known in the previous round. After that, we use the associated objects of $v_l$ to update the current nearest $k$ objects in $HP_k$, and $v_l$ is set accomplished after processed. 

{\bf (2)} Otherwise, if there is any unprocessed border vertices in $n_c$,  then the shortest distance from $v_l$ to $v_q$ cannot be inferred immediately and further exploration is required utilizing the ODIN-KNN-Init, as discussed in Section \ref{sec:incremental-computation:step3}.

{\bf For each processed live vertex $v_l\in \mathcal{V}_p$ from the previous round}, if it is {still a live vertex} in the current round, we consider it accomplished. In this case, all associated objects of $v_l$ could potentially be part of the final $k$NN result, including those objects that are currently associated with $v_l$ as well as those that were formerly associated with $v_l$ and inserted into $HP_k$ in the previous round (and thus cached in $S_v$ in {ODIN-KNN-Init}). We As the formerly associated objects may have changed their locations in the current round, we have to examine such objects and update $HP_k$ accordingly. {In the other case where $v_l$ has become obsolete in the current round, we simply remove its formerly associated objects from $HP_k$.}

{\bf Finally, for each processed border vertex $v_b\in\mathcal{V}_b$}, if its all adjacent vertices have been processed, it is marked as accomplished; otherwise, it is considered unaccomplished {and kept for the next step}. 
{In the above process, we use a set ${S_v}'$ to cache the processed vertices for the incremental exploration in the next round.}  

\vspace{-4pt}
\subsubsection{Iteratively expanding the search scope} 
\label{sec:incremental-computation:step3}
The final $k$NNs of $v_q$ are identified and the algorithm terminates once the following termination condition is met: $HP_k$ is full and the shortest distance from the top object $o^*$ in $HP_k$ to $v_q$ is not greater than the shortest distance from 
the nearest {unaccomplished} vertex to $v_q$. 
When the algorithm terminates, the objects in $HP_k$ are reported as the $k$NNs of $v_q$ (Lines 6-7).

If the termination condition has not been met, we insert the unaccomplished vertices {that have been captured during the exploration of candidate nodes in the last step} into the priority queue $Q_d$ used by the {ODIN-KNN-Init} algorithm to prioritize vertex visits, and then continue to explore the remaining part of the global skeleton graph (excluding the accomplished vertices), denoted by $G^{s'}$, utilizing {ODIN-KNN-Init} ($G^{s'}$, $HP_k$, $Q_d$, $S'_v$, $v_q$) till the termination condition is met or $Q_d$ is empty (Line 8).

\begin{figure}[ht!]
  \centering
    \begin{subfigure}{0.235\textwidth}
      \centering   
\includegraphics[width=\linewidth]{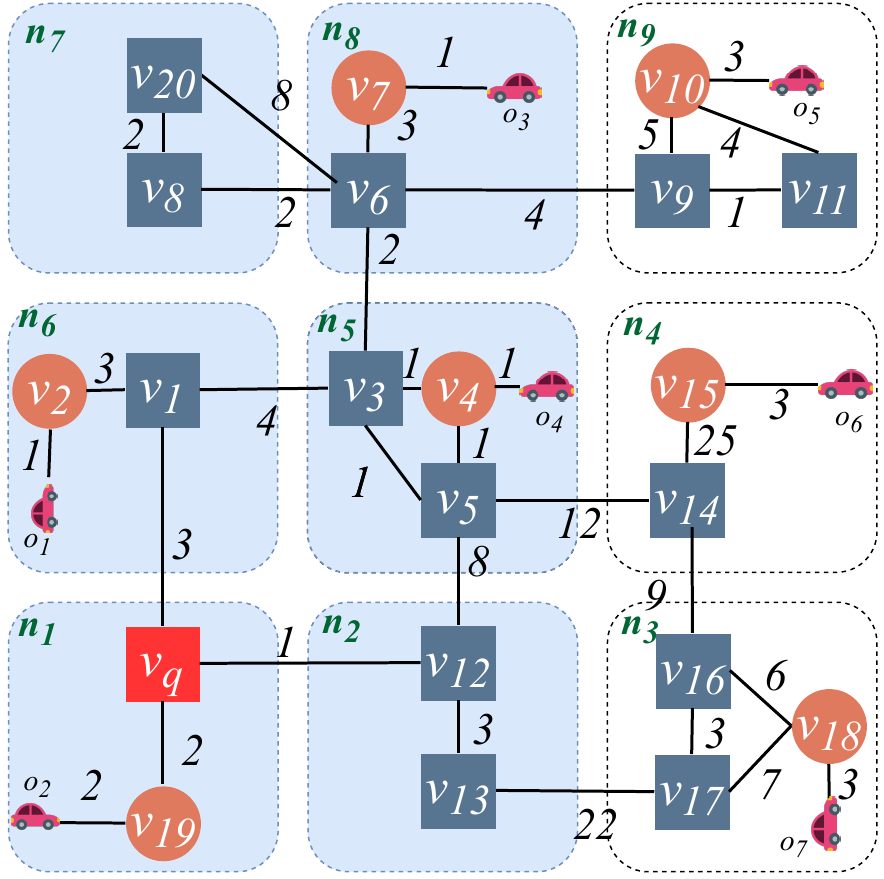}       \caption{{$G^s$}}
        \label{fig:skeleton-graph-previous}
    \end{subfigure}
    \hspace{0.05in}
    \begin{subfigure}{0.225\textwidth}
\includegraphics[width=\linewidth]{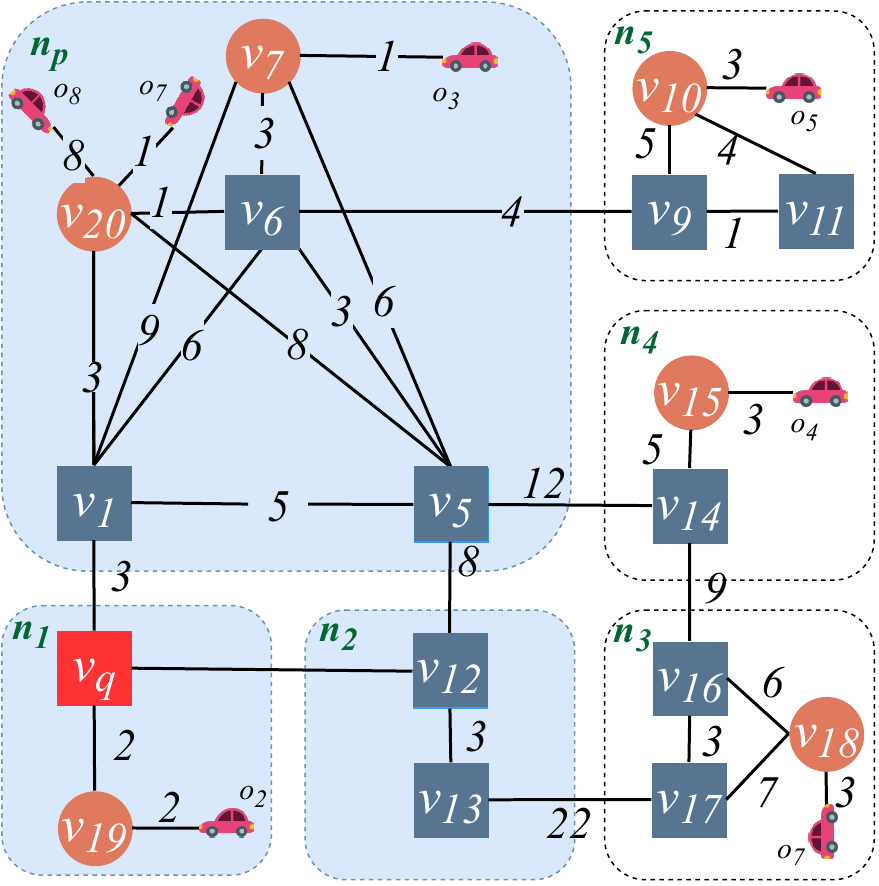}
\captionsetup{justification=raggedleft,width=1.5in}
\caption{{$G^{s'}$}}
        \label{fig:skeleton-graph-current}
    \end{subfigure}
\vspace{-0.2cm}
\caption{Evolving Global Skeleton Graph}\label{exam:Inc-knn}
\end{figure}

\begin{example}
In illustrating {ODIN-KNN-Inc} using global skeleton graphs $G^s$ and $G^{s'}$ (Fig.~\ref{exam:Inc-knn}), let's consider a scenario. Initially, 4 NNs of $v_q$ are determined based on the global skeleton graph $G^s$ from the previous round, where shaded nodes represent processed vertices. Now, $G^s$ has evolved into $G^{s'}$ (Fig.~\ref{fig:skeleton-graph-current}). In the current round, nodes $n_5$, $n_6$, $n_7$, and $n_8$ have merged into $n_p$, causing changes in live vertices. {ODIN-KNN-Inc} incrementally seeks the new 4NNs of $v_q$ based on $G^{s'}$. First, it identifies candidate nodes $n_1$, $n_2$, and $n_p$. For $n_1$ and $n_2$, since their vertices were processed in the previous round and the object $o_1$ didn't move, we only update the status of $v_{12}$ and $v_{19}$ as accomplished and $v_{13}$ as unaccomplished. When exploring $n_p$, it removes objects $o_1$ and $o_4$ from $HP_k$ because live vertices $v_2$ and $v_4$ disappeared. The algorithm computes $SD(v_{20},v_q)$ as the minimum value among the distances between $v_{20}$ and the remaining vertices. It inserts the associated objects $o_7$ and $o_8$ of $v_{20}$ into $HP_k$ and marks $v_1$ and $v_7$ as accomplished, while $v_5$ is labeled unaccomplished. At this stage, the current 4NNs are $\lbrace o_2, o_7, o_3, o_8\rbrace$, and the unaccomplished vertices include $v_{13}$, $v_5$, and $v_6$. In the third step, since $SD(o_8, v_q)>SD(v_{13}, v_q)$, unaccomplished processed vertices are added to $Q_d$, and {ODIN-KNN-Init} is applied to traverse $G^{s'}$, excluding the accomplished vertices in the shaded nodes. This process continues until the final 4NNs are identified.

 \end{example}

{{\bf Cost analysis.} For simplicity, we approximate the cost of ODIN-KNN-Inc {by the cost} of expanding the remaining part of $G^s$ with ODIN-KNN-Init, under the assumption that the cost of processing the live vertices in the explored candidate nodes can be considered constant, given that their shortest distances to $v_q$ can be inferred without traversing $G^s$. Suppose the number of border and live vertices in the remaining part of $G^s$ are $b'$ and $l'$ respectively, and the average degree of each border vertex is $d_g$. We can show that the cost of traversing the remaining part of $G^s$ is $O(b'\cdot d_g\cdot\log(b'+l')+l')$. This is consistent with how the cost of traversing $G^s$ is computed in Section~\ref{subsec:odin-knn-init}.}

\vspace{-0.3cm}
\section{Experiments}\label{sec:exp}
\subsection{Experiment setup and datasets}
The experiments are conducted on a cloud server with 32 virtual cores and 128GB RAM. All programs are implemented in Java. 

{\bf Datasets.} 
This evaluation employs multiple synthetic datasets and one real dataset. To generate the synthetic datasets, we adopt five real road network datasets including New York (NY), Florida (FLA), California and Nevada (CAL), East USA (EUSA), and Central USA (CUSA)~\cite{DIMACS}. For each road network, we simulate three sets of objects that respectively follow the Zipfian Distribution (ZD), the Gaussian Distribution (GD), and the Uniform Distribution (UD) to be used in the experiments. 
Additionally, we transform T-Drive~\cite{yuan2011driving}, a real taxi trajectory dataset collected from Beijing (BJ), into a moving object dataset which we call TD. T-Drive contains timestamped location records taken from 10,357 taxis, and each taxi's location is sampled every 177 seconds on average. 

In all experiments, a snapshot of the objects is taken once per ten seconds by letting each object in the dataset
report its current location and update the $k$NNs of queries at the same frequency based on the new snapshot. 
Our experiments with other snapshot frequencies reveal the same trends. In each set of experiments, all results shown are the average of 20 runs by default. Finally, we summarize the information of each road network in Table~\ref{table:size}{, and the value ranges of parameters used in the evaluation are given in Table~\ref{tab:expri-parameters}. The following evaluations involving those parameters will take their default values unless otherwise specified}. 

{\bf Baselines.} We implement S-Grid~\cite{huang2007s}, ER-$k$NN~\cite{demiryurek2009efficient}, {INSQ~\cite{7498339}}, V-Tree~ \cite{shen2017v}, SIM$k$NN~\cite{cao2018sf}, and TEN*-Index~\cite{ouyang2020progressive} as baseline methods, which have been discussed in Section~\ref{sec:related-work}. {When building ODIN and V-tree, we divide the index computation for each node into parallel tasks to reduce the construction time for subsequent evaluations. In order to compare fairly with other baselines on construction cost, we introduce two additional baselines: ODIN-Single and V-tree-Single. These baselines construct ODIN and V-tree respectively without the parallel acceleration. Additionally, we introduce a simplified version of ODIN called S-ODIN, which removes the shortcuts connecting the border and live vertices {and retains only} the shortcuts between border vertices within each materialized node. This is done to test the benefit of the removed shortcuts.

{We have also included OCP~\cite{li2010efficient} as a baseline which updates the $k$NNs incrementally based on object movement (i.e., closer to or farther from the query point). However, it relies on precomputed shortcuts between all pairs of vertices on road networks, incurring a significantly higher overhead, nearly three orders of magnitude greater than ODIN's construction time (e.g., 9,870 seconds for OCP vs. 2 seconds for ODIN on NY). Therefore, comparing construction times is not relevant. Since OCP does not involve maintenance costs, our primary focus is on query performance, depicted in Fig~10(a), (d), and (g).}

\vspace{0.05in}
\begin{table}[h]
\centering
\caption{\small {Statistics on the Road Network Datasets}}\label{table:size}
\vspace{-0.1in}
\resizebox{\linewidth}{!}{
\begin{tabular}{cccccc}
\hline
{\bf Road Networks} & {\bf \#Vertices} & {\bf \#Edges} & {\bf Road Networks} & {\bf \#Vertices} & {\bf \#Edges} \\
\hline
NY & 264,346 & 733,846 & BJ & 2,280,770 & 4,829,430 \\
\hline
FLA & 1,070,376 & 2,712,798 & EUSA & 3,598,623 & 8,778,114 \\
\hline
CAL & 1,890,815 & 4,657,742 & CUSA & 14,081,816 & 34,292,496\\
\hline
\end{tabular}}
\end{table}

\begin{table}[h]
\newcommand{\tabincell}[2]{\begin{tabular}{@{}#1@{}}#2\end{tabular}}
\centering
\caption{\small{Summary of Parameters Used in Evaluation}}\label{tab:expri-parameters}
\vspace{-0.1in}
\smallskip\noindent
 \fontsize{7pt}{6pt}\selectfont
\begin{tabular}{|c|c|c|}
\hline
{\bf Parameters} & {\bf Meaning} & {\bf \makecell{Value range\\(bold number is default value)}}\\
\hline
$k$ & \tabincell{c}{number of nearest neighbors}& {\bf 10}, 20, 30, 40, 50\\
\hline
$m$ & branches of a node & 2, {\bf 4}, 6, 8, 10  \\
\hline
$z$ & subgraph size threshold &100, 200, {\bf 300}, 400, 500\\
\hline
$\mu$ & \tabincell{c}{underfilled threshold}& {\bf 5}, 10, 15, 20, 25\\
\hline
$\Gamma_o$ & number of objects & {\bf 30k}, 60k,90k, 12k, 15k\\
\hline
$P_o$ &  \tabincell{c}{percentage of objects moving}& {\bf 25\%}, {50\%}, 75\%, 100\%\\
\hline
\end{tabular}
\end{table}

\vspace{-0.1in}
\subsection{Evaluating {ODIN}}
Our evaluation of ODIN involves its construction time, memory consumption, and maintenance cost. 

\begin{figure*}[thb!]
  \centering
  \captionsetup{font={small}}
  \begin{subfigure}{0.162\linewidth}
      \centering         \includegraphics[width=\textwidth]{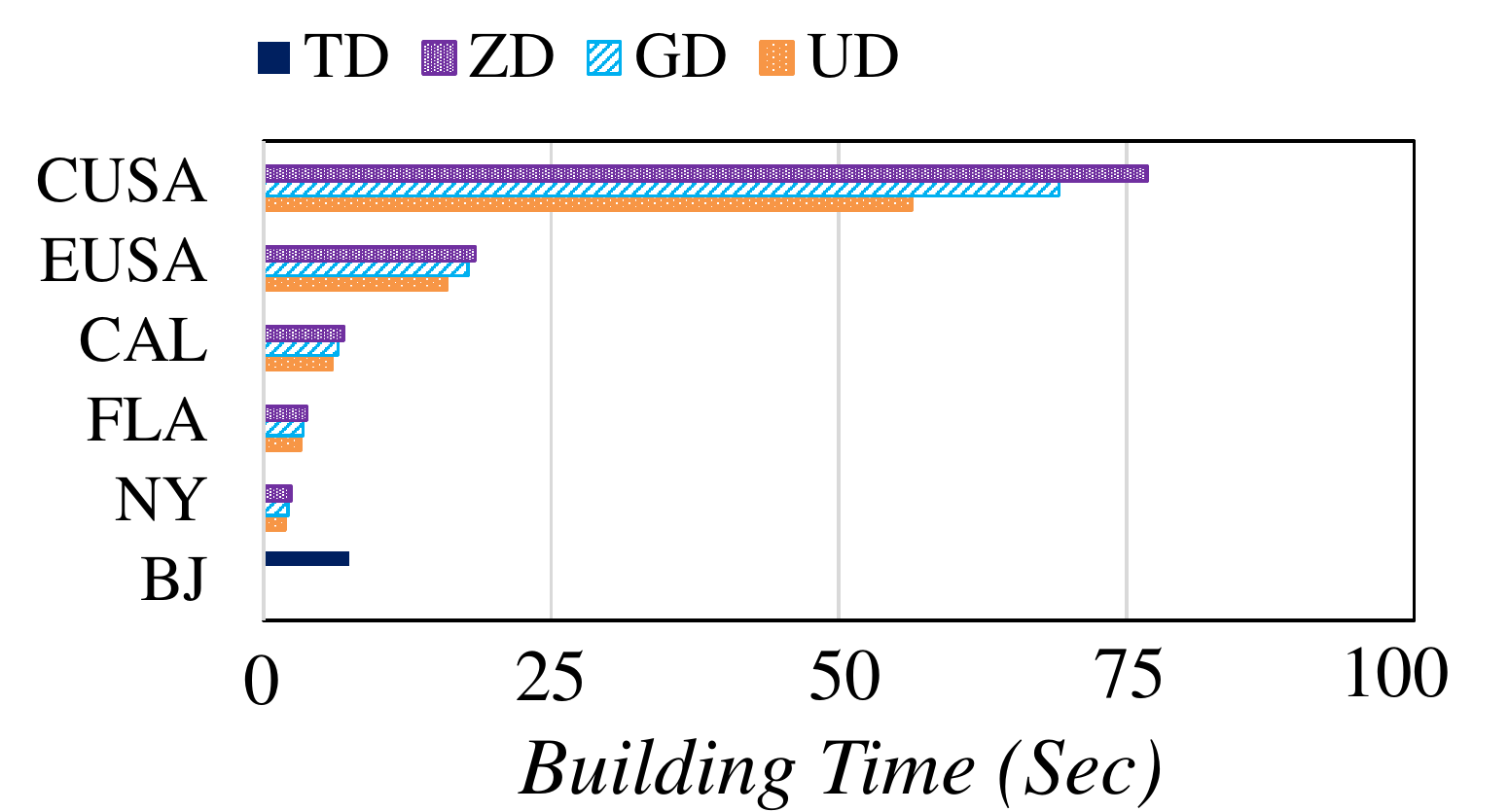}
      \vspace{-0.6cm}
\caption{\scriptsize{Construction Time}}\label{building-time-all}
    \end{subfigure}      
    \begin{subfigure}{0.162\linewidth}
      \centering   
\includegraphics[width=\textwidth]{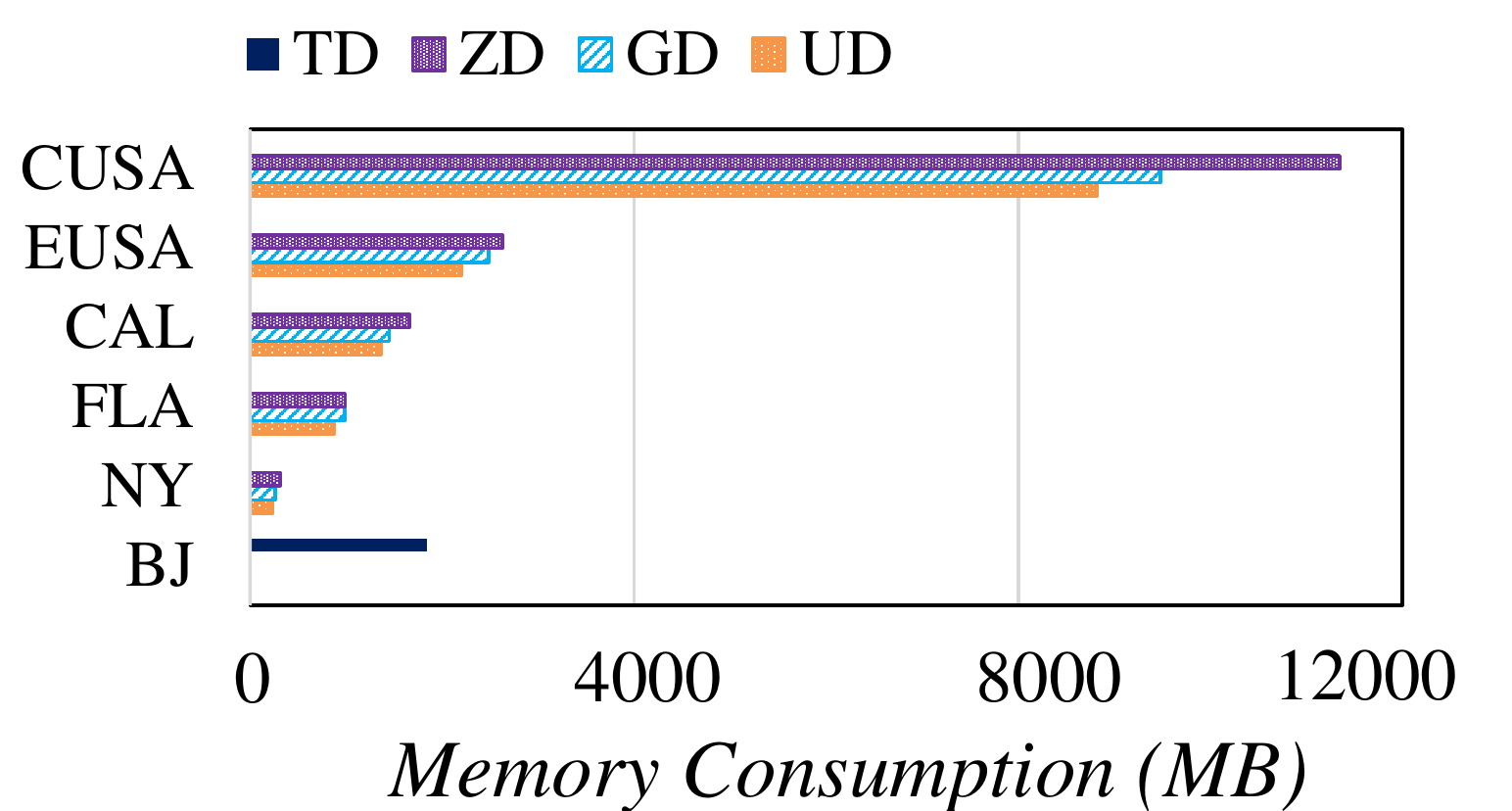}
      \vspace{-0.6cm}
      \caption{\scriptsize{Memory Consumption}}\label{memory-consumption-all}
    \end{subfigure}
    \begin{subfigure}{0.162\linewidth}
      \centering   
\includegraphics[width=\linewidth]{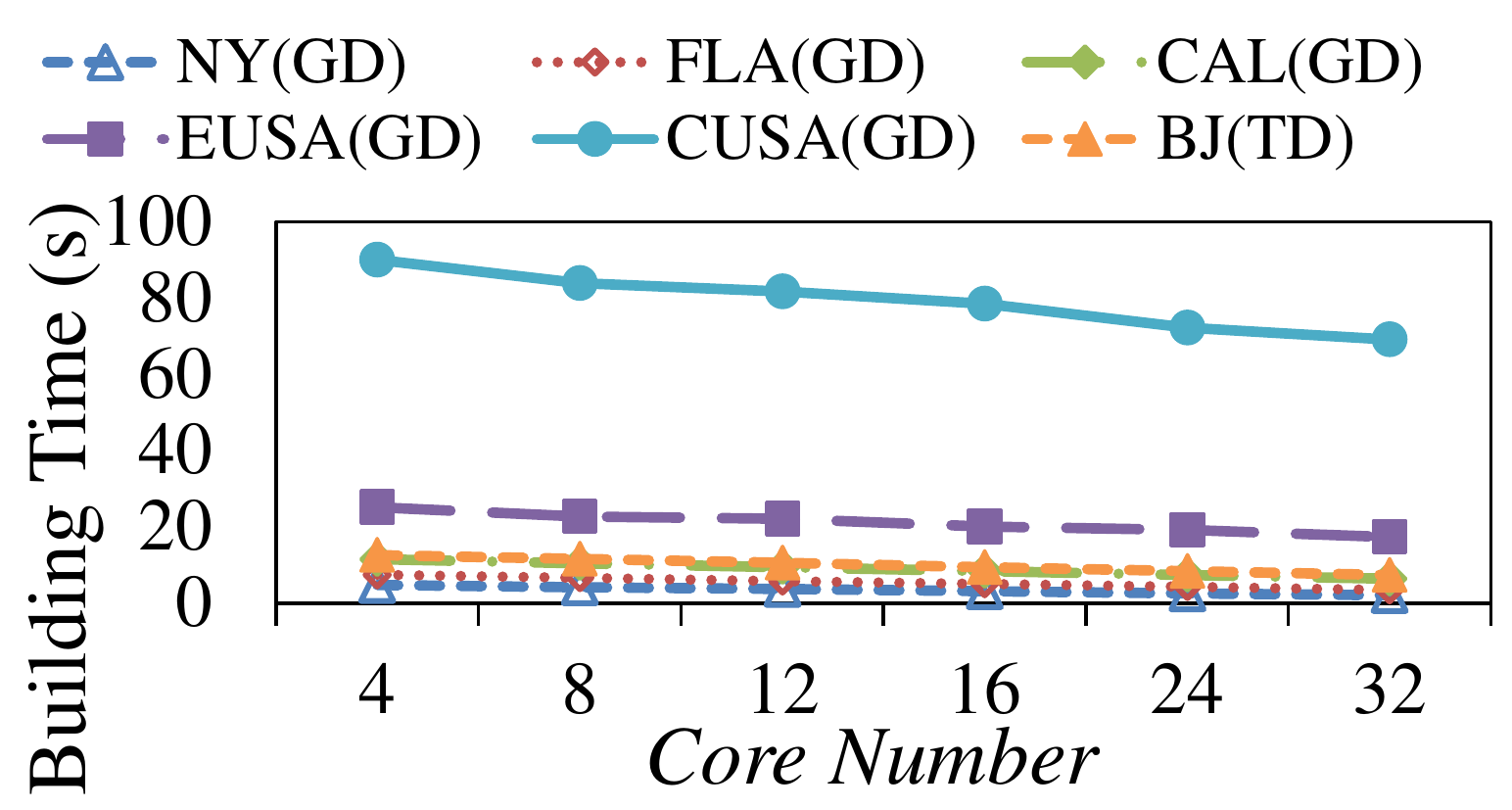}
      \vspace{-0.6cm}
    \caption{\scriptsize{Influence of $\Gamma_c$}}
        \label{fig:core-number-building-cost}
    \end{subfigure}
    \begin{subfigure}{0.162\linewidth}
      \centering   
\includegraphics[width=\linewidth]{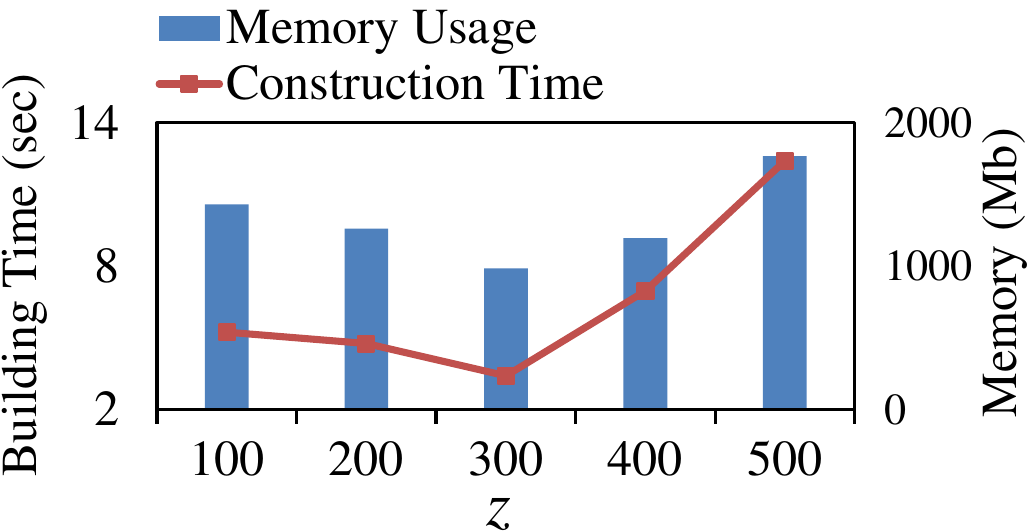}
      \vspace{-0.6cm}
    \caption{\scriptsize{Influence of $z$ (FLA)}}
        \label{fig:l-CUSA}
    \end{subfigure}
     \begin{subfigure}{0.162\linewidth}
      \centering       \includegraphics[width=\textwidth]{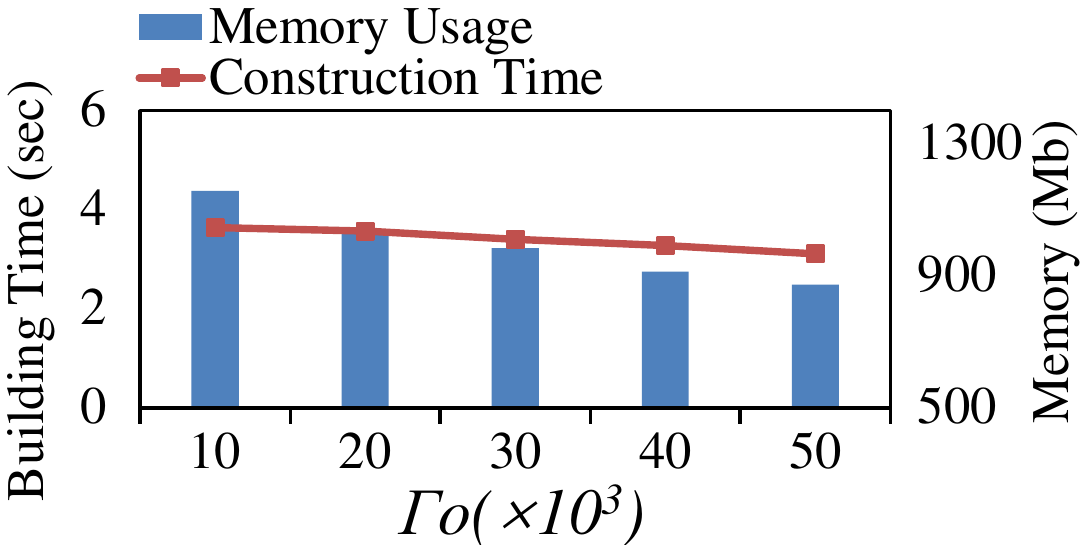}
      \vspace{-0.6cm}
    \caption{\scriptsize{Influence of $\Gamma_{o}$} (FLA)}\label{exp:build-cost-object-OD}
    \end{subfigure}
    \begin{subfigure}{0.162\linewidth}
      \centering   
\includegraphics[width=\textwidth]{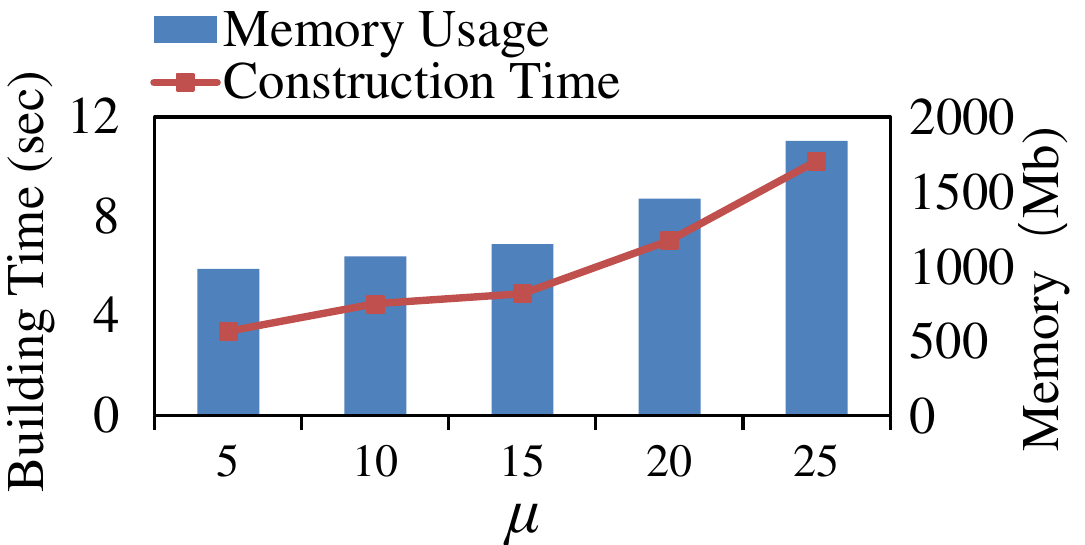}
      \vspace{-0.6cm}
    \caption{\scriptsize{Influence of  $\mu$} (FLA)}\label{exp:build-cost-mu}
    \end{subfigure}
\vspace{-0.5cm}
\caption{
{Construction Time and Memory Consumption of ODIN}
}\label{exp:query-cost}
 \vspace{-0.4cm}
\end{figure*}


\begin{figure*}[h!]
\vspace{0.2cm}
  \centering
  \captionsetup{font={small}}
    \begin{subfigure}{0.162\textwidth}
      \centering         \includegraphics[width=\textwidth]{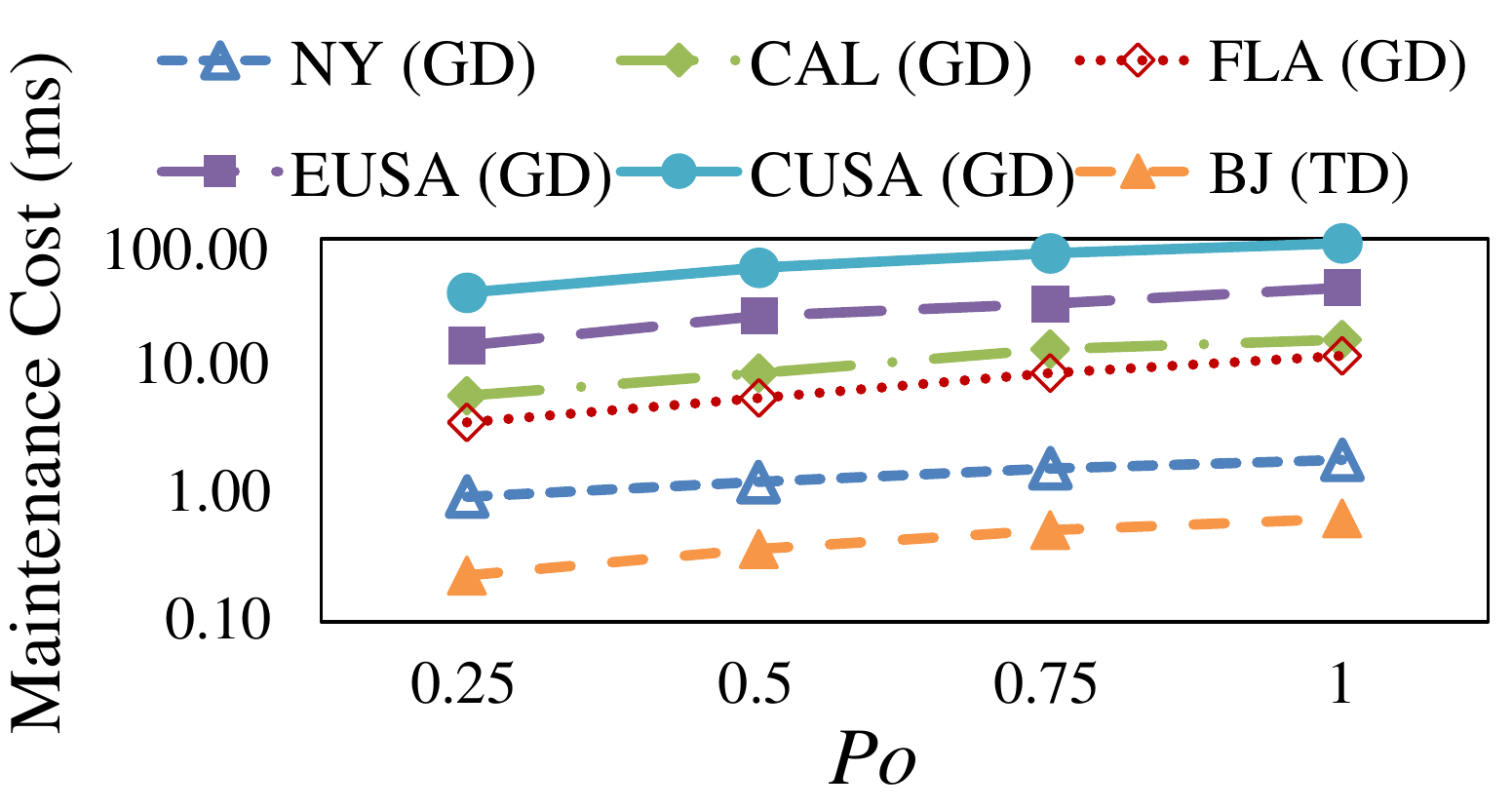}
      \vspace{-0.6cm}
\caption{\scriptsize{Maintenance Cost ($P_o$)}}
        \label{fig:maintenance-po}
    \end{subfigure} 
     \begin{subfigure}{0.162\textwidth}
      \centering   
\includegraphics[width=\textwidth]{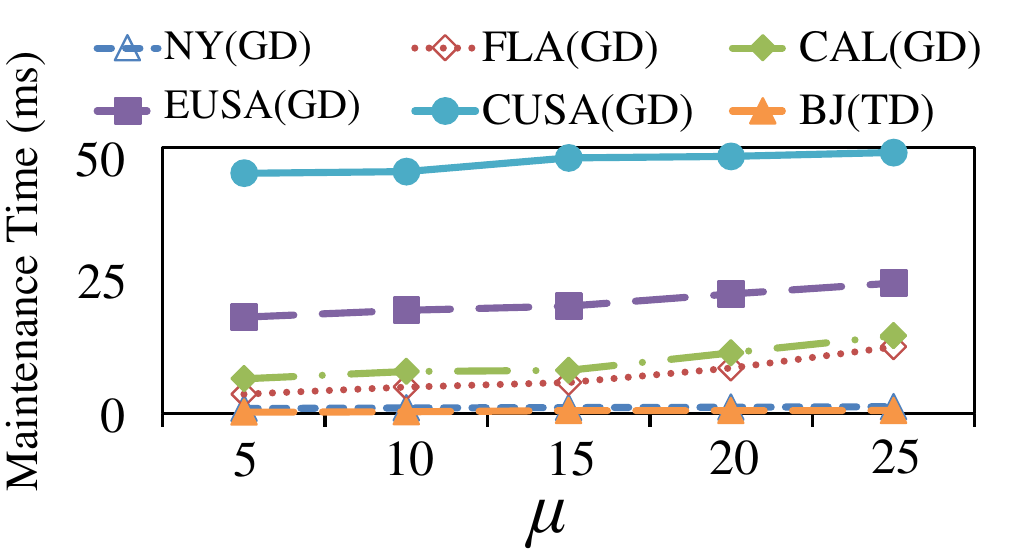}
    \vspace{-0.6cm}
\caption{\scriptsize{Maintenance Cost ($\mu$)}}\label{fig:Maintenance-cost-mu}
    \end{subfigure}
    \begin{subfigure}{0.162\textwidth}
      \centering   
\includegraphics[width=\linewidth]{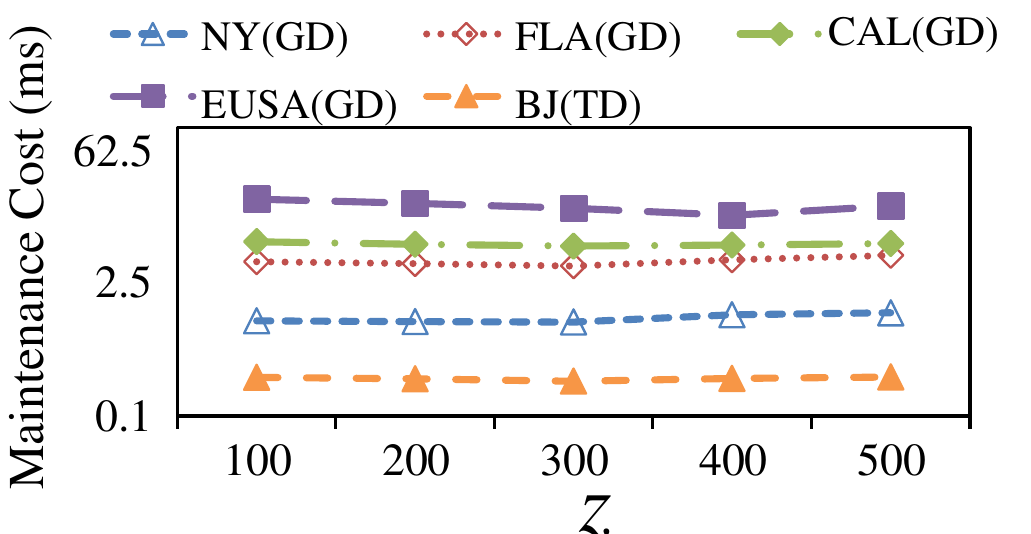}
      \vspace{-0.6cm}
\caption{\scriptsize{Maintenance Cost ($z$)}}
        \label{fig:maintenance-z}
    \end{subfigure}
    \begin{subfigure}{0.162\textwidth}
      \centering   
\includegraphics[width=\textwidth]{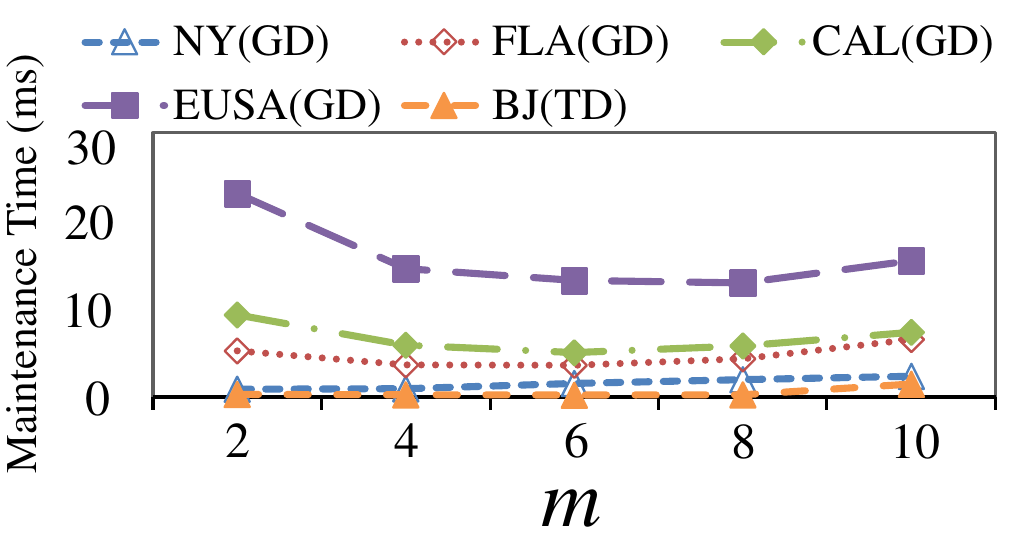}
     \vspace{-0.6cm}
\caption{\scriptsize{Maintenance Cost ($m$)}}\label{fig:Maintenance-m}
    \end{subfigure}
        \begin{subfigure}{0.162\textwidth}
      \centering   
\includegraphics[width=\linewidth]{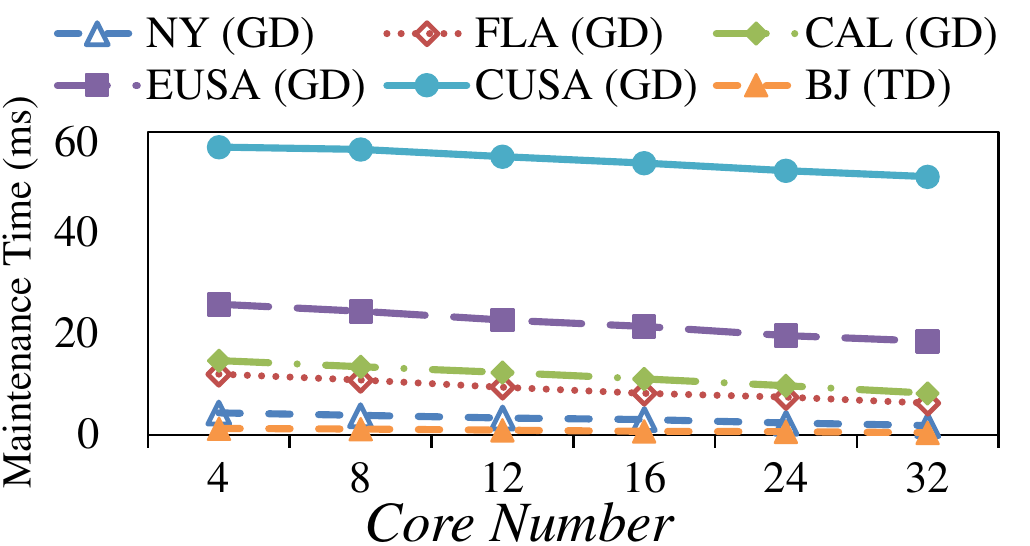}
      \vspace{-0.6cm}
\caption{\scriptsize{Maintenance Cost ($\Gamma_c$)}}
        \label{fig:maintenance-core-number}
    \end{subfigure}
        \begin{subfigure}{0.162\textwidth}
      \centering   
\includegraphics[width=\linewidth]{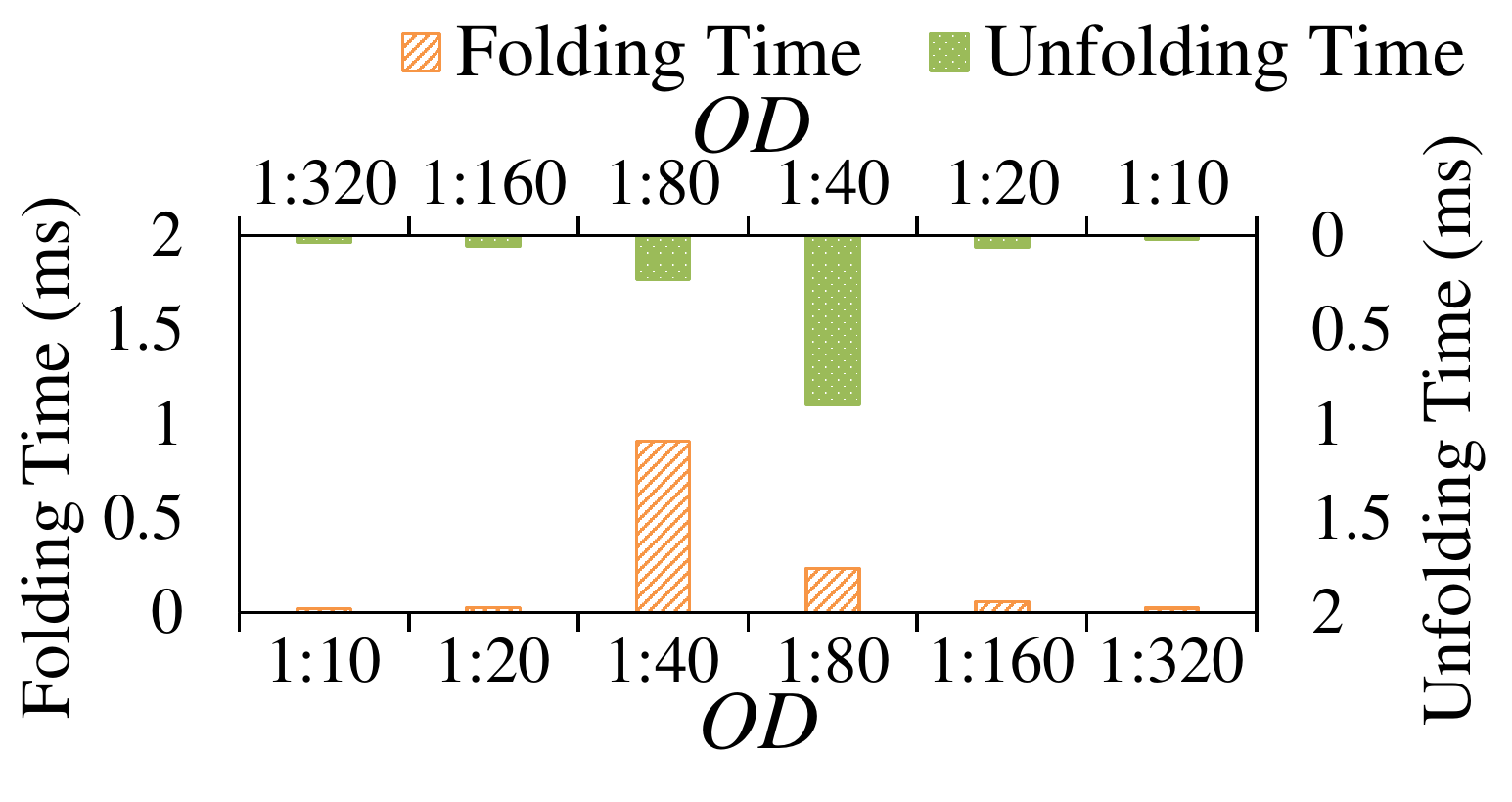}
      \vspace{-0.6cm}
\caption{\scriptsize{Maintenance Cost (OD)}}
        \label{fig:fold-unfold-density}
    \end{subfigure}
\vspace{-0.5cm}
\caption{Effects of Parameters on Maintenance Cost 
}\label{exp:maintenance-cost}
\vspace{-0.3cm}
\end{figure*}

\begin{figure*}[h!]
  \centering
  \captionsetup{font={small}}
      \begin{subfigure}{0.162\linewidth}
      \centering   
\includegraphics[width=\linewidth]{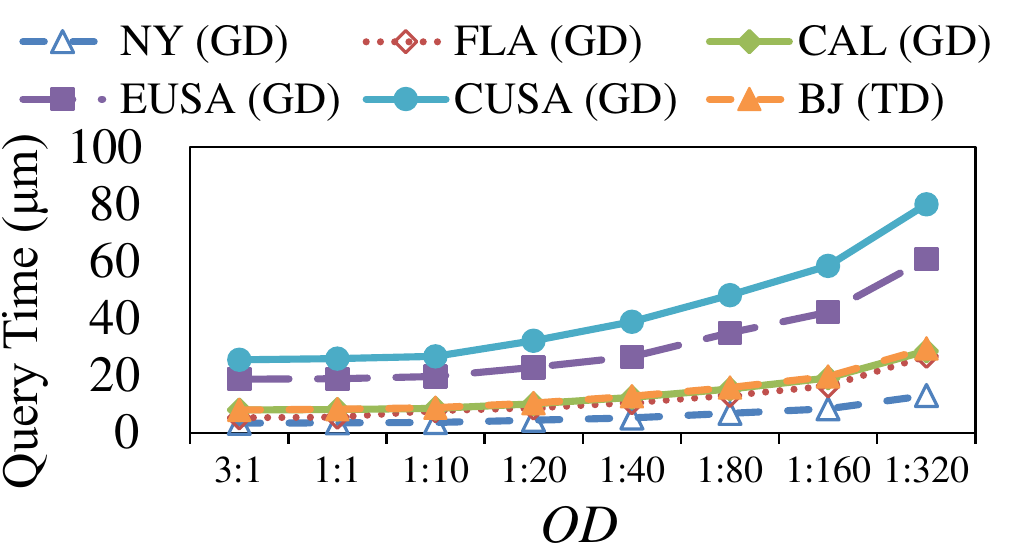}
      \vspace{-0.6cm}
        \caption{\scriptsize{Query Time w.r.t. OD}}
        \label{fig:query-cost-OD}
    \end{subfigure}
  \begin{subfigure}{0.162\linewidth}
      \centering         \includegraphics[width=\linewidth]{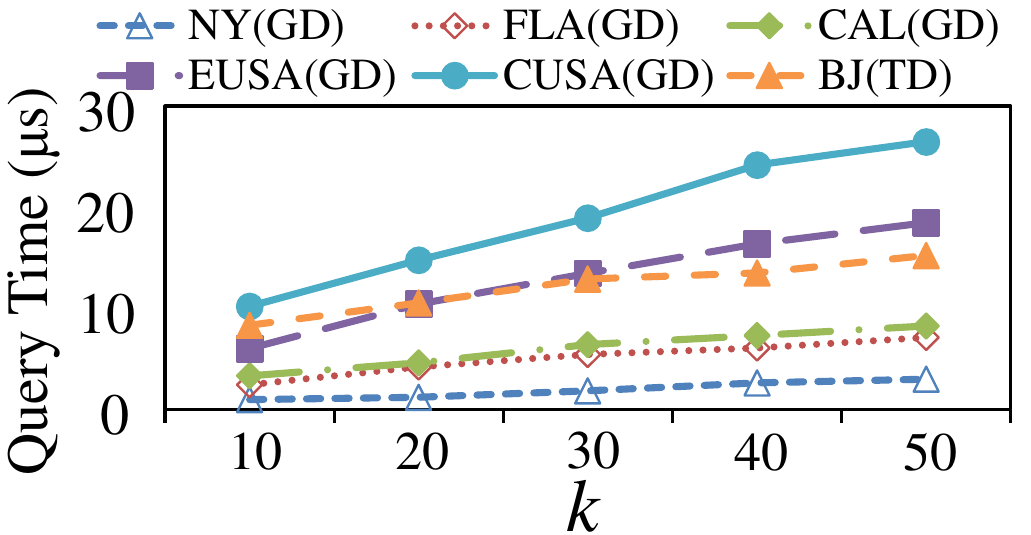}
      \vspace{-0.6cm}
        \caption{\scriptsize{Query Time w.r.t. $k$}}
        \label{fig:query-cost-k}
    \end{subfigure}       
    \begin{subfigure}{0.162\linewidth}
      \centering   
\includegraphics[width=\linewidth]{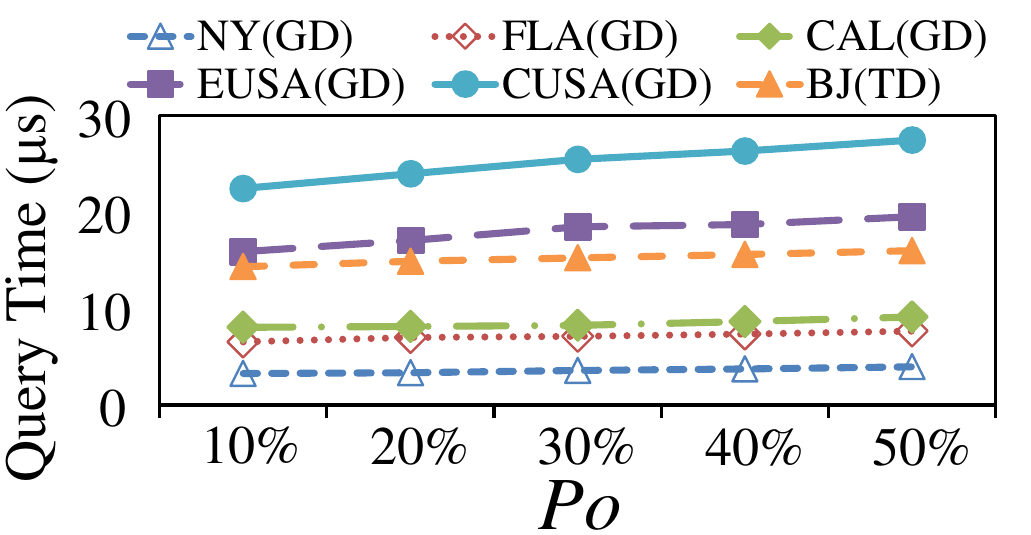}
      \vspace{-0.6cm}
        \caption{\scriptsize{Query Time w.r.t. $P_o$}}
        \label{fig:query-cost-Po}
    \end{subfigure}
    \begin{subfigure}{0.162\linewidth}
      \centering   
\includegraphics[width=\linewidth]{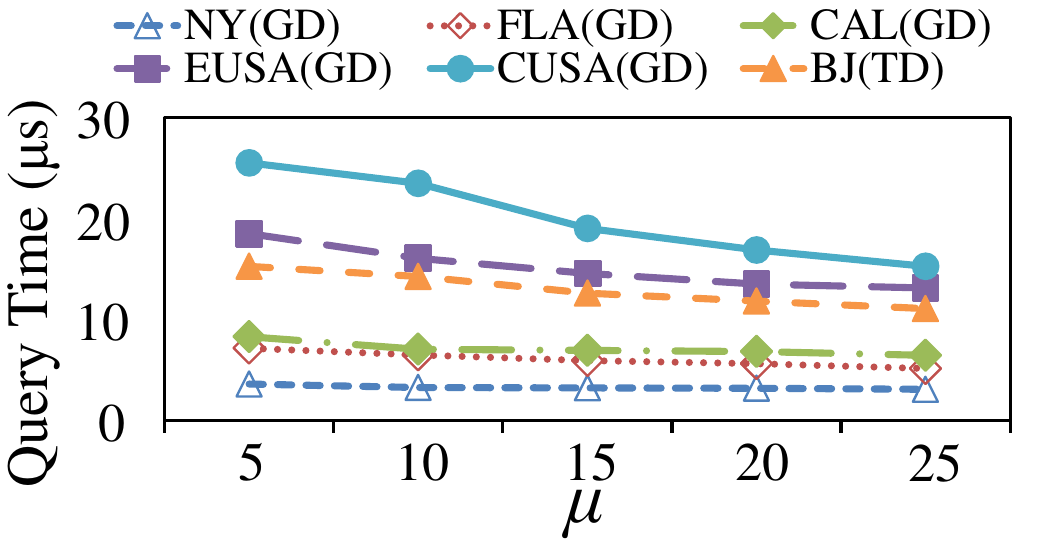}
      \vspace{-0.6cm}
        \caption{\scriptsize{Query Time w.r.t. $\mu$}}
        \label{fig:query-cost-mu}
    \end{subfigure}
     \begin{subfigure}{0.162\linewidth}
      \centering   
\includegraphics[width=\linewidth]{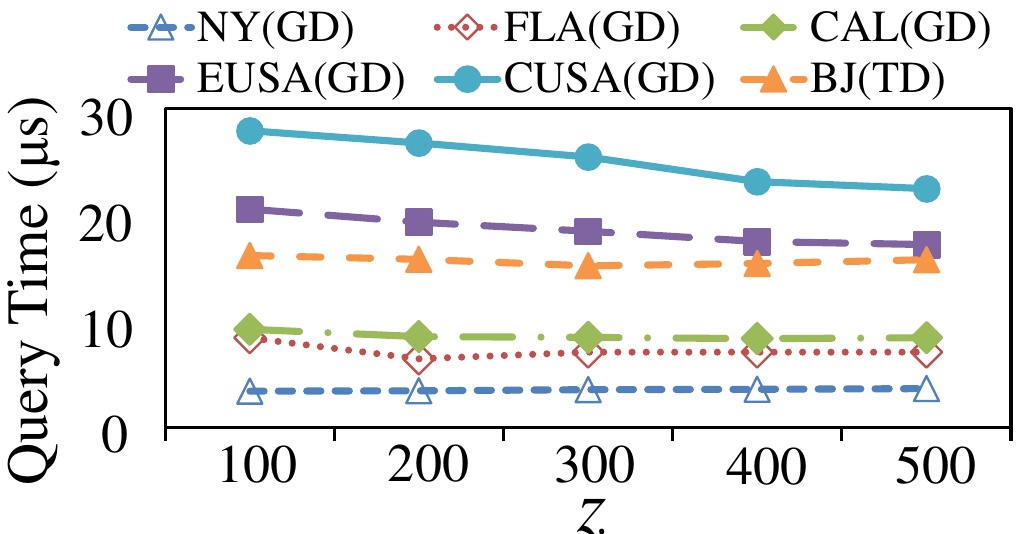}
      \vspace{-0.6cm}
        \caption{\scriptsize{Query Time w.r.t. $z$}}
        \label{fig:query-cost-z}
    \end{subfigure}
    \begin{subfigure}{0.162\linewidth}
      \centering         \includegraphics[width=\linewidth]{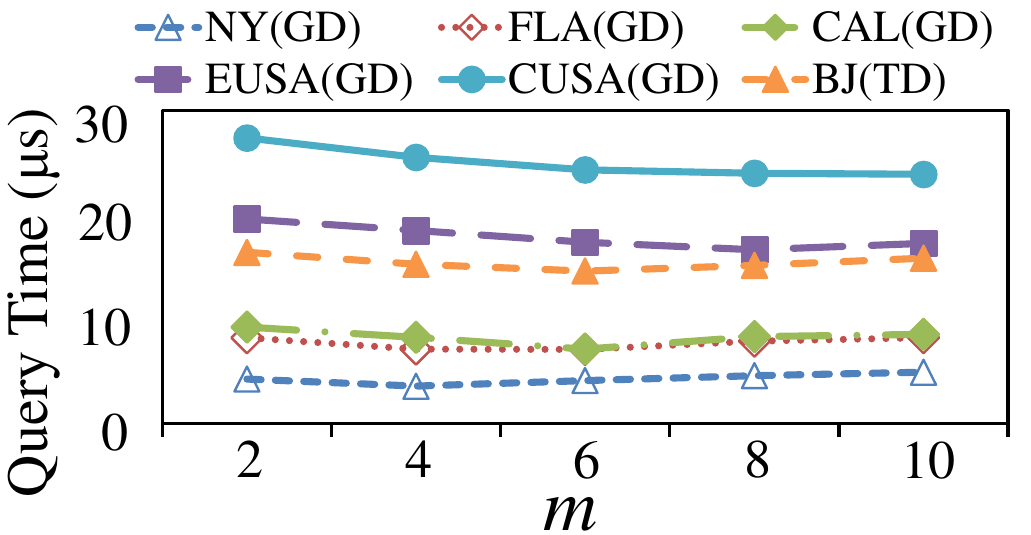}
      \vspace{-0.6cm}
        \caption{\scriptsize{Query Time w.r.t. $m$}}
        \label{fig:query-cost-m}
    \end{subfigure}
\vspace{-0.4cm}
\caption{Effects of Parameters on Query Cost} \label{exp:query-cost}
\vspace{-0.3cm}
\end{figure*}

{\bf Index construction cost.} 
Fig.~\ref{building-time-all} depicts the elapsed time of constructing {ODIN}. More time is required to construct {ODIN} on larger road networks, in line with our intuition, as {ODIN} on larger road networks contains more leaf nodes and the shortest distance between any pair of vertices within each leaf node has to be precomputed.
The effect of the object distribution is much more pronounced on CUSA, as CUSA is much larger than the other road networks.
Compared with UD, a greater part of the road network is sparse in GD and ZD on CUSA, and thus more nodes in {ODIN} covering that part of the road network are likely to be folded during the construction, which increases the construction cost. For the same reason, the memory consumption on each road network has a similar trend shown in Fig.~\ref{memory-consumption-all}. 

{Fig.~\ref{fig:core-number-building-cost} illustrates the decrease in construction cost of ODIN as the number of cores increases. The construction cost is more significantly reduced on CUSA, indicating the effectiveness of parallel acceleration and the optimization of MPBS in computing shortcuts between live and border vertices within each materialized non-leaf node.}


{{Figs.}~\ref{fig:l-CUSA}-\ref{exp:build-cost-mu} depict the size and construction time of {ODIN} under different values of $z$, $\Gamma_o$, and $\mu$. These evaluations reveal similar trends on all road networks, so we present the results on FLA only.} From Fig.~\ref{fig:l-CUSA}, we observe that the construction time first decreases and then increases as $z$ grows. The reason is that the number of leaf nodes in {ODIN} drops dramatically as $z$ starts to grow in the beginning, which also leads to a decrease in construction time. However, as $z$ continues to grow, the size of the leaf subgraphs starts to increase. Consequently, the cost of computing the shortest distances for all pairs of vertices within each leaf node also increases, which outweighs the decrease in the number of leaf nodes when $z$ is beyond a certain level (e.g., 300 on FLA). For the same reason, the memory consumption has a similar trend to that of the construction time.  

Fig.~\ref{exp:build-cost-object-OD} shows that both the construction time and the memory usage of {ODIN} decrease slowly with increasing $\Gamma_o$ (the objects number). In general, $\Gamma_o$ is proportional to the number of live vertices. Too small a value of $\Gamma_o$ will result in many active nodes being underfilled and thus having to be folded during the construction, leading to the relatively large construction time and memory usage. As $\Gamma_o$ increases, the rising number of live vertices leads to a  decrease in the number of active nodes involved in the folding operations, resulting in a slow decline in both construction time and memory usage. Additionally, we evaluate the influence of $\mu$ for a fixed $\Gamma_o$. As shown in Fig.~\ref{exp:build-cost-mu}, both construction time and memory storage increase as $\mu$ grows. The reason is that with a larger $\mu$, more active nodes will meet the Folding Criteria, leading to an increasing  number of folding operations. \looseness=-1
\vspace{0.1cm}
\begin{table}[h]
\captionof{table}{\small {Impact of OD on Folding and Unfolding Operations}}\label{Tab:OD-Fold-Unfold}
\vspace{-0.03in}
\begin{minipage}[c]{0.235\textwidth}
\centering
    \fontsize{7pt}{6pt}\selectfont
    \begin{tabular}{|c|c|c|}
  \hline
  \multirow{2}{*}{{OD decreases}}  &  \multicolumn{2}{c|}{{ folding operations}}  \\
  \cline{2-3} & {number} & {ratio} \\
  \hline
  \scriptsize{1:10} & \scriptsize{0} & \scriptsize{0} \\
  \hline
  \scriptsize{1:20} & \scriptsize{4} & \scriptsize{1.6\%} \\
  \hline
  \scriptsize{1:40} & \scriptsize{252} & \scriptsize{99.6\%} \\
  \hline
  \scriptsize{1:80} & \scriptsize{60} & \scriptsize{93.8\%} \\
  \hline
  \scriptsize{1:160} & \scriptsize{18} & \scriptsize{94.7\%} \\
  \hline
  \scriptsize{1:320} & \scriptsize{6} & \scriptsize{93.9\%} \\
  \hline
  \end{tabular}
\end{minipage}
\hspace{0.01cm}
\vspace{-0.03in}
\begin{minipage}[c]{0.235\textwidth}
\centering
   \fontsize{7pt}{6pt}\selectfont
  \begin{tabular}{|c|c|c|}
  \hline
  \multirow{2}{*}{{{OD increases}}}  &  \multicolumn{2}{c|}{{ unfolding operations}}  \\
  \cline{2-3} & {number} & {ratio} \\
  \hline
  \scriptsize{1:320} & \scriptsize{1} & \scriptsize{6.3\%} \\
  \hline
  \scriptsize{1:160} & \scriptsize{15} & \scriptsize{78.95\%} \\
  \hline
  \scriptsize{1:80} & \scriptsize{60} & \scriptsize{93.8\%} \\
  \hline
  \scriptsize{1:40} & \scriptsize{240} & \scriptsize{98.4\%} \\
  \hline
  \scriptsize{1:20} & \scriptsize{4} & \scriptsize{0.4\%} \\
  \hline
  \scriptsize{1:10} & \scriptsize{0} & \scriptsize{0} \\
  \hline
  \end{tabular}
\end{minipage}
\end{table}

\vspace{0.05in}
{\bf Maintenance cost.} {Fig.~\ref{exp:maintenance-cost} shows the maintenance cost of {ODIN} as we vary $P_o$, $\mu$, $z$, $m$, $\Gamma_c$, and OD, {where OD is measured by the ratio of the number of objects to the number of vertices.} 
On each dataset, we observe that the maintenance cost grows roughly linearly w.r.t. $P_o$, as more obsolete and newly appeared live vertices have to be processed, resulting in more folding and unfolding operations.

Next, we fix $P_o$ and vary $\mu$, $z$, and $m$. As shown in Fig.~\ref{fig:Maintenance-cost-mu}, the maintenance cost for each road network increases approximately linearly with growing $\mu$, since a greater $\mu$ leads to more underfilled active nodes to be folded.
Fig.~\ref{fig:maintenance-z} shows that the maintenance cost first slightly decreases and then increases as $z$ grows. When $z$ is very small, the sizes of active nodes are small on average, which makes the fluctuation in the number of live vertices in each active node more pronounced and thus increases the possibility of nodes being folded and unfolded.
When $z$ is beyond a certain level (e.g., $z$=400 on EUSA), the number of nodes affected by folding and unfolding operations tends to be stable but the cost of maintaining the larger skeleton graph for each active node becomes greater, which dominates the maintain cost and makes it start to increase slightly. Since $m$ has a similar relationship with the size of the active node as $z$, the effect of $m$ on the maintenance cost is also comparable to that of $z$ and is shown in Fig.~\ref{fig:Maintenance-m}. 

{Fig.~\ref{fig:maintenance-core-number} {shows that} the maintenance time slightly decreases with an increasing number of cores as more cores {speed up} MPBS in computing the shortcuts between live and border vertices}. Fig.~\ref{fig:fold-unfold-density} shows the impact of OD on folding and unfolding time, which dominate the maintenance cost per snapshot. We systematically decrease OD from 1:10 to 1:320 (shown in the bottom horizontal axis) at each snapshot in FLA and observe a sharp increase in folding time (represented by the left vertical axis) when OD is halved to 1:40 from 1:20, as a majority of active nodes need to be folded during this transition. {This is reflected in the left sub-table of Table~\ref{Tab:OD-Fold-Unfold}, where the ``ratio'' column indicates the percentage of active nodes involved in folding operations for each OD change.} After this, the number of folding operations and the folding time stabilize after a sharp {drop} as OD further declines from 1:40 to 1:320. When we increase OD from 1:320 to 1:10 (depicted by the top horizontal axis), the unfolding time (shown by the right vertical axis) initially increases and then stabilizes after reaching its peak, mirroring the number of unfolding operations, as illustrated in the right sub-table of Table~\ref{Tab:OD-Fold-Unfold}. 

}

\vspace{-6pt}
\subsection{Evaluating the $K$NN Processing Algorithms} \label{subsec:amt-knn-evaluation}
We use {ODIN-KNN} to collectively refer to our search solution including the {ODIN-KNN-Init} and the {ODIN-KNN-Inc} algorithms. 
For {ODIN-KNN}, the cost of processing one query is measured by taking the average response time over 10 consecutive search rounds, starting from the first round for each query, on a set of 1,000 queries ($\Gamma_q=1,000$). 

\vspace{-0.05in}
\subsubsection{Query processing cost}\label{subsec:qc-evaluation}
Fig.~\ref{exp:query-cost} depicts the elapsed time for identifying the $k$NNs for one query with varying OD, $k$, $P_o$, $\mu$, $z$, and $m$. 
{Fig.~\ref{fig:query-cost-OD} shows that the query cost increases as the OD decreases, but the rate of increase is much lower compared to the rate of decrease in OD. This is because a smaller OD implies fewer objects around the query vertex, causing ODIN-KNN to explore a larger region to identify the $k$NNs. Nevertheless, the expansion cost is limited by the adaptive index granularity in the explored regions provided by ODIN, resulting in a relatively small growth rate in the query time. Specifically, OD decreases by a factor of 960 from 3:1 (where almost every edge has an object) to 1:320, yet the query cost in each graph increases only by a maximum of 4 times.} Fig.~\ref{fig:query-cost-k} indicates that the query processing time grows with the increasing value of $k$, but the growth rate is sub-linear across a wide range of $k$ values, for the following reason. {Since the shortest distances from all objects in an active node to the query vertex $v_q$ can be identified when that node is explored and very often there is a large number of objects in the node, a slight increase in $k$ may not necessarily lead to an immediate increase in the number of active nodes to be explored by the search algorithm. As the number of active nodes explored is a dominating factor in the query processing cost, the processing time shows a sub-linear trend w.r.t. $k$.} \looseness=-1 

Fig.~\ref{fig:query-cost-Po} shows that the query processing time on each dataset increases slightly with a growing $P_o$. 
As more objects move on the road network, more live vertices that are evaluated in the previous round are likely to become obsolete and are thus not able to be reused in the current round. Meanwhile, more new live vertices appear in the current round to be explored. Both factors contribute to increased query processing time.


Fig.~\ref{fig:query-cost-mu}-\ref{fig:query-cost-m} depict the impact of varying $\mu$, $z$, $m$ on the query processing time. All of them share very similar trends.
The query processing time first declines as the values of these parameters increase, but the descending trend slows down when the parameter is above a certain threshold (e.g., {m=6}). The reason is that these parameters have direct or indirect relationships with the number of active nodes to be explored for given queries, while the number of active nodes evaluated is generally proportionate to the query processing time. In particular, the parameter $\mu$ denotes the lower bound on the number of live vertices in each set of active nodes with the same parent node, which dominates the number of explored active nodes for given queries. 
A smaller $\mu$ leads to more active nodes to be explored. Hence, the query processing time decreases with growing $\mu$, as shown in Fig.~\ref{fig:query-cost-mu}. As other parameters ($z$, $m$) all have positive correlations with $\mu$, they have similar influence on the query processing time as $\mu$. 

\begin{figure}[htbp] 
\centering
\vspace{-0.01cm}
\begin{subfigure}{0.225\textwidth}
      \centering   
\includegraphics[width=\linewidth]{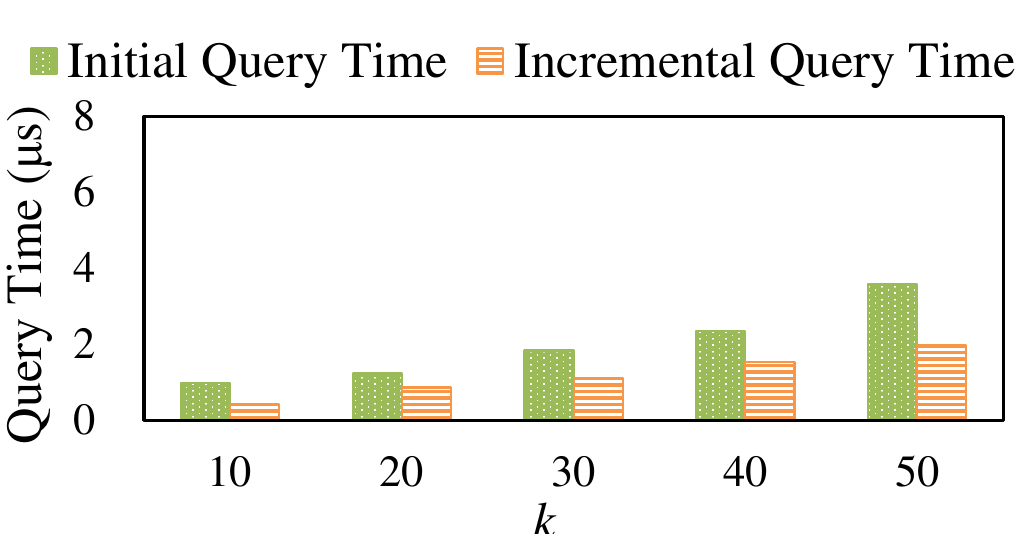}
      \vspace{-0.6cm}
        \caption{\scriptsize{NY (GD)}}
        \label{fig:query-cost-NY}
    \end{subfigure}       
    \begin{subfigure}{0.225\textwidth}
      \centering   
\includegraphics[width=\linewidth]{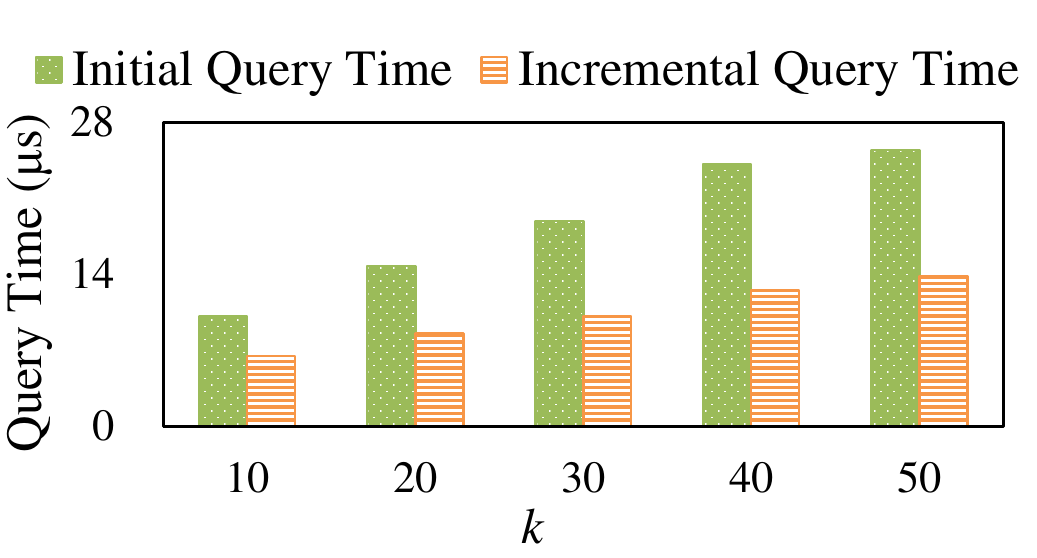}
      \vspace{-0.6cm}
        \caption{\scriptsize{CUSA (GD)}}
        \label{fig:query-cost-EUSA}
    \end{subfigure}
\vspace{-0.15cm}
\caption{
Initial and Incremental Query Costs
} \label{exp:initial-incremental}
\vspace{-0.2cm}
\end{figure}

\begin{figure}
\begin{subfigure}{0.243\textwidth}
      \centering         \includegraphics[width=\linewidth]{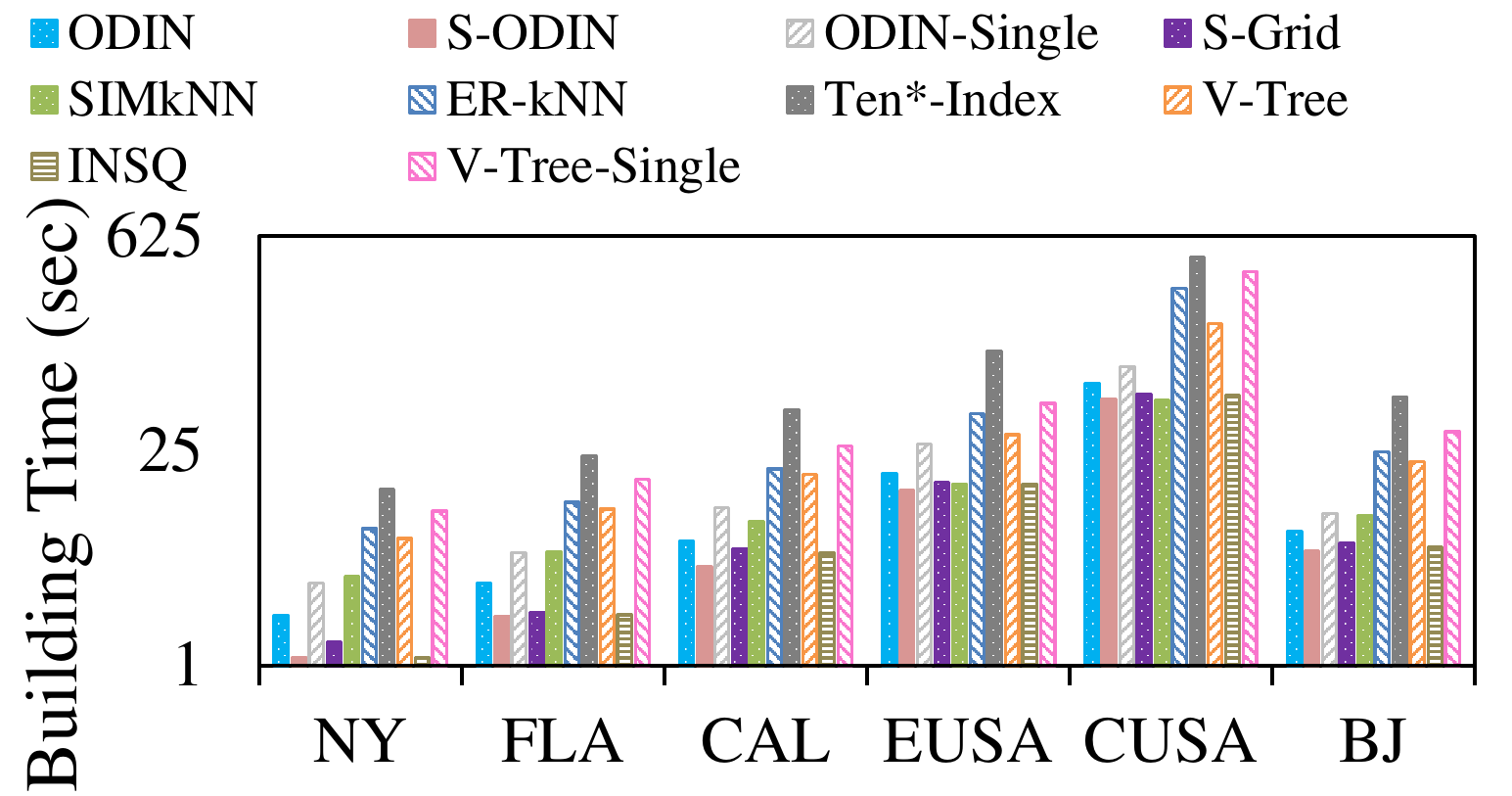}
      \vspace{-0.6cm}
    \caption{\scriptsize{Building Time}}
        \label{fig:consruction-cost-com}
         \vspace{-0.35cm}
    \end{subfigure} 
    \begin{subfigure}{0.243\textwidth}
      \centering     \includegraphics[width=\linewidth]{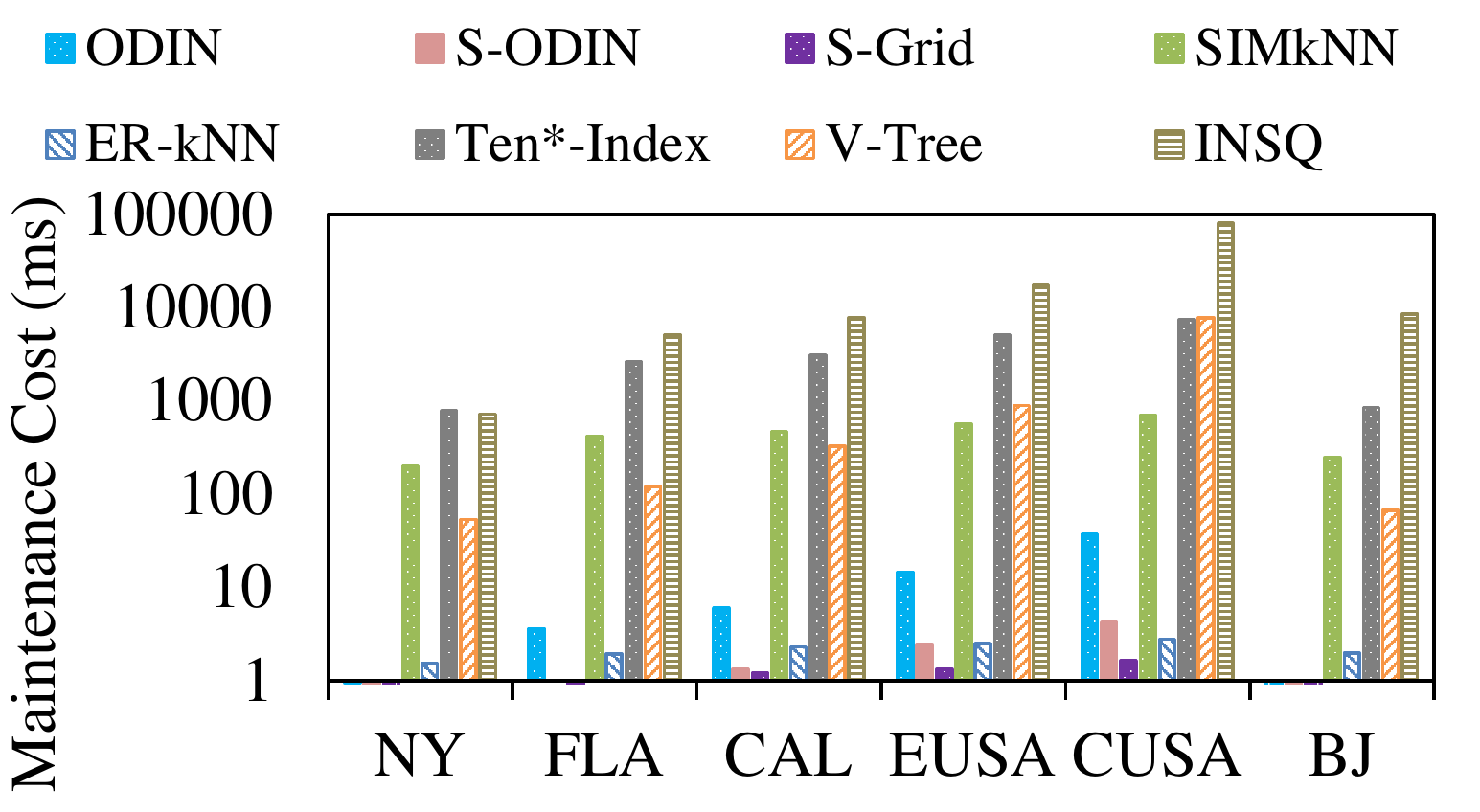}
      \vspace{-0.6cm}
\caption{\scriptsize{Maintenance Time}}
        \label{fig:maintenance-cost-com}
        \vspace{-0.35cm}
    \end{subfigure}
     \begin{subfigure}{0.243\textwidth}
      \centering   
\includegraphics[width=\linewidth]{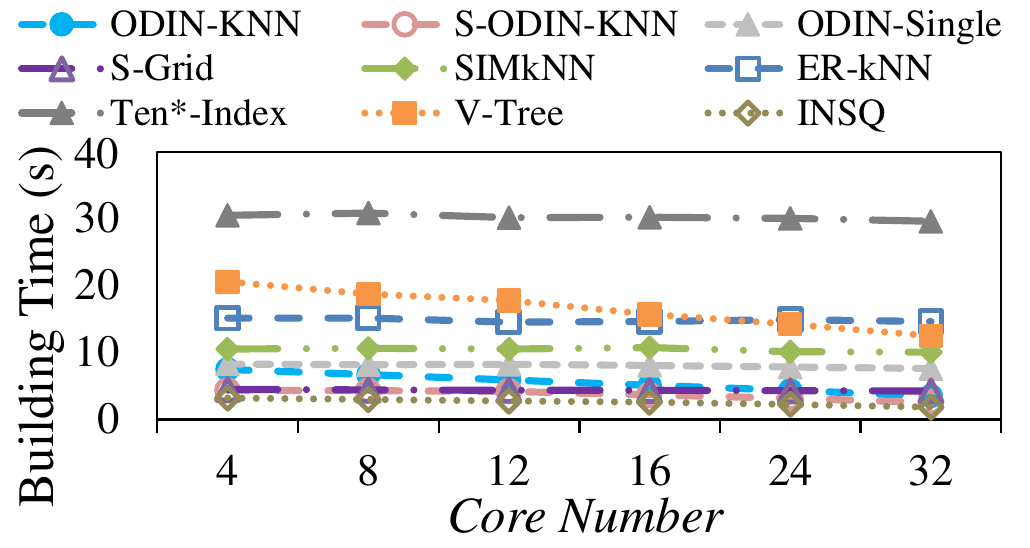}
      \vspace{-0.6cm}
    \caption{\scriptsize{Building time ( $\Gamma_c$)}}
        \label{fig:tu9c}
        \vspace{-0.35cm}
    \end{subfigure}  
     \begin{subfigure}{0.243\textwidth}
      \centering   \includegraphics[width=\linewidth]{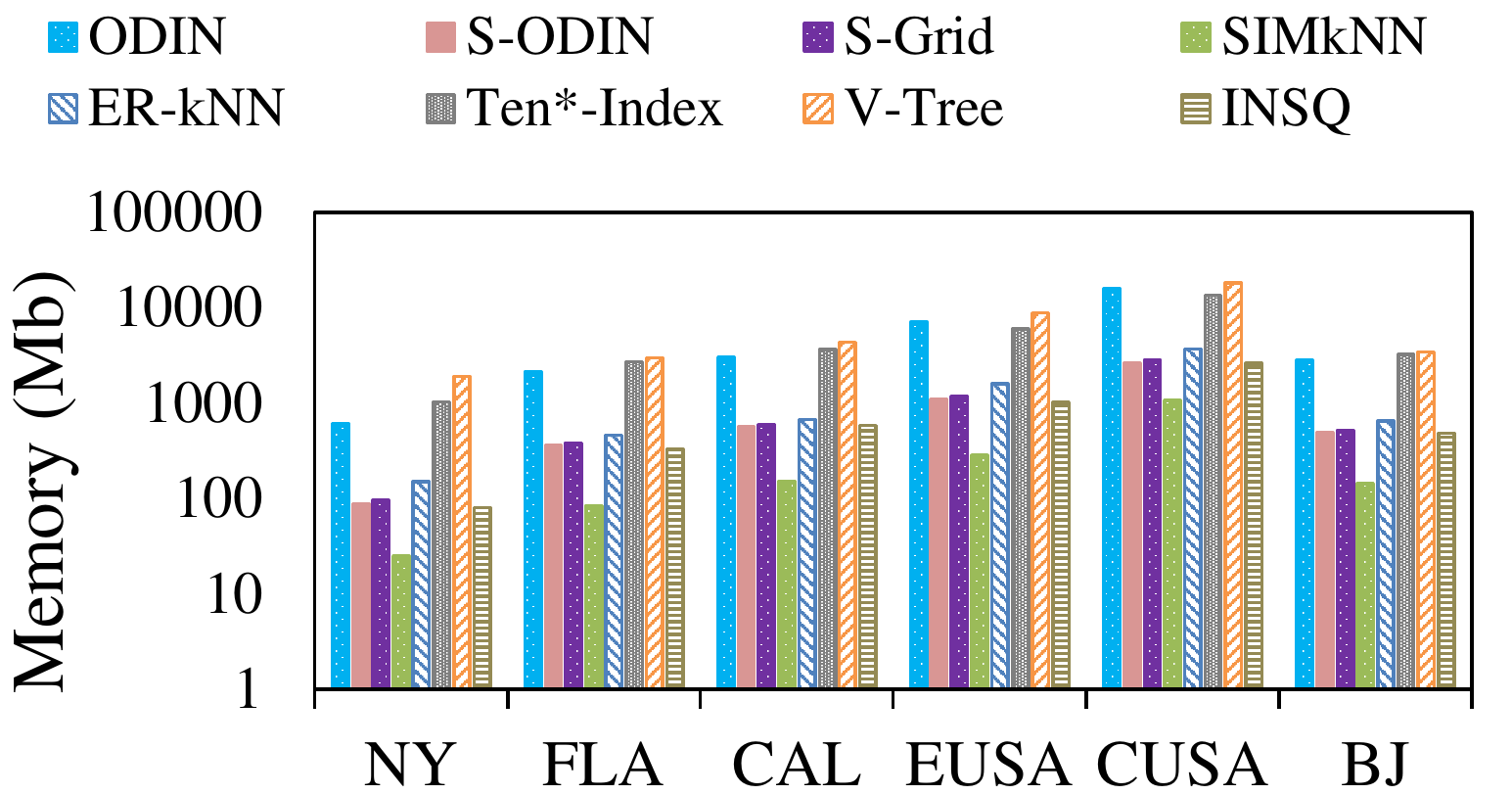}
       \vspace{-0.6cm}
\caption{\scriptsize{Memory Usage}}
        \vspace{-0.35cm}
        \label{fig:tu9d}
    \end{subfigure}   
\vspace{-0.2cm}
\caption{
Comparison of Index Building/Maintenance Time with Baselines
}\label{exp:construction-maintenance-comparison}
\vspace{0.2cm}
\end{figure}

\begin{figure*}[ht!]
  \centering
  \captionsetup{font={small}}
\begin{subfigure}{0.162\textwidth}
      \centering         \includegraphics[width=\linewidth]{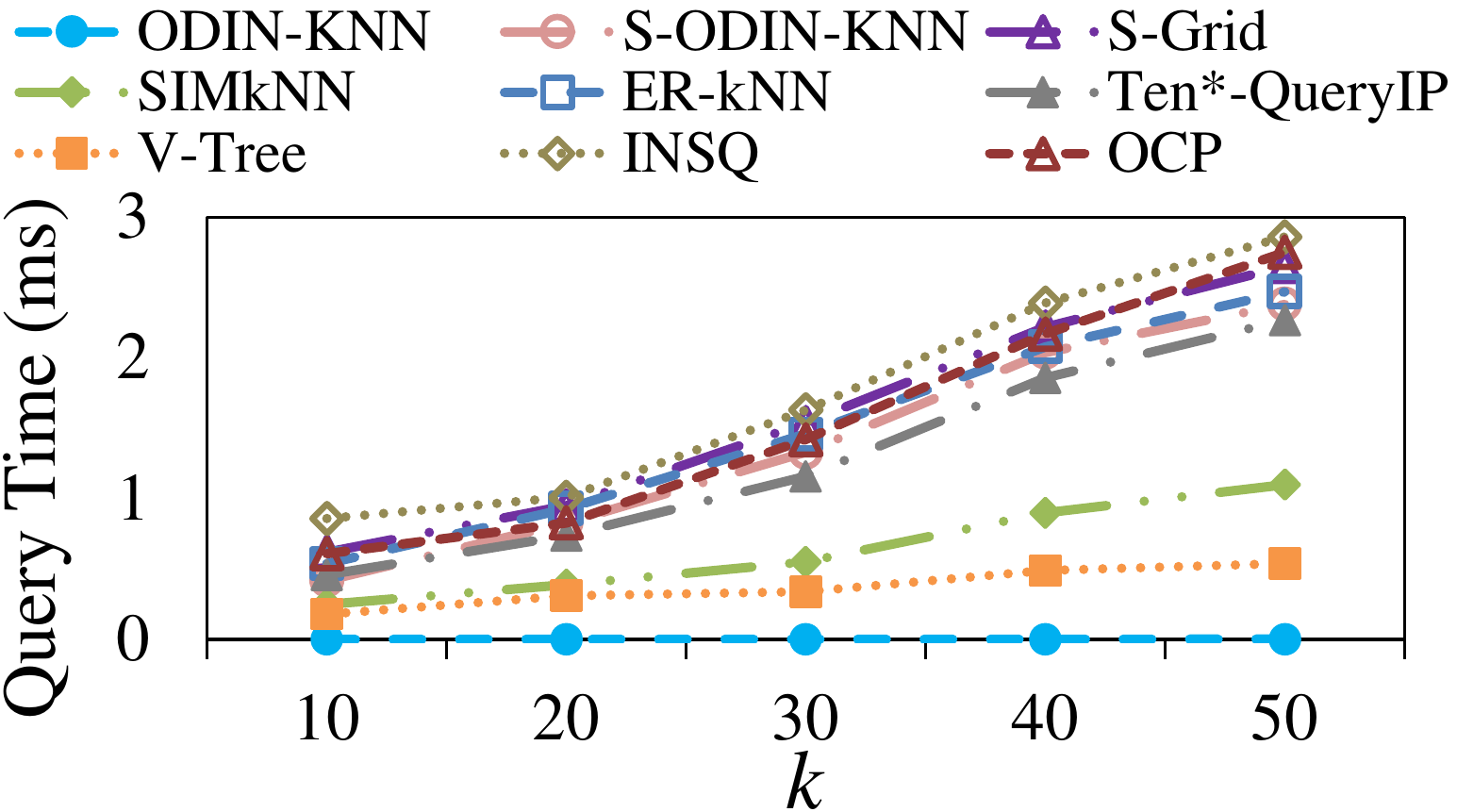}
      \vspace{-0.6cm}
    \caption{\scriptsize{Varying $k$ (NY)}}
        \label{fig:query-cost-comparison-k}
        \vspace{0.15cm}
    \end{subfigure}
\begin{subfigure}{0.162\textwidth}
      \centering         \includegraphics[width=\linewidth]{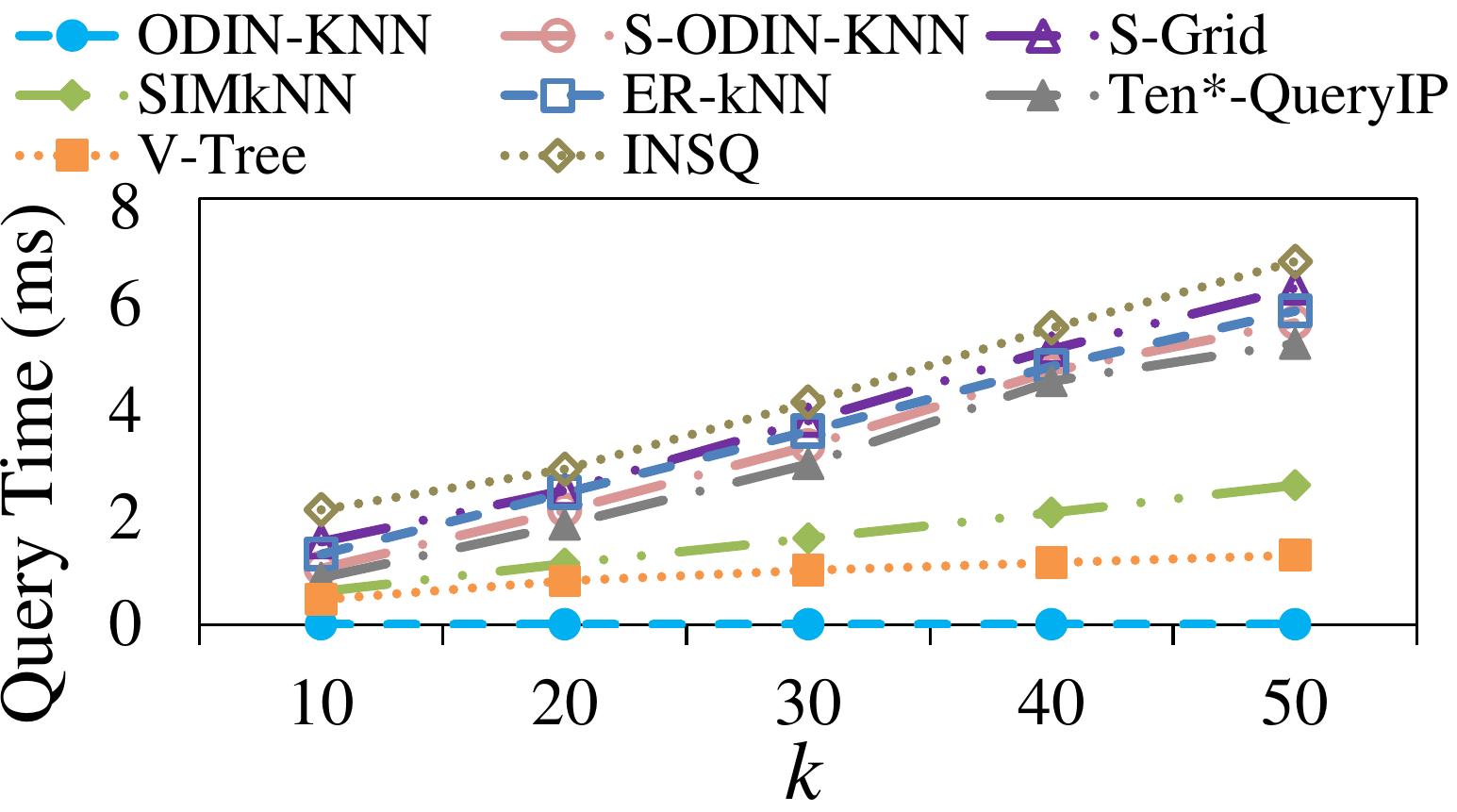}
      \vspace{-0.6cm}
    \caption{\scriptsize{Varying $k$ (FLA)}}
        \label{fig:query-cost-comparison-k}
        \vspace{0.15cm}
\end{subfigure}
\begin{subfigure}{0.162\textwidth}
      \centering   
\includegraphics[width=\linewidth]{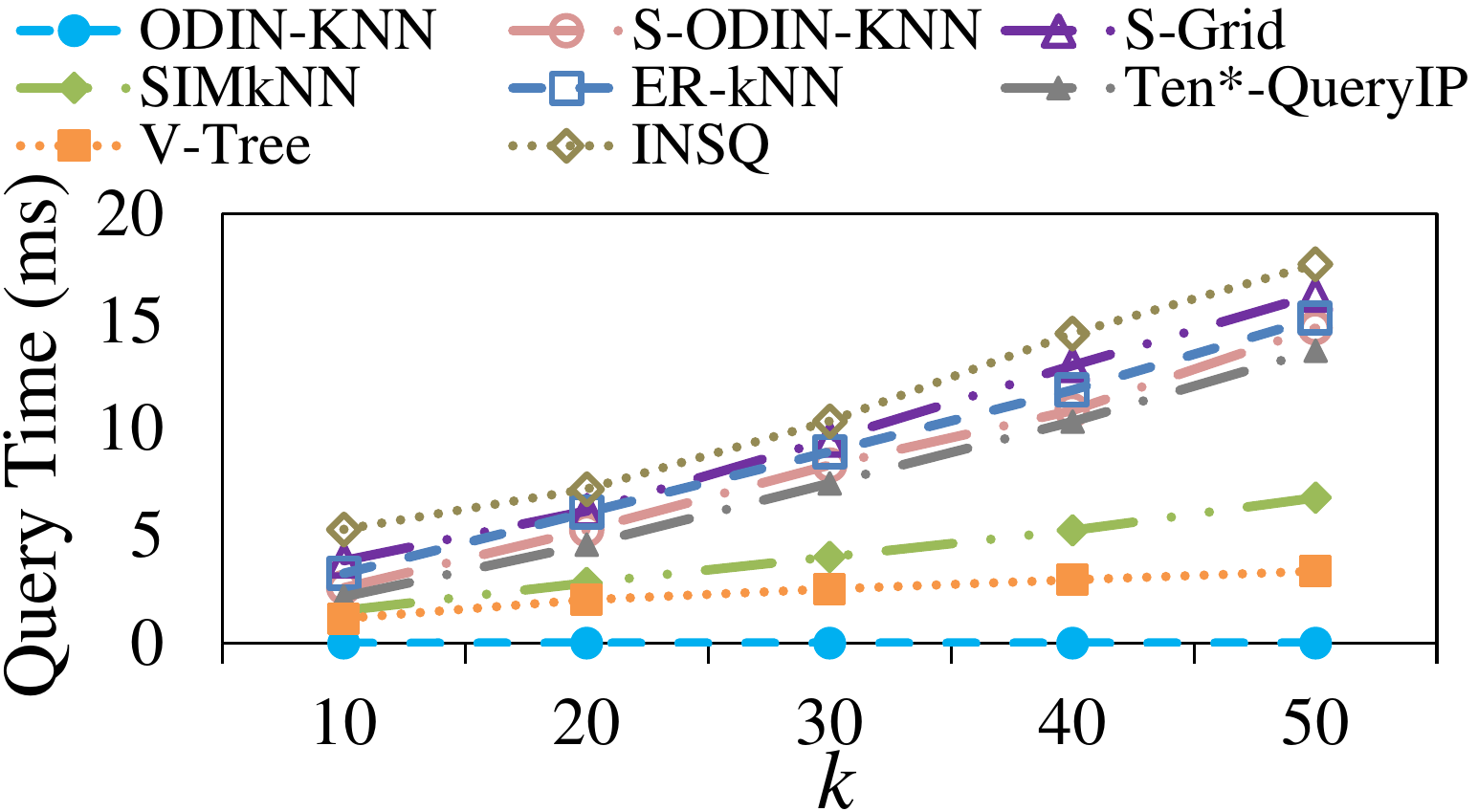}
      \vspace{-0.6cm}
    \caption{\scriptsize{Varying $k$ (EUSA)}}
        \label{fig:query-cost-comparison-k}
        \vspace{0.15cm}
    \end{subfigure}
\begin{subfigure}{0.162\textwidth}
      \centering   
\includegraphics[width=\linewidth]{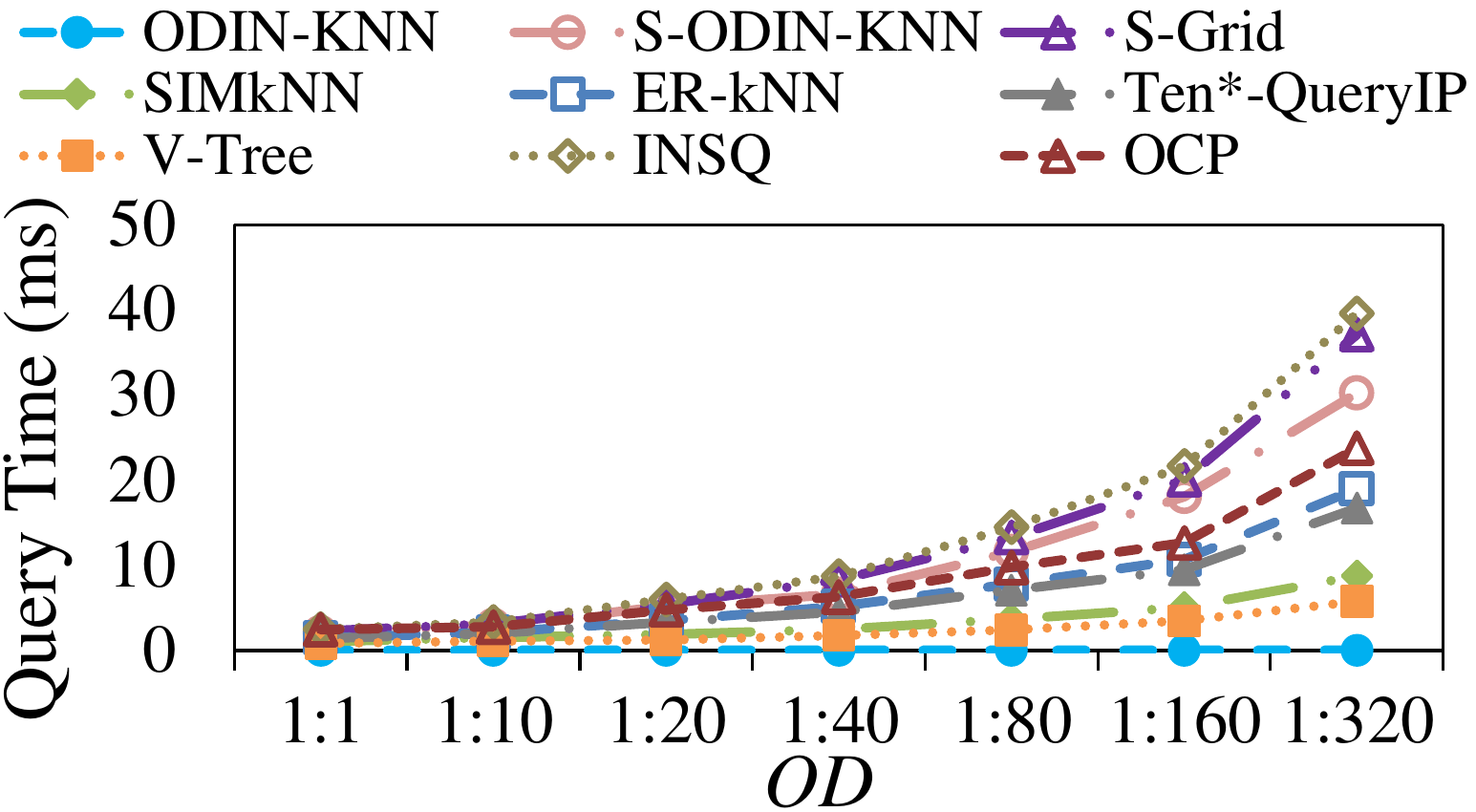}
      \vspace{-0.6cm}
    \caption{\scriptsize{Varying OD (NY)}}
        \label{fig:query-cost-comparison-d}
        \vspace{0.15cm}
    \end{subfigure} 
\begin{subfigure}{0.162\textwidth}
      \centering   
\includegraphics[width=\linewidth]{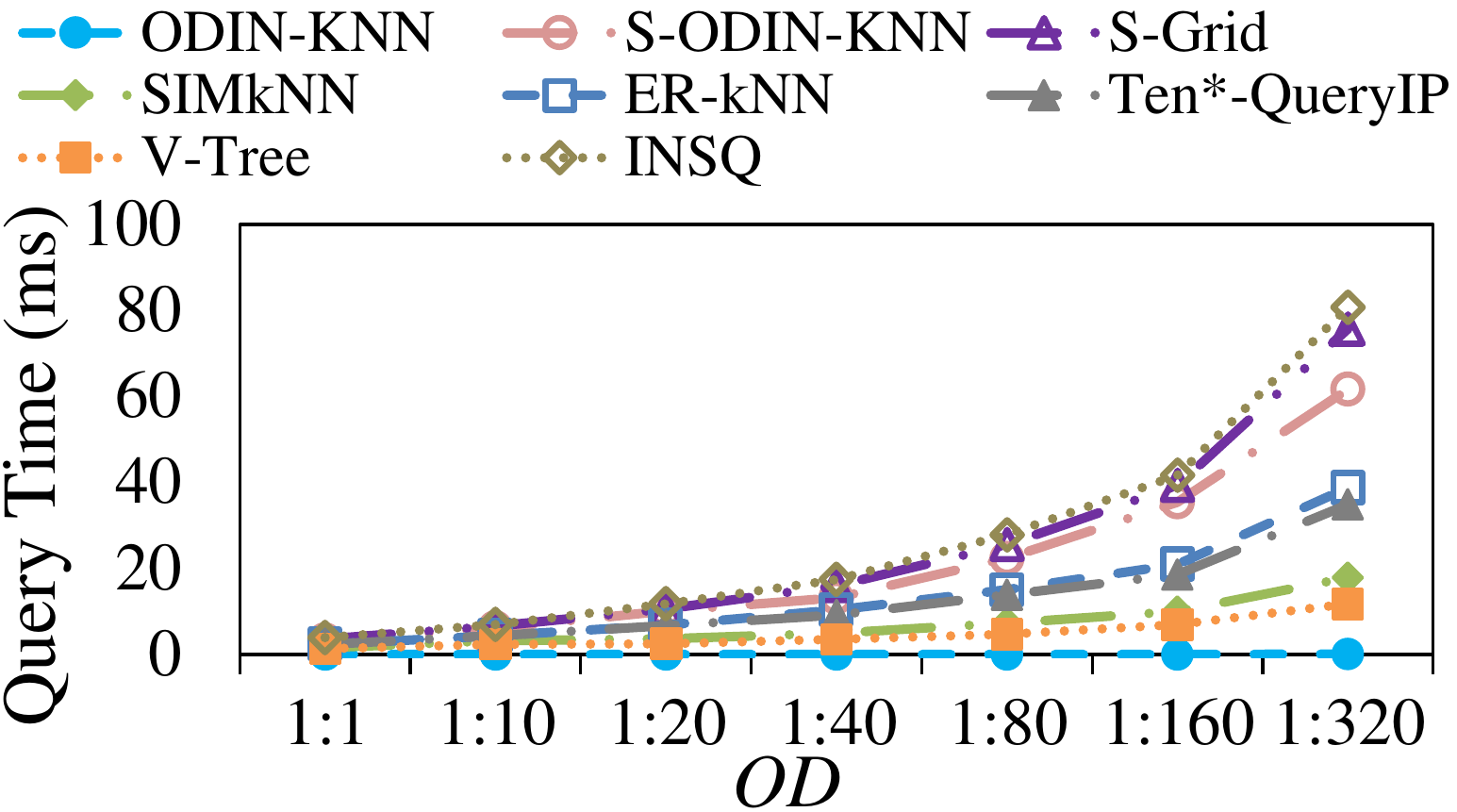}
      \vspace{-0.6cm}
    \caption{\scriptsize{Varying OD (FLA)}}
        \label{fig:query-cost-comparison-e}
        \vspace{0.15cm}
    \end{subfigure}   
\begin{subfigure}{0.162\textwidth}
      \centering   
\includegraphics[width=\linewidth]{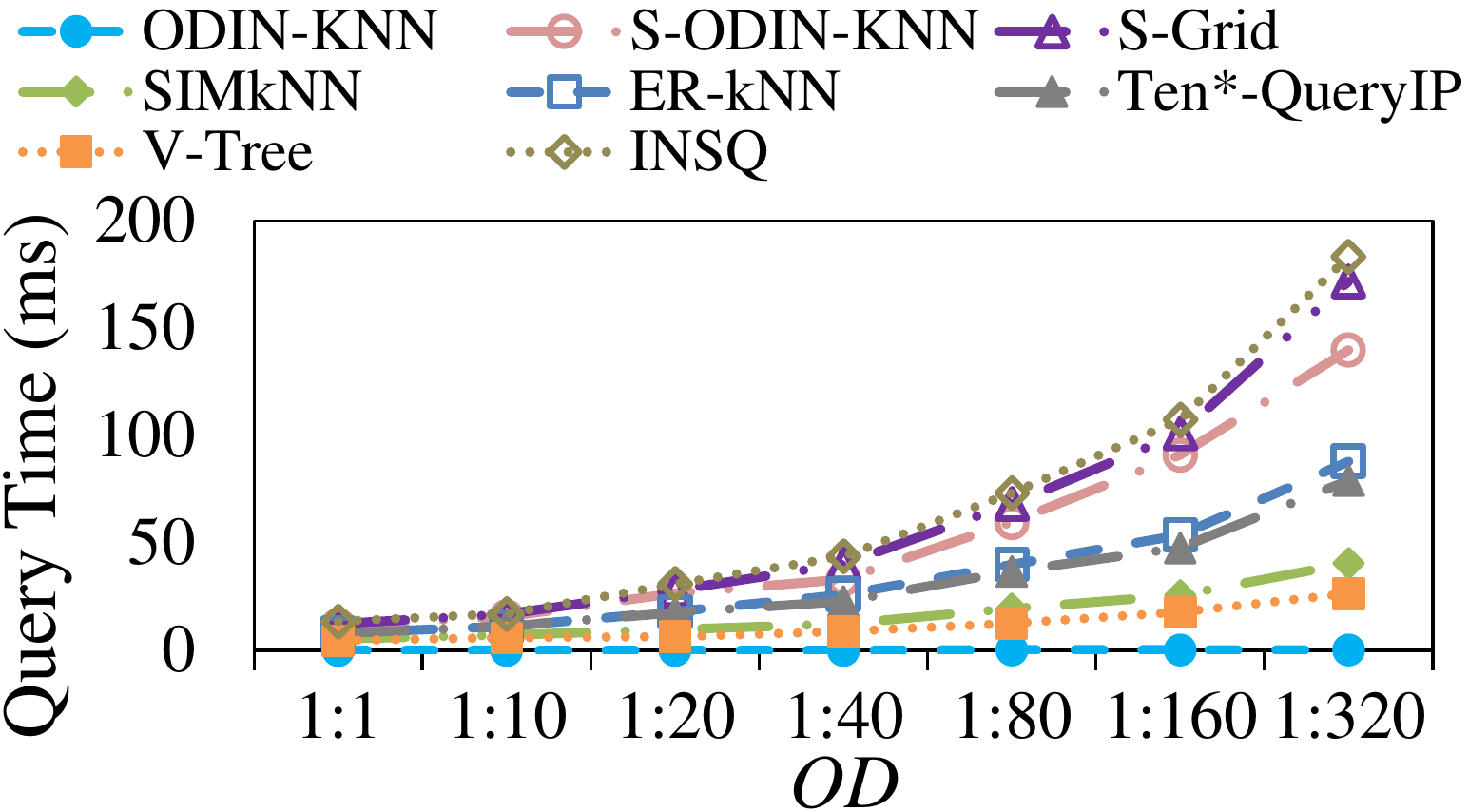}
      \vspace{-0.5cm}
\caption{\scriptsize{Varying OD (EUSA)}}
        \label{fig:query-cost-comparison-f}
        \vspace{0.15cm}
    \end{subfigure} 
\begin{subfigure}{0.162\textwidth}
      \centering         \includegraphics[width=\linewidth]{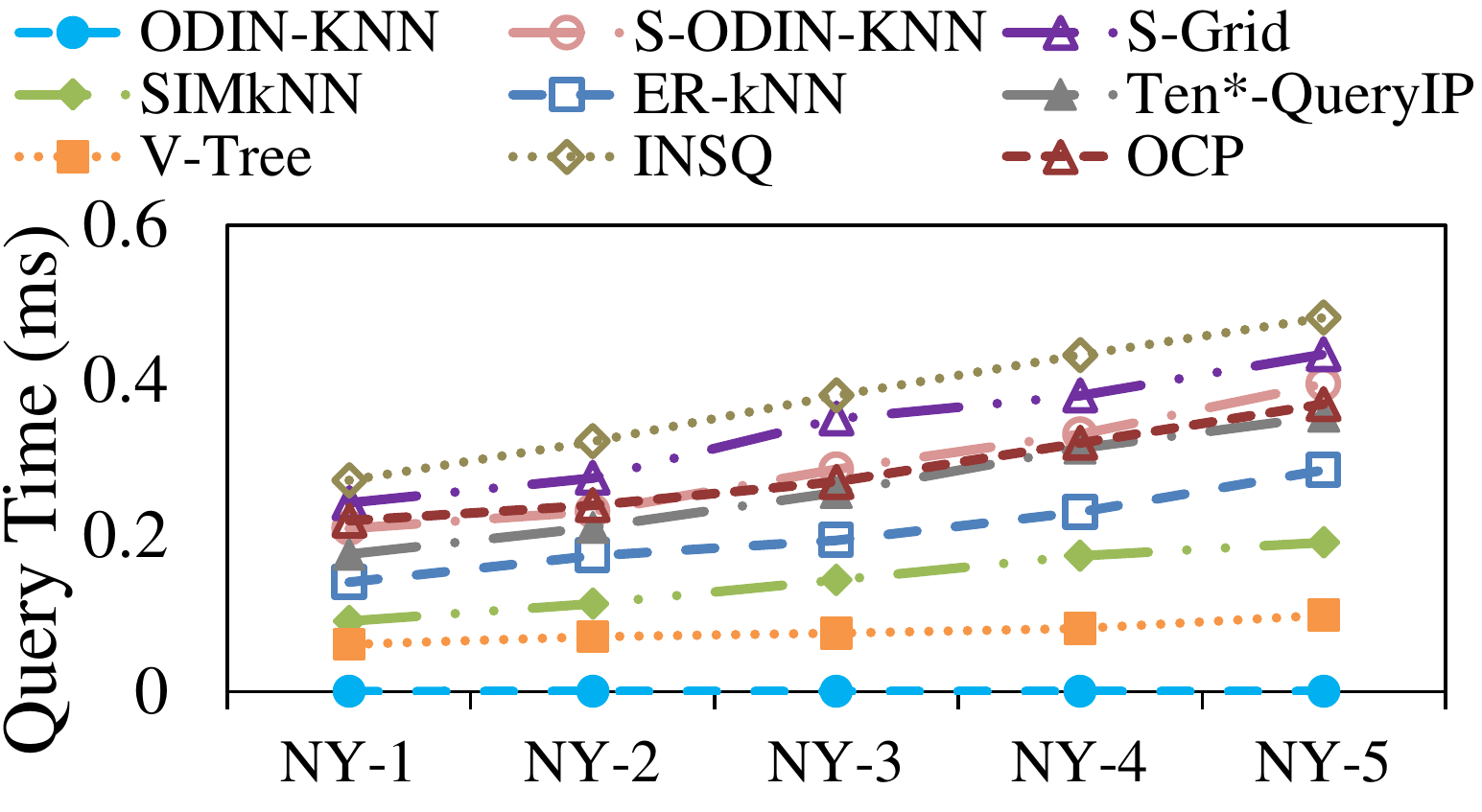}
      \vspace{-0.6cm}
    \caption{\scriptsize{Varying $\Gamma_c$ (NY)}}
        \label{fig:query-cost-comparison-scale-g}
    \end{subfigure}
\begin{subfigure}{0.162\textwidth}
      \centering   
\includegraphics[width=\linewidth]{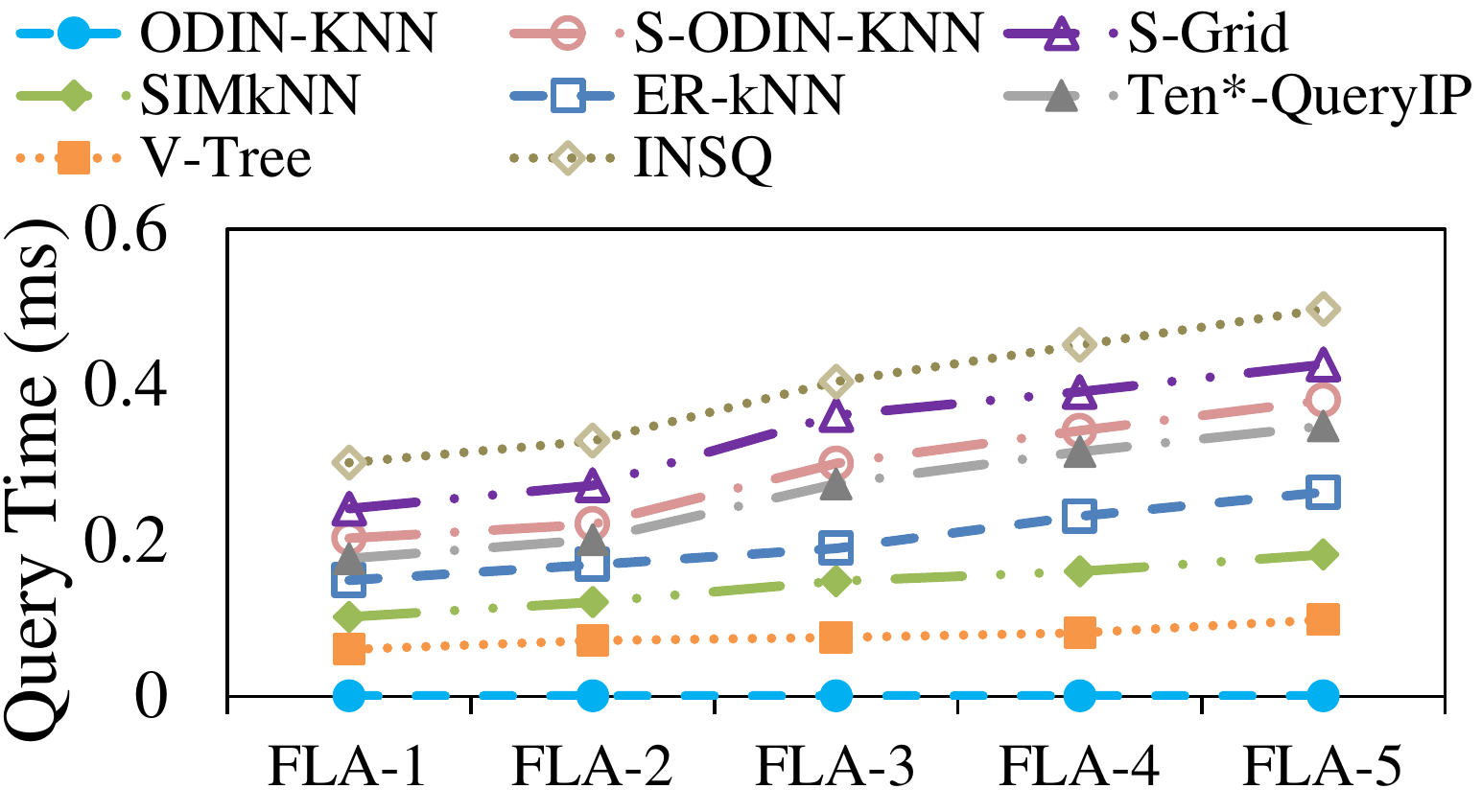}
      \vspace{-0.6cm}
    \caption{\scriptsize{Varying $\Gamma_c$ (FLA)}}
        \label{fig:query-cost-comparison-scale-h}
    \end{subfigure}
\begin{subfigure}{0.162\textwidth}
      \centering   
\includegraphics[width=\linewidth]{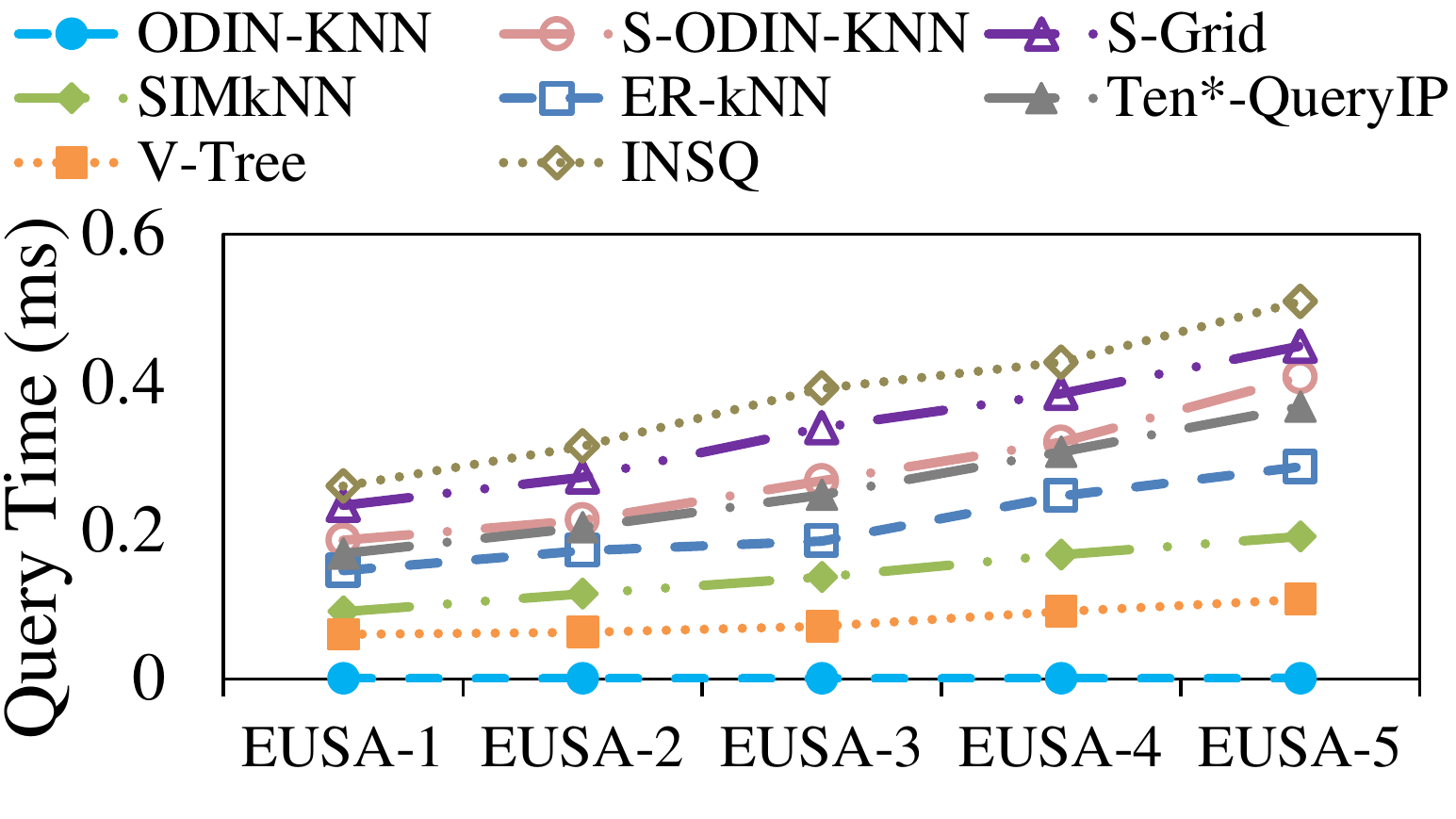}
      \vspace{-0.6cm}
    \caption{\scriptsize{Varying $\Gamma_c$ (EUSA)}}
        \label{fig:query-cost-comparison-scale-i}
    \end{subfigure}
\begin{subfigure}{0.162\textwidth}
      \centering   
\includegraphics[width=\linewidth]{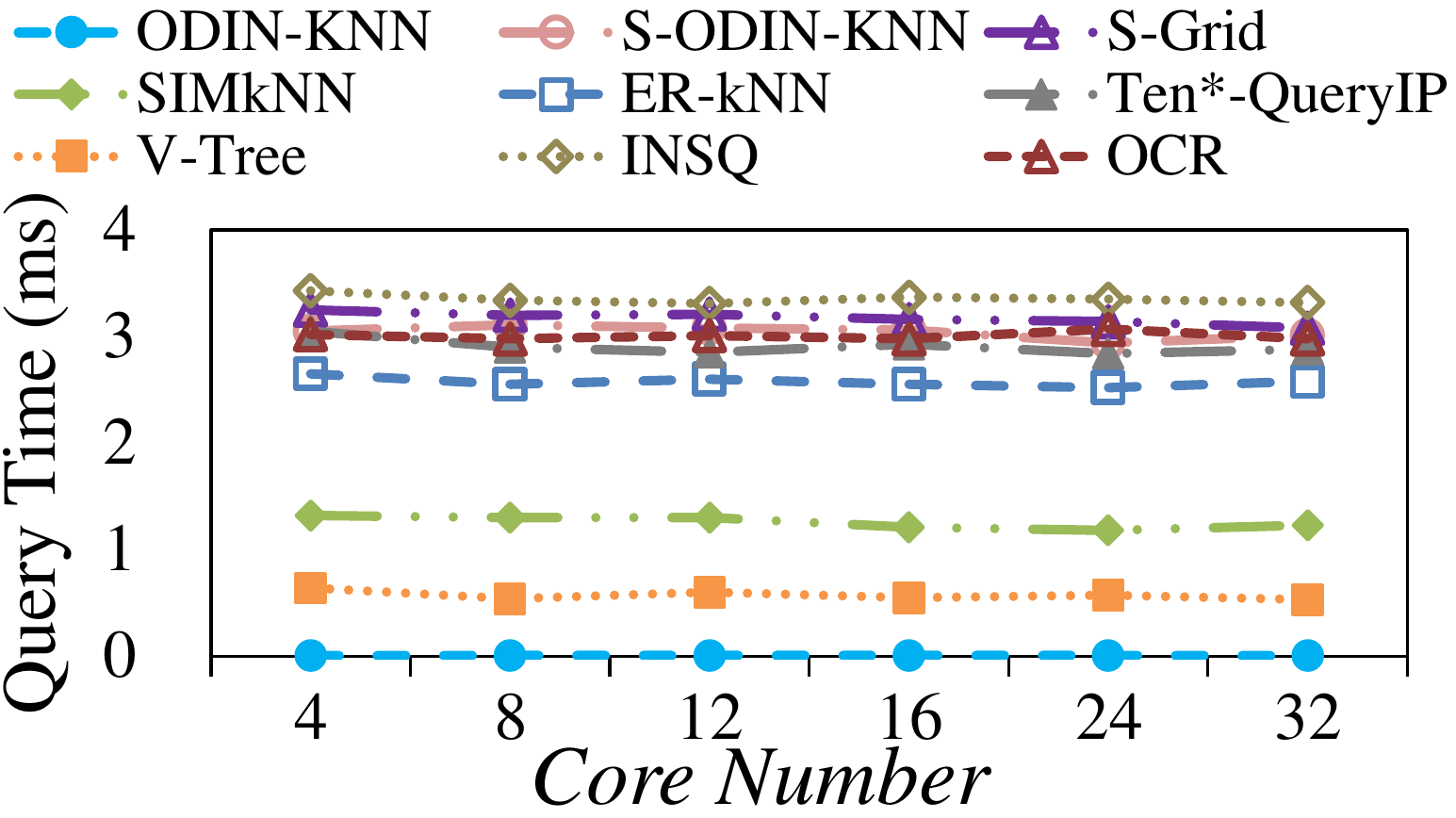}
      \vspace{-0.6cm}
    \caption{\scriptsize{Different Scales (NY)}}
        \label{fig:query-cost-comparison-core-number-j}
    \end{subfigure}
\begin{subfigure}{0.162\textwidth}
      \centering         \includegraphics[width=\linewidth]{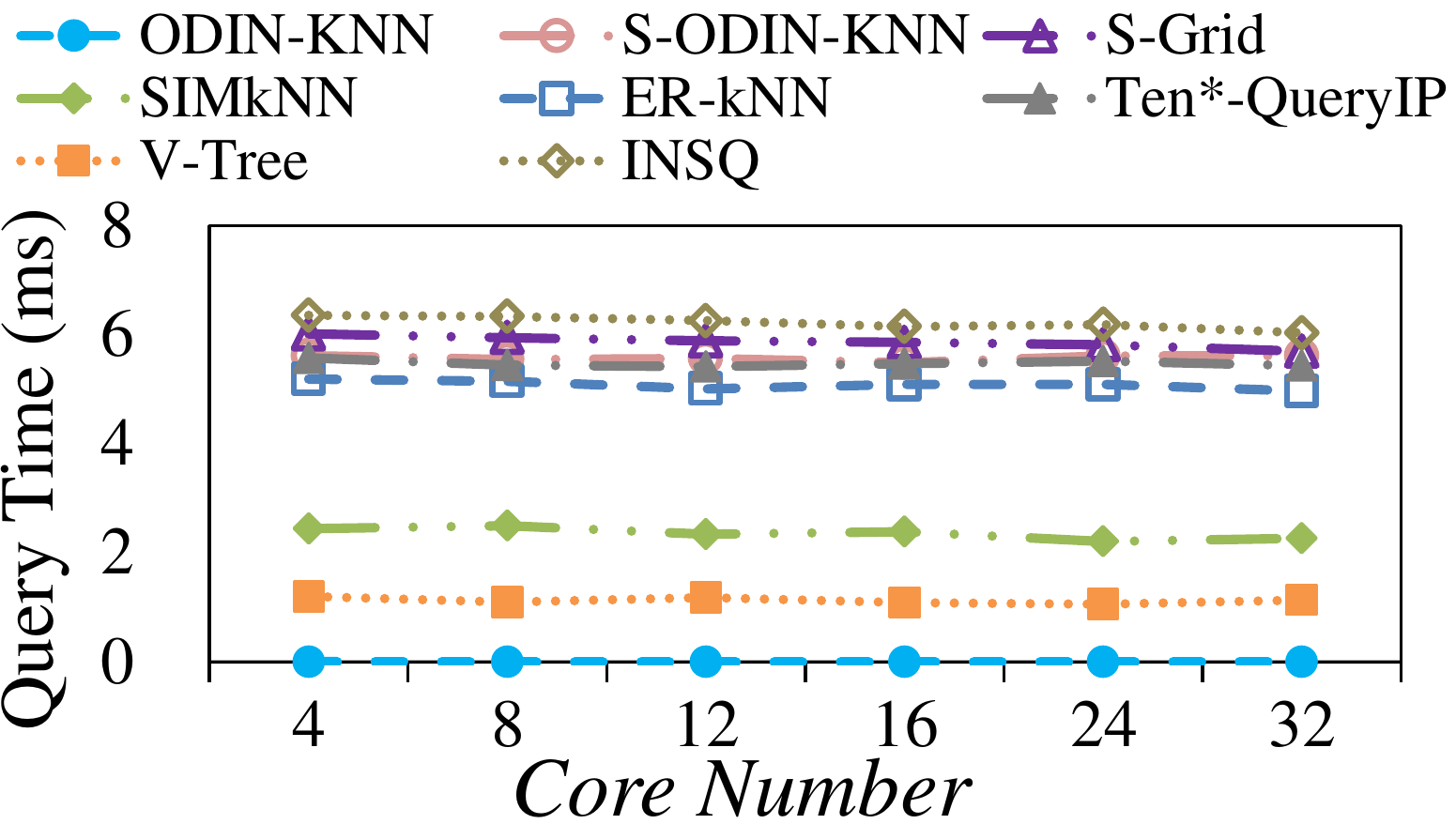}
      \vspace{-0.6cm}
    \caption{\scriptsize{Different Scales (FLA)}}
        \label{fig:query-cost-comparison-core-number-k}
    \end{subfigure}   
\begin{subfigure}{0.162\textwidth}
      \centering         \includegraphics[width=\linewidth]{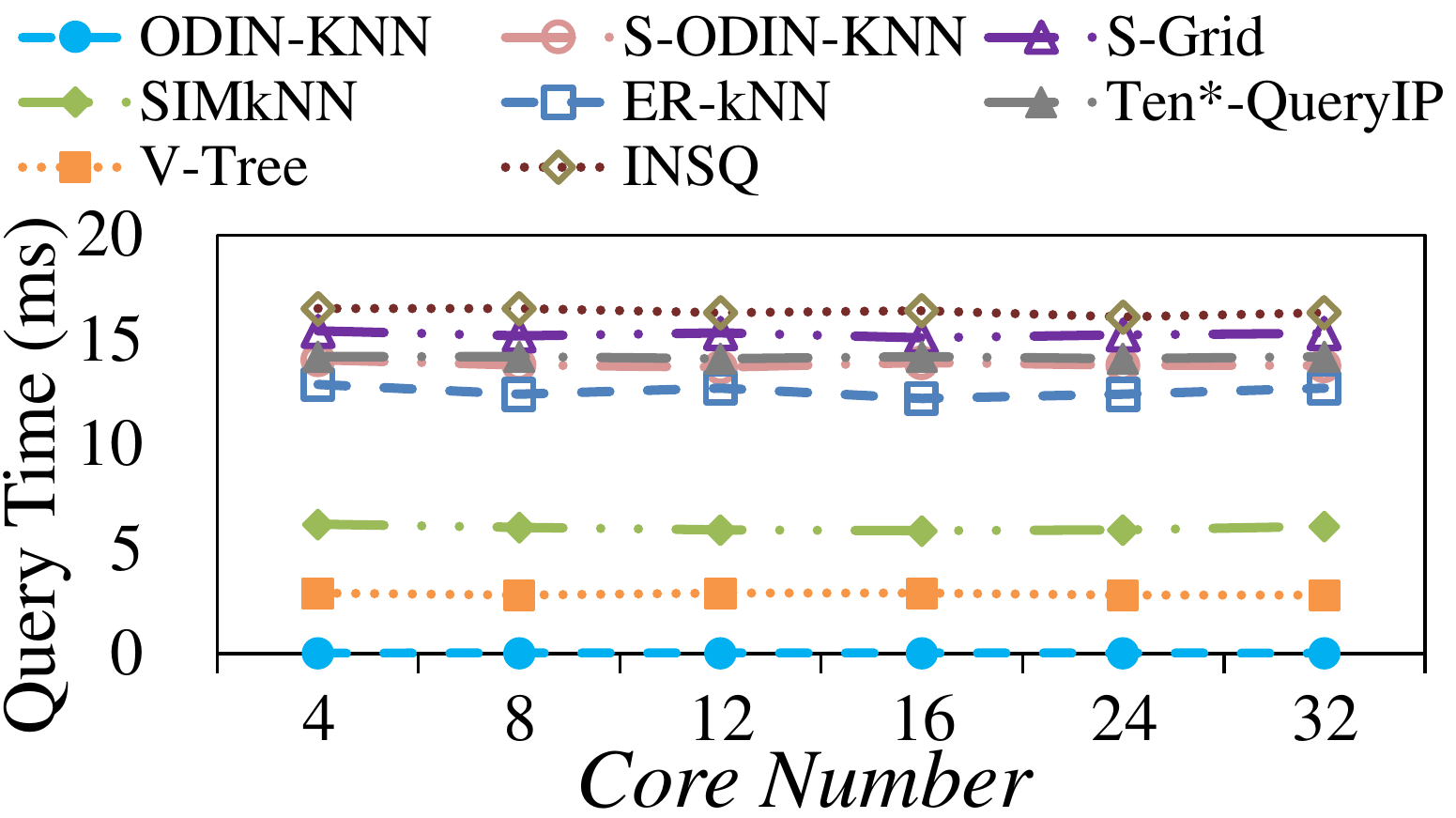}
      \vspace{-0.6cm}
    \caption{\scriptsize{Different Scales (EUSA)}}
        \label{fig:query-cost-comparison-core-number-l}
    \end{subfigure}
\vspace{-0.5cm}
\caption{
Comparison of Query Time with Baselines
}\label{exp:query-time-comparison}
\vspace{-0.5cm}
\end{figure*}

\vspace{-0.1cm}
\subsubsection{Initial vs. incremental processing}\label{subsec:ini-inc}

We compare the performance of {ODIN-KNN-Init} and {ODIN-KNN-Inc}, with results shown
in Fig.~\ref{exp:initial-incremental}. 
{We refer to the processing time of {ODIN-KNN-Init} on 1,000 queries as the {\em initial query time}, and the average time of updating $k$NNs over multiple search rounds for each query as the {\em incremental query time}.}
The results demonstrate that the incremental query time is significantly less than the initial query time on each dataset and the distinction is magnified as $k$ increases, thanks to the incremental mechanism that can reuse results from the previous search round. 
With larger $k$ values, the explored region in each round also expands, leading to more results to be reused by subsequent search rounds, which renders the advantage of {ODIN-KNN-Inc} more apparent.  \looseness=-1 



\vspace{-6pt}
\subsection{Comparison with baselines}\label{subsec:comparison-baseline}
We first evaluate the time to construct and maintain various indexes with the same set of objects on each graph {as well as the memory usage}, and the results are shown in Fig.~\ref{exp:construction-maintenance-comparison}. 
Compared with tree-based indexes including V-Tree and Ten$^*$-Index, ODIN {and ODIN-Single} perform significantly better on both construction and maintenance costs. This is because ODIN { and ODIN-Single} only materialize the skeleton graphs for a subset of the nodes and only a subset of the materialized skeleton graphs need to be maintained, while the other two indexes have to construct and maintain all index entities as objects evolve. ER-$k$NN has to traverse the entire road network to encode each edge no matter whether the edge is useful in a given cell for the query processing during the construction, leading to higher construction cost than ODIN. {When we compare the construction cost of all indexes with a varying number of cores, we find that except for ODIN and V-tree, for which the building time slightly declines, the performance of other indexes remains stable. }
For the maintenance cost, SIM$k$NN must continuously split and merge the partitioned cells with a complex strategy, {whereas INSQ requires recomputation of the Voronoi cells centered around each object as they evolve, so both of them underperform ODIN on the maintenance cost.} S-Grid have lower maintenance cost than ODIN due to the simplicity of its structure. However, since ODIN outperforms S-Grid by several orders of magnitude in query time as will be shown below, the difference in building and maintenance cost has minimal impact on the overall performance. For example, under the typical setting on FLA in previous work \cite{shen2017v} where the ratio of updates to queries is approximately 30:1, the maintenance cost accounts for less than 1\% of the total computation cost in a search round. {Without the shortcuts between live and border vertices within materialized nodes, S-ODIN {offers} less than an order of magnitude {of saving over ODIN} on building time, maintenance cost, and memory usage, as illustrated in Fig. \ref{exp:construction-maintenance-comparison}. However, S-ODIN-KNN, the $k$NN search algorithm based on S-ODIN, is less efficient than ODIN-KNN by almost three orders of magnitude as depicted by Fig.~\ref{exp:query-time-comparison}. This clearly validates the benefits of the shortcuts omitted by S-ODIN.}
{Fig.~\ref{exp:query-time-comparison} shows the impact of different parameters on the query performance of all algorithms. As a special case, we just conduct the comparison with OCP in NY as its two matrices keeping the shortest distances between vertices and the ``distance determinate'' edges do not fit in the memory for the other two graphs. The query processing time of all algorithms increases as $k$ grows in each dataset. Among them {ODIN-KNN} performs the best irrespective of $k$ values. The runner-up, V-Tree, is slower than {ODIN-KNN} by two orders of magnitude. Fig.~\ref{fig:query-cost-comparison-e}-\ref{fig:query-cost-comparison-f} shows that the query time for all algorithms decreases as OD increases. Yet, ODIN-KNN consistently outperforms and exhibits the lowest growth rate, benefiting from ODIN's adaptive index granularity in response to varying OD. Next, we evaluate the scalability of each algorithm concerning the scale of the graph. We choose five subgraphs from NY, FLA, and EUSA with 10K--50K vertices respectively and equip each subgraph with the same number of objects. As such, a larger subgraph has a smaller OD, which makes the query time of all algorithms increase with the growing size of the graph, but ODIN-KNN has the lowest rate of increase, as depicted by Fig.~\ref{fig:query-cost-comparison-scale-g}-\ref{fig:query-cost-comparison-scale-i}. Fig.~\ref{fig:query-cost-comparison-core-number-j}-\ref{fig:query-cost-comparison-core-number-l} shows that ODIN always outperforms the baselines by at least two orders of magnitude regardless of the number of cores used.} 

\vspace{-0.1cm}
\section{Conclusions}\label{sec:con}
This work centers on processing C$k$NN queries for moving objects on road networks. We introduce ODIN, an elastic tree index that adapts to varying ODs through folding/unfolding operations. The indexing of live vertices and border vertices as well as the associated distance information facilitate the computation of the shortest distance from objects to the given query point. We also present two algorithms: ODIN-KNN-Init, for initial $k$NN computation, and ODIN-KNN-Inc, for incremental $k$NN updates based on cached results from the previous round. Finally, extensive experiments demonstrate our approach's superiority over state-of-the-art methods. 


\section{Acknowledgement}
This work was supported by NSFC Grants [No.62172351], NSERC Discovery Grants [RGPIN-2022-04623], Populus Innovation Research Funding [CCF-HuaweiDB202301].
\balance
\bibliographystyle{IEEEtran}
\bibliography{reference}
\vspace{-30pt}
\begin{IEEEbiography}[{\vspace{-35pt}\includegraphics[width=1in,height=1in,clip,keepaspectratio]{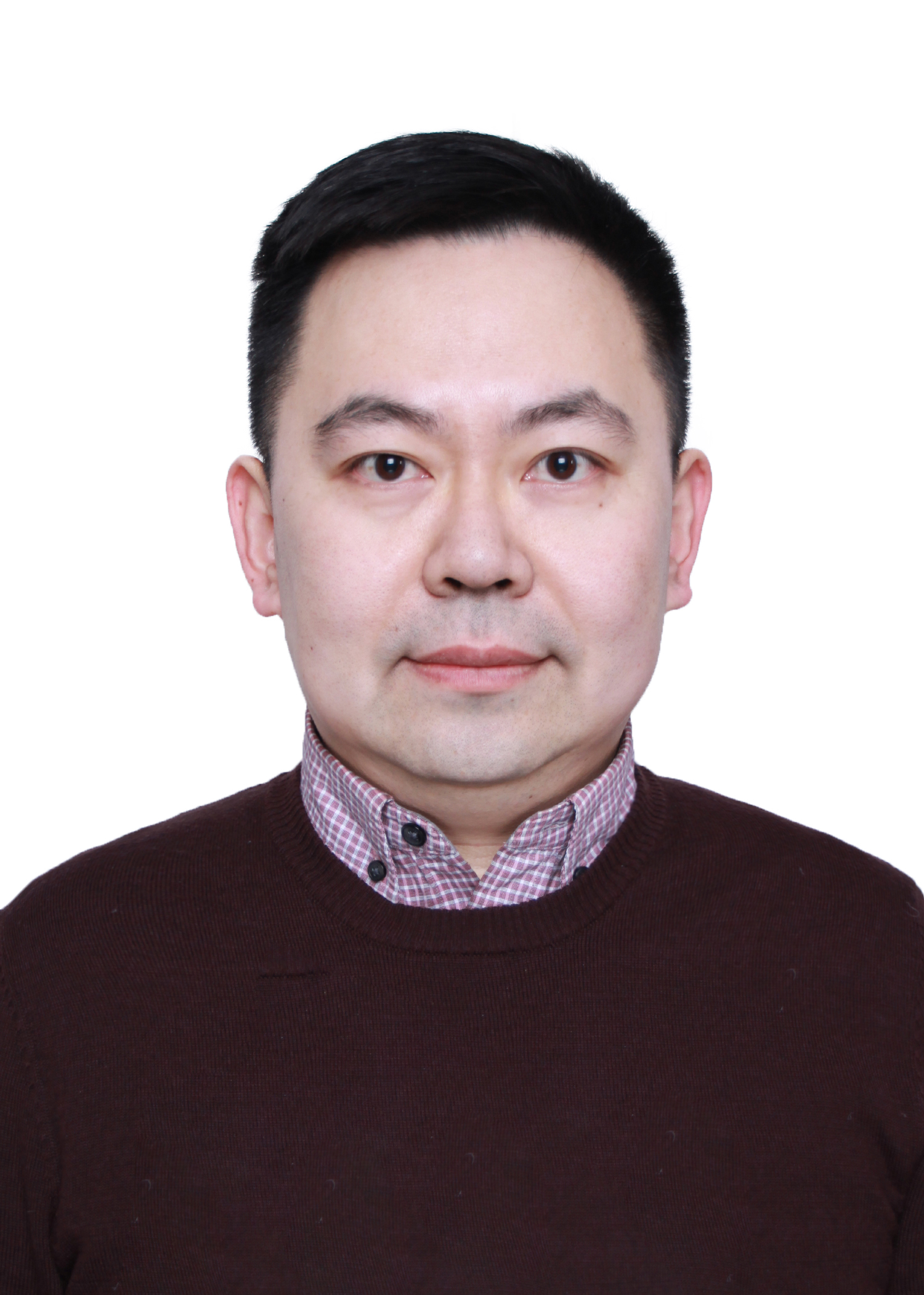}}]
{Ziqiang Yu} received his Ph.D. degree in computer science in 2015 from Shandong University, China. He is currently an associate professor in the School of Computer and Control Engineering, Yantai University, China. His research interests mainly focus on spatial-temporal data processing and graph computing. 
\end{IEEEbiography}

\vspace{-60pt}
\begin{IEEEbiography}[{\vspace{-35pt}\includegraphics[width=1in,height=1in,clip,keepaspectratio]{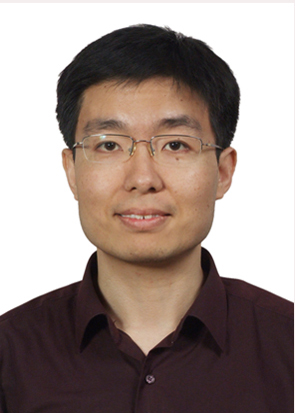}}]{Xiaohui Yu}
received his Ph.D. degree in computer science in 2006 from the University of Toronto, Canada. He is  an associate professor in the School of Information Technology, York University, Toronto.  His research interests are in the areas of database systems and data mining.
\end{IEEEbiography}

\vspace{-56pt}
\begin{IEEEbiography}[{\vspace{-35pt}\includegraphics[width=1in,height=1in,clip,keepaspectratio]{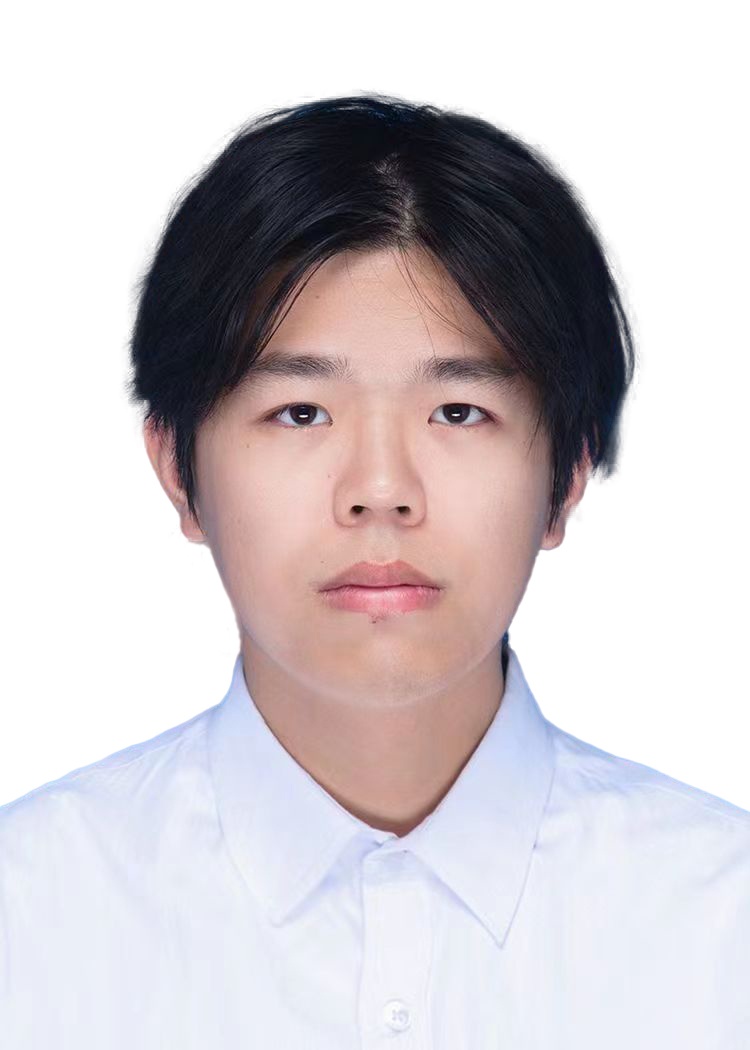}}]{Tao Zhou} is studying for a master's degree in school of software, University of Science and Technology of China. He is also an intern at Microsoft Suzhou Technology Co., Ltd. His research interests mainly focus on the query processing over spatial-temporal data and database systems.
\end{IEEEbiography}

\vspace{-58pt}
\begin{IEEEbiography}[{\vspace{-35pt}\includegraphics[width=1in,height=1in,clip,keepaspectratio]{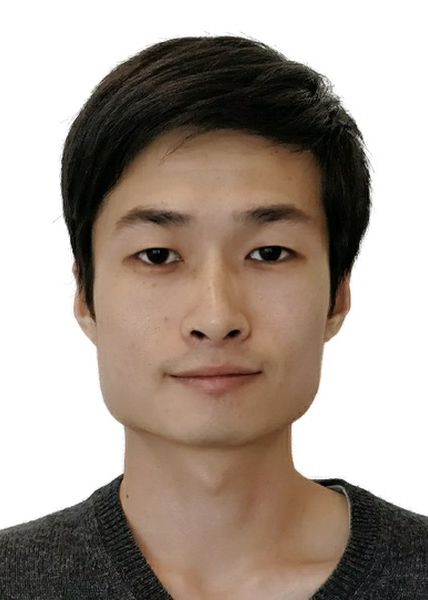}}]{Yueting Chen}
received his Master degree in computer science in 2017 from Shandong University, China. He is now a Ph.D. candidate in Computer Science at York university starting in 2017, supervised by Xiaohui Yu. His research interests mainly focus on databases and data management systems.
\end{IEEEbiography}

\vspace{-56pt}
\begin{IEEEbiography}[{\vspace{-35pt}\includegraphics[width=1in,height=1in,clip,keepaspectratio]{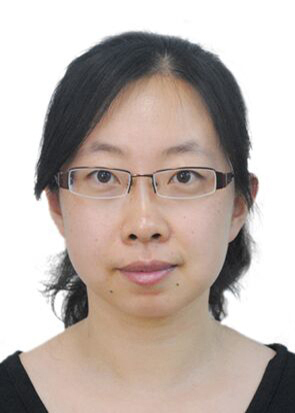}}]{Yang Liu}
received her Ph.D. degree in computer science and engineering in 2008 from York University, Canada. She is currently an associate professor in the Department of Physics and Computer Science, Wilfrid Laurier University, Canada. Her main areas of research are data mining and information retrieval.
\end{IEEEbiography}

\vspace{-56pt}
\begin{IEEEbiography}[{\vspace{-35pt}\includegraphics[width=1in,height=1in,clip,keepaspectratio]{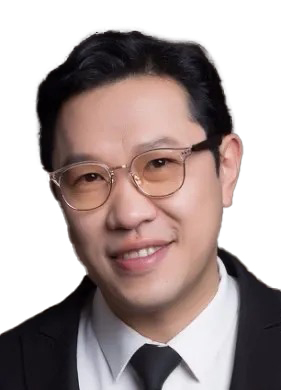}}]{Bohan Li}
received his Ph.D. degree from Harbin University of Science and Technology, China. He is currently an associate professor at Nanjing University of Aeronautics and Astronautics, China. His research interests include spatial-temporal database and knowledge graph.
\end{IEEEbiography}

\ifCLASSOPTIONcompsoc


\ifCLASSOPTIONcaptionsoff
  \newpage
\fi

\end{document}